\documentclass[nojss,shortnames]{jss}

\usepackage{thumbpdf,lmodern}

\usepackage{framed}
\usepackage{graphicx}

\usepackage{amsmath}
\usepackage{verbatim}

\usepackage{setspace}
\usepackage{subcaption}
\usepackage{booktabs}
\usepackage{multirow}
\usepackage{amssymb}
\usepackage{makecell}

\def\mm#1{\ensuremath{\boldsymbol{#1}}} 

\def\tcircled#1{%
	\leavevmode\hbox{\setbox0=\hbox{#1}\relax
	\dimen0=\the\wd0\dimen1=\the\ht0
	\ifdim\dimen0>\dimen1\dimen2=\dimen0\else\dimen2=\dimen1\fi
	\divide\dimen0 by 2\divide\dimen1 by 2\advance\dimen2 by 5pt
	\put(\strip@pt\dimen0, \strip@pt\dimen1){\circle{\strip@pt\dimen2}}
	\box0}%
}

\newcommand{\gX}{\mathcal X} 
\newcommand{\gL}{\mathcal L} 
\newcommand{\gN}{\mathcal N} 
\newcommand{\gD}{\mathcal D} 
\newcommand{\gE}{\mathcal E} 
\newcommand{\gP}{\mathcal P} 

\newcommand{\Var}{\text{Var}} 

\newcommand{\bmu}{{\mm \mu}}

\newcommand{\btheta}{{\mm \theta}}

\newcommand{\bn}{\mm{n}}

\author{Lisa Amrhein\\Helmholtz Zentrum M\"unchen \\ Technical University Munich
   \And Christiane Fuchs\\ Bielefeld University\\Helmholtz Zentrum M\"unchen\\ Technical University Munich }
\Plainauthor{Lisa Amrhein, Christiane Fuchs}

\title{stochprofML: Stochastic Profiling Using Maximum Likelihood Estimation in \proglang{R}}
\Plaintitle{stochprofML: Stochastic Profiling Using Maximum Likelihood Estimation in R}
\Shorttitle{stochprofML in \proglang{R}}

\Abstract{
\noindent Tissues are often heterogeneous in their single-cell molecular expression, and this can govern the regulation of cell fate. For the understanding of development and disease, it is important to quantify heterogeneity in a given tissue. We introduce the \proglang{R} package \pkg{stochprofML} which is designed to parameterize heterogeneity from the cumulative expression of small random pools of cells. This method outweighs the demixing of mixed samples with a saving in cost and effort and less measurement error. The approach uses the maximum likelihood principle and was originally presented in \cite{bajikar_parameterizing_2014}; its extension to varying pool sizes was used in \cite{tirier_pheno-seq_2018}. We evaluate the algorithm's performance in simulation studies and present further application opportunities.
}

\Keywords{stochprofML, stochastic profiling, gene expression, cell-to-cell hetereogeneity, mixture models, \proglang{R}}
\Plainkeywords{stochprofML, stochastic profiling, gene expression, cell-to-cell hetereogeneity, mixture models, R}

\Address{
Lisa Amrhein\\
  Institute of Computational Biology\\
  Helmholtz Zentrum M\"unchen, Germany\\
  \emph{and}\\
  Department of Mathematics\\
  Technical University Munich, Germany\\
  Email: \email{lisa.amrhein@helmholtz-muenchen.de}\\
  \newline
  Christiane Fuchs\\
  Faculty of Business Administration and Economics\\
  Bielefeld University, Germany\\
  \emph{and}\\
  Institute of Computational Biology\\
  Helmholtz Zentrum M\"unchen, Germany\\
  \emph{and}\\
  Department of Mathematics\\
  Technical University Munich, Germany\\
  Email: \email{christiane.fuchs@uni-bielefeld.de}\\
URL: \url{ https://tinyurl.com/CFuchslab }\\
}
\begin{document}


\section[Introduction: Stochastic Profiling]{Introduction: Stochastic Profiling} \label{sec:intro}

Tissues are built of cells which contain their genetic information on DNA strings, so-called \emph{genes}. These genes can lead to the generation of \emph{messenger RNA (mRNA)} which transports the genetic information and induces the production of \emph{proteins}. Such mRNA molecules and proteins are modes of expression by which a cell reflects the presence, kind and activity of its genes. In this paper, we consider such \emph{gene expression} in terms of quantities of mRNA molecules. 

Gene expression is stochastic. It can differ significantly between, e.g., types of cells or tissues, and between individuals. In that case, one refers to \emph{differential gene expression}. In particular, cells can be differentially expressed between healthy and sick tissue samples from the same origin. Moreover, cells can differ even within a small tissue sample, e.g. within a tumour that consists of several mutated cell populations. Mathematically, we regard two populations to be different if their mRNA counts follow different probability distributions. If there is more than one population in a tissue, we call it heterogeneous. The expression of such tissues is often described by mixture models. Detecting and parameterising heterogeneities is of utmost importance for understanding development and disease.

The amount of mRNA molecules of a gene in a tissue sample can be assessed by various techniques such as microarray measurements \citep{kurimoto_improved_2006,tietjen_single-cell_2003} or sequencing \citep{sandberg_entering_2014,ziegenhain_comparative_2017}. Measurements of single cells yield the highest possible resolution. They are best suited for identification and description of heterogeneity in large and error-free datasets. In practice, however, single-cell data often comes along with high cost, effort and technical noise \citep{grun_validation_2014}. Instead of considering single-cell data, we analyze the cumulative gene expression of small pools of randomly selected cells \citep{janes_identifying_2010}. The pool size should be large enough to substantially reduce measurement error and cost, and at the same time small enough such that heterogeneity is still identifiable.

We developed the algorithm \pkg{stochprofML} to infer single-cell regulatory states from such pools 
\citep{bajikar_parameterizing_2014}. 
In contrast to previously existing methods, we neither require a priori knowledge about the mixing weights \citep[such as][]{shen-orr_cell_2010} nor about expression profiles \citep[such as][]{erkkila_probabilistic_2010}; other than most bulk deconvolution methods, like CIBERSORT \citep{newman_robust_2015}, so-called signature matrices for the populations are not needed to infer population fractions.

In \cite{bajikar_parameterizing_2014}, we applied \pkg{stochprofML} to measurements from human breast epithelial cells and revealed the functional relevance of the heterogeneous expression of a particular gene. In a second study, we applied the algorithm to 
clonal tumor spheroids of colorectal cancer \citep{tirier_pheno-seq_2018}. Here, a single tumor cell was cultured, and after several rounds of replication, each resulting spheroid was imaged and sequenced. However, pool sizes differed between tissue samples as each spheroid contained a different number of cells ranging from less than ten to nearly~200 cells. 
Therefore, we extended \pkg{stochprofML} to be able to handle pools of different sizes. 

In this work, we explain the statistical reasoning and \pkg{R} implementation of \pkg{stochprofML} \citep{stochProfPkg}. In Section~\ref{sec:models}, we derive the statistical model. After a first description of the nomenclature, we introduce basic statistical descriptions of continuous univariate single-cell gene expression. The complexity of the model is increased step by step: First, we account for cell-to-cell heterogeneity through the use of mixture distributions. Then, we extend the modeling from single-cell to small-pool measurements by introducing convolutions of statistical distributions. Finally, we calculate the likelihood and present ways for model selection.
Section~\ref{sec:Il} shows how the \pkg{R} package can be used to generate data and to infer model parameters. This is followed by simulation studies in Section~\ref{sec:simstudies}, investigating the influence of pool sizes, differences in parameter settings and uncertainty about cell counts on the resulting parameter inference. In Section~\ref{sec:interpr}, we elaborate the interpretation of inferred heterogeneity.
Section~\ref{sec:sum_dis} concludes the work.


\section{Models and software}\label{sec:models}

The \pkg{stochprofML} algorithm aims at maximum likelihood estimation of the corresponding model parameters. Hence, we derive the likelihood functions of the parameters and show details of the estimation and its implementation. The new elements of the most recent version of the algorithm are introduced along the line.\newline

\subsection{Single-cell models of heterogeneous gene expression}\label{sec:statdistr}

\begin{figure}[!h]
\centering\includegraphics[width=\textwidth]{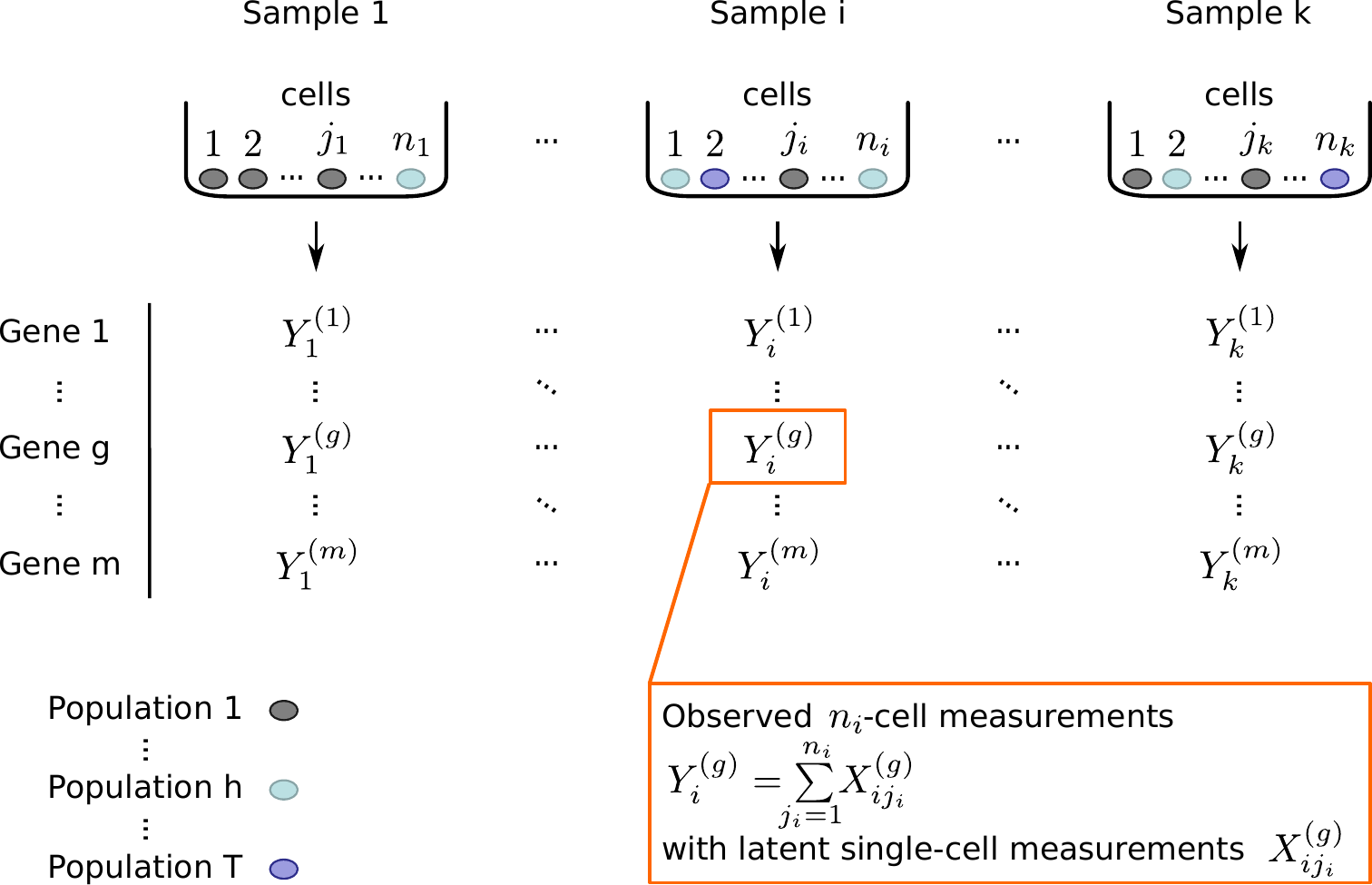}
\caption{Experimental design of pooling cells into samples, measuring the pooled gene expression across several genes for which identical population structures are assumed. The table illustrates the index notation of (tissue) samples, single cells, populations and genes as well as observed and latent measurements.}
\label{Fig:Naming}
\end{figure}

Suppose there are $k$ (tissue) samples, indexed by~$i \in \{ 1,\ldots,k\}$. From each tissue sample~$i$, we collect a pool of a known number of cells.
The cells are either indexed by~$j \in \{ 1,\ldots,n\}$ if the cell pool size is the same in all measurements, or, as possible in the latest implementation, by~$j_i \in \{ 1,\ldots,n_i\}$ in case cell pool sizes vary between measurements. In the latter, more general case, the cell numbers are variable over the~$k$ cell pools and summarized by~$\vec{n} = (n_1,\ldots, n_k )$.
From each sample, the gene expression of $m$ genes is measured, indexed by~$g \in \{ 1,\ldots,m\}$.
We assume that each cell stems from one out of~$T$ cell populations, indexed by~$h \in \{ 1,\ldots,T\}$. If~$T>1$ in the set of all cells of interest, the tissue is called heterogeneous. The notation is illustrated in Figure~\ref{Fig:Naming}.

Biologically, the different cell populations correspond to different regulatory states or~---~especially in the context of cancer~---~to different (sub-)clones. For example, there might be two populations within a considered tissue: one occupying a basal regulatory state, where the expression of genes is at a low level, and one from a second regulatory state, where genes are expressed at a higher level.

As described in Section~\ref{sec:intro}, there are various technologies to measure gene expression. In case of microarray techniques \citetext{as done in previous applications of stochastic profiling, see \citealp{janes_identifying_2010} and \citealp{bajikar_parameterizing_2014}}, the measured gene expression is typically described in terms of continuous probability distributions. Conditioned on the cell population, \pkg{stochprofML} provides two choices for the distribution of the expression of one gene:

\paragraph{Lognormal distribution.}\label{paragraph:ln}

The two parameters defining a univariate lognormal distribution $\gL\gN(\mu,\sigma^2)$ are called log-mean $\mu \in \mathbb{R}$ and log-standard deviation $\sigma>0$. These are the mean and the standard deviation of the normally distributed random variable $\log(X)$, the natural logarithm of~$X$. The probability density function (PDF) of~$X$ is given by
\[
f_\text{LN}(x|\mu, \sigma^2) 
= \frac1{\sqrt{2\pi}\sigma x}\,\exp\left(-\frac{(\log x-\mu)^2}{2\sigma^2}\right) \quad\text{for }x>0.
\]
A random variable $X\sim\gL\gN(\mu,\sigma^2)$ has expectation and variance
\begin{equation}
\E(X)= \exp\left(\mu+\frac{\sigma^2}{2}\right)\qquad\text{and}\quad\Var(X)=\exp\left(2\mu+\sigma^2\right)\left(\exp\left(\sigma^2\right)-1\right).
\label{E_Var_lognormal}
\end{equation}

\paragraph{Exponential distribution.}\label{paragraph:exp}
An exponential distribution $\gE \! \gX \! \gP(\lambda)$ is defined by the rate parameter $\lambda>0$. The PDF is given by
\[
f_{E\! X\! P}(x| \lambda)
= \lambda \exp\left( -\lambda x \right) \quad\text{for }x \geq 0.
\]
A random variable $X\sim\gE \! \gX \! \gP(\lambda)$ has expectation and variance
\[
\E(X)= \frac{1}{\lambda}\qquad\text{and}\quad\Var(X)= \frac{1}{\lambda^2}.
\]
In general, the lognormal distribution is an appropriate description of continuous gene expression. With its two parameters, it is more flexible than the exponential distribution. However, the lognormal distribution cannot model zero gene expression. In case of zeros in the data, it could be modified by adding very small values such as 0.0001, or one uses the exponential distribution to model this kind of expression.

In case of~$T$ cell populations, we describe the expression of one gene by a stochastic mixture model. Let $\left( p_1, \ldots, p_T\right)$ with~$p_1+\ldots+p_T=1$ denote the fractions of populations in the overall set of cells. \pkg{stochprofML} offers the following three mixture models:
\begin{enumerate}
\item \textbf{Lognormal-lognormal (LN-LN)}: Each population $h$ is represented by a lognormal distribution with population-specific parameter $\mu_h$ (different for each population~$h$) and identical~$\sigma$ for all~$T$ populations. The single-cell expression~$X$ that originates from such a mixture of populations then follows
\[
	X \sim \begin{cases}
		 \gL\gN(\mu_1,\sigma^2) & \text{with probability } p_1\\
		 \vdots &\\		 
		 \gL\gN(\mu_h,\sigma^2) & \text{with probability } p_h\\
		 \vdots &\\
 		 \gL\gN(\mu_T,\sigma^2) & \text{with probability } \left(1-\sum_{h=1}^{T-1} p_h\right).
		 \end{cases}
\]
\item \textbf{Relaxed lognormal-lognormal (rLN-LN)}: This model is similar to the LN-LN model, but each population~$h$ is represented by a lognormal distribution with a different parameter set ($\mu_h$, $\sigma_h$). The single-cell expression~$X$ follows
\[
	X \sim \begin{cases}
		 \gL\gN(\mu_1,\sigma_1^2) & \text{with probability } p_1\\
		\vdots &\\		 
		 \gL\gN(\mu_h,\sigma_h^2)& \text{with probability } p_h\\
		 \vdots &\\
 		 \gL\gN(\mu_T,\sigma_T^2) & \text{with probability } \left(1-\sum_{h=1}^{T-1} p_h\right).
		 \end{cases}
\]
\item \textbf{Exponential-lognormal (EXP-LN)}: Here, one population is represented by an exponential distribution with parameter $\lambda$, and all remaining~$T-1$ populations are modeled by lognormal distributions analogously to LN-LN, i.e. with population-specific parameters $\mu_h$ and identical $\sigma$. The single-cell expression $X$ then follows
\[
	X \sim \begin{cases}
		 
		 \gL\gN(\mu_1,\sigma^2)& \text{with probability } p_{1}\\
		 \vdots &\\		
		 \gL\gN(\mu_h,\sigma^2)& \text{with probability } p_{h}\\
		 \vdots &\\
 		 \gL\gN(\mu_{T-1},\sigma^2) & \text{with probability } p_{T-1} \\
 		 \gE \! \gX \! \gP(\lambda) & \text{with probability } \left(1-\sum_{h=1}^{T-1} p_h\right).\\
		 \end{cases}
\]
\end{enumerate}
The LN-LN model is a special case of the rLN-LN model. It assumes identical~$\sigma$ across all populations. Biologically, this assumption is motivated by the fact that, for the lognormal distribution, identical~$\sigma$ lead to identical coefficient of variation
$$ \text{CV}(X) = \frac{\sqrt{\text{Var}(X)}}{\E(X)}= \sqrt{\exp(\sigma^2)-1}$$
even for different values of~$\mu$. In other words, the linear relationship between the mean expression and the standard deviation is maintained across cell populations in the LN-LN model. The appropriateness of the different mixture models can be discussed both biologically and in terms of statistical model choice (see Section~\ref{subsec:modelchoice}). 

Within one set of genes under consideration, we assume that the same type of model (LN-LN, rLN-LN, EXP-LN) is appropriate for all genes. The parameter values, however, may differ. In case of~$T$ cell populations, we describe the single-cell gene expression~$X^{(g)}$ for gene~$g$ by a mixture distribution with PDF
\begin{align*}
f_\text{T-pop}\left(x^{(g)} | \right. &\, \left.  \mm{\theta}^{(g)}, \mm{p}\right) = \\
 &\, p_1 f_1\left(x^{(g)}|\theta_1^{(g)}\right) + \ldots+ p_h f_h\left(x^{(g)}|\theta_h^{(g)}\right)+\ldots+ \left(1-\sum_{h=1}^{T-1} p_h\right)f_T\left(x^{(g)}|\theta_T^{(g)}\right),
 \end{align*}
where $ f_h$ with $ h\in\{1,\ldots, T\}$ represents the PDF of population~$h$ that can be either lognormal or exponential, and $\mm{\theta}^{(g)}=\{\theta_1^{(g)},\ldots,\theta_T^{(g)}\}$ are the (not necessarily disjoint) distribution parameters of the~$T$ populations for gene~$g$.

\paragraph{Example: Mixture of two populations - Part 1.}\label{Example_mix2_Part1}

We exemplify the two-population case. Here, the PDF of the mixture distribution for gene~$g$ reads
\[
f_\text{2-pop} (x^{(g)} |\mm{\theta}^{(g)})= p f_1 (x^{(g)} |\theta_1^{(g)})+(1-p)f_2 (x^{(g)} |\theta_2^{(g)}),
\]
where $p$ is the probability of the first population. The univariate distributions $f_1^{(g)}$ and $f_2^{(g)}$ depend on the chosen model :
\begin{enumerate}
\item \textbf{LN-LN}:  $f_1 (x^{(g)} |\theta_1^{(g)}) = f_\text{LN}(x^{(g)} | \mu_1^{(g)}, \sigma^2) $ and $f_2(x^{(g)} |\theta_2^{(g)}) = f_\text{LN}(x^{(g)}  | \mu_2^{(g)}, \sigma^2) $,
i.e.\ there are four unknown parameters: $p, \mu_1^{(g)}, \mu_2^{(g)}$ and $ \sigma^2$.

\item \textbf{rLN-LN}: $f_1 (x^{(g)} |\theta_1^{(g)}) = f_\text{LN}(x^{(g)}  | \mu_1^{(g)}, {\sigma_1}^2) $ and $f_2(x^{(g)} |\theta_2^{(g)}) = f_\text{LN}(x^{(g)}  | \mu_2^{(g)}, {\sigma_2}^2) $
i.e.\ there are five unknown parameters: $p, \mu_1^{(g)}, \mu_2^{(g)}, {\sigma_1}^2$ and ${\sigma_2}^2$.

\item \textbf{EXP-LN}: $f_1(x^{(g)} |\theta_1^{(g)}) = f_\text{LN}(x^{(g)}  | \mu^{(g)}, {\sigma}^2) $ and $f_2(x^{(g)} |\theta_2^{(g)}) =  f_\text{EXP}(x^{(g)}  | \lambda^{(g)})$.
i.e.\ there are four unknown parameters: $p, \mu^{(g)}$, $\sigma^2$ and~$\lambda^{(g)}$.

\end{enumerate}
Note that although each lognormal population has its individual~$\sigma$, these $\sigma$-values remain identical across genes.

\subsection{Small-pool models of heterogeneous gene expression}\label{subsubsec:spmodel}

\begin{figure}[p]
\centering\includegraphics[width=0.9\linewidth]{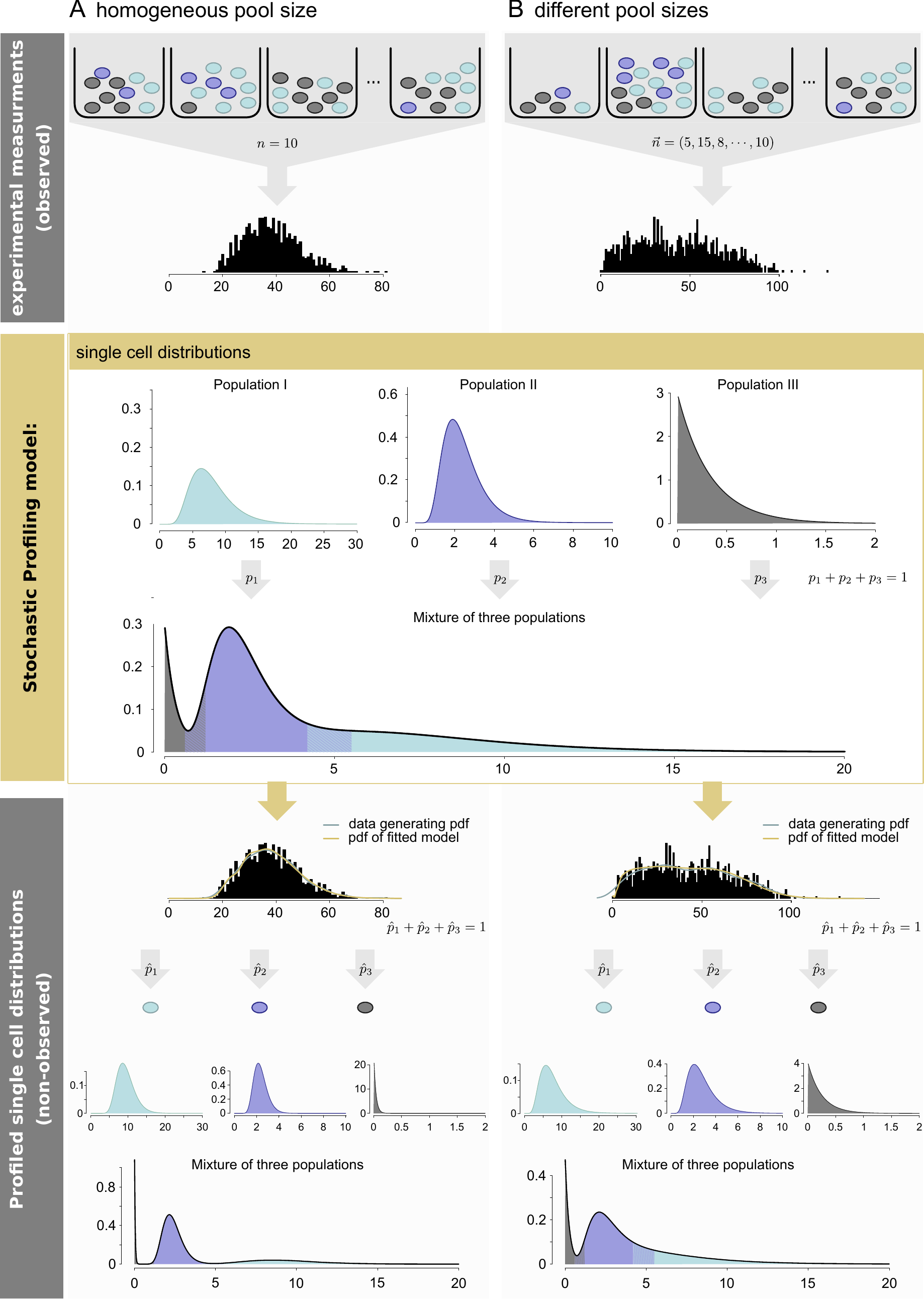}
\caption{Stochastic Profiling can be performed either on measurements of (A) homogeneous pool size of $n$ cells or of (B) different pool sizes given by the cell number vector $\vec{n}$. In both cases, the \pkg{stochprofML} algorithm estimates the parameters for the specified number of populations from pooled data, leading to inferred single-cell distributions for each population. Appendix~\ref{mixed n_densities_dens} describes how this density is visualized here.}
\label{Fig:OverviewSPML}
\end{figure}

\pkg{stochprofML} is tailored to analyze gene expression measurements of small pools of cells, beyond the analysis of standard single-cell gene expression data. In other words, the single-cell gene expression~$X_{ij_i}^{(g)}$ described in Section~\ref{sec:statdistr} is assumed latent. Instead, we consider observations
\begin{equation}
Y_i^{(g)} = \sum_{j_i=1}^{n_i} X_{ij_i}^{(g)}     
\label{eq:sum_of_sc}
\end{equation}
for~$i=1,\ldots,k$, which represent the overall gene expression of the $i$th cell pool for gene~$g$. In the first version of \pkg{stochprofML}, pools had to be of equal size $n$, i.e.\ for each measurement~$Y_{i}^{(g)}$ one had to extract the same number of cells from each tissue sample. This was a restrictive assumption from the experimental point of view. The recent extension of \pkg{stochprofML} allows each cell pool~$i$ to contain a different number~$n_i$ of cells (see also Figures~\ref{Fig:Naming} and \ref{Fig:OverviewSPML}). 

The algorithm aims to estimate the single-cell population parameters despite the fact that measurements are available only in convoluted form. To that end, we derive (in Section~\ref{sec:lf}) the likelihood function of the parameters in the convolution model~\eqref{eq:sum_of_sc}, where we assume the gene expression of the single cells to be independent within a tissue sample. For better readability, we suppress for now the superscript~$(g)$ and introduce it again in Section~\ref{sec:lf}. 

The derivation of the distribution of $Y_i$ 
is described in Appendix~\ref{App:Trinom}.
The corresponding PDF $f_{n_i}(y_i|\mm{\theta},\mm{p})$ 
of an observation~$y_i$ 
which represents the overall gene expression from sample~$i$ (consisting of~$n_i$ cells) is given by
\begin{align}
f_{n_i}\left(y_i| \! \right. &\, \left.\mm{\theta}, \mm{p}\right) = \notag \\
 &\,\sum_{\ell_1=0}^{n_i}\sum_{\ell_2=0}^{{n_i}-\ell_1}\cdots\sum_{\ell_{T-1}=0}^{{n_i}-\sum_{h=1}^{T-2} \ell_h} {{n_i} \choose {\ell_1,\ell_2,\ldots,\ell_T}}p_1^{\ell_1}p_2^{\ell_2}\cdots p_T^{\ell_T} f_{ ( \ell_1,\ell_2,\ldots,\ell_T)}\left(y_i|\mm{\theta} \right),
\label{multdistr}
\end{align}
where $\ell_T = n_i-\sum_{h = 1}^{T-1} \ell_h$ and $p_T = 1-\sum_{h = 1}^{T-1} p_h$. 
Here, $f_{(\ell_1,\ell_2,\ldots,\ell_T)}$ 
describes the PDF of a pool of~$n_i$ cells with \emph{known} composition of the single populations, i.e.\ it is known that there are~$\ell_1$~cells from population~$1$, $\ell_2$~cells from population~$2$ etc.
${n_i \choose {\ell_1,\ell_2,\ldots,\ell_T} } p_1^{\ell_1}p_2^{\ell_2}\cdots p_T^{\ell_T} $~represents the multinomial probability of obtaining exactly this composition~$(\ell_1,\ldots,\ell_T)$ using the multinomial coefficient ${n_i \choose {\ell_1,\ell_2,\ldots,\ell_T} }=n_i!/(\ell_1! \ldots \ell_T!)$. Equation~\eqref{multdistr} sums up over all possible compositions~$(\ell_1,\ldots,\ell_T)$ with~$\ell_1,\ldots,\ell_T\in\mathbb{N}_0$ and~$\ell_1+\ldots+\ell_T=n_i$.
Taken together, $f_{n_i}(y_i|\mm{\theta}, \mm{p})$ 
determines the PDF of~$y_i$ 
with respect to each possible combination of $n_i$ cells of~$T$~populations. 

Thus, the calculation of~$f_{n_i}(y_i|\mm{\theta}, \mm{p})$
requires knowledge of~$f_{(\ell_1,\ell_2,\ldots,\ell_T)}(y_i|\mm{\theta})$
. The derivation of this PDF depends on the choice of the single-cell model (LN-LN, rLN-LN, or EXP-LN; see Section~\ref{sec:statdistr}) that was made for~$X_{ij_i}$
.
\begin{enumerate}

\item \textbf{LN-LN: } 
$$f_{(\ell_1,\ldots,\ell_h,\ldots,\ell_T)}(y_i|\mm{\theta}) = f^\text{LN-LN}_{(\ell_1,\ldots,\ell_h,\ldots,\ell_T)}(y_i|\mu_1,\ldots,\mu_h,\ldots, \mu_T,\sigma^2) $$
 
is the density of a sum~$Y_i=X_{i1}+\ldots+X_{in_i}$ of ${n_i}$ independent random variables with
\[
	X_{ij_i} \sim \begin{cases}
		 \gL\gN(\mu_1,\sigma^2) & \text{if } 1\leq j_i\leq J_1\\
		\vdots &\\		 
		 \gL\gN(\mu_h,\sigma^2) & \text{if } J_{h-1}< j_i\leq J_h\\
		 \vdots &\\
 		 \gL\gN(\mu_T,\sigma^2) & \text{if } J_{T-1}< j_i\leq J_T=n_i,
		 \end{cases}
\]
with $J_1 = \ell_1,\ldots, J_h = \ell_1+\ell_2 + \ldots + \ell_h, \ldots, J_T = \ell_1 + \ell_2 + \ldots + \ell_T = n_i $. $Y_i$~is the convolution of random variables~$X_{i1},\ldots,X_{in_i}$, which is here the convolution of~$T$ sub-convolutions: a convolution of $\ell_1$ times $\gL\gN(\mu_1,\sigma^2)$, plus a convolution of $\ell_2$ times $\gL\gN(\mu_2,\sigma^2)$, and so on, up to a convolution of $\ell_T$ times $\gL\gN(\mu_T,\sigma^2)$.

There is no analytically explicit form for the convolution of lognormal random variables. Hence, $f^\text{LN-LN}_{(\ell_1,\ldots,\ell_h,\ldots,\ell_T)}$ is approximated using the method by \citet{fenton_sum_1960}. That is, the distribution of the sum~$A_1+\ldots+A_m$ of independent random variables\linebreak $A_i\sim\gL\gN(\mu_{A_i},\sigma_{A_i}^2)$ is approximated by the distribution of a random variable~$B\sim\gL\gN(\mu_B,\sigma_B^2)$ such that
\[
\E(B)=\E(A_1+\ldots+A_m) \quad\text{and}\quad
\Var(B)=\Var(A_1+\ldots+A_m).
\]
According to Equation~\eqref{E_Var_lognormal}, that means that~$\mu_B$ and~$\sigma_B$ are chosen such that the following equations are fulfilled:
\[
\exp\left(\mu_B+\frac{\sigma_B^2}{2}\right)
= \exp\left(\mu_{A_1}+\frac{\sigma_{A_1}^2}{2}\right)
+ \ldots
+ \exp\left(\mu_{A_m}+\frac{\sigma_{A_m}^2}{2}\right)=:\Gamma
\]
and
\begin{eqnarray*}
&& \exp\left(2\mu_B+\sigma_B^2\right)\left(\exp\left(\sigma_B^2\right)-1\right) = \\
&& \quad \exp\left(2\mu_{A_1}+\sigma_{A_1}^2\right)\left(\exp\left(\sigma_{A_1}^2\right)-1\right)
+ \ldots
+ \exp\left(2\mu_{A_m}+\sigma_{A_m}^2\right)\left(\exp\left(\sigma_{A_m}^2\right)-1\right)
=:\Delta.
\end{eqnarray*}
That is achieved by choosing
\[
\mu_B = \log(\Gamma)-\frac12 \sigma_B^2 \quad\text{and}\quad
\sigma_B^2 = \log\left(\frac{\Delta}{\Gamma^2}+1\right).
\]
This approximation is implemented in the function \code{d.sum.of.lognormals()}. The overall PDF is computed through \code{d.sum.of.mixtures.LNLN()}.

\item \textbf{rLN-LN:}
$$f_{(\ell_1,\ldots,\ell_h,\ldots,\ell_T)}(y_i|\mm{\theta}) = f^\text{rLN-LN}_{(\ell_1,\ldots,\ell_h,\ldots,\ell_T)}(y_i|\mu_1,\ldots,\mu_h,\ldots, \mu_T,\sigma_1^2,\ldots,\sigma_h^2,\ldots,\sigma_T^2) $$
is the PDF of a sum~$Y_i=X_{i1}+\ldots+X_{in_i}$ of ${n_i}$ independent random variables with
\[
	X_{ij_i} \sim \begin{cases}
		 \gL\gN(\mu_1,\sigma_1^2) & \text{if } 1\leq j_i\leq J_1\\
		\vdots &\\		 
		 \gL\gN(\mu_h,\sigma_h^2) & \text{if } J_{h-1}< j_i\leq J_h\\
		 \vdots &\\
 		 \gL\gN(\mu_T,\sigma_T^2) & \text{if } J_{T-1}< j_i\leq J_T=n_i,
		 \end{cases}
\]
with $J_1 = \ell_1,\ldots, J_h = \ell_1+\ell_2 + \ldots + \ell_h, \ldots, J_T = \ell_1 + \ldots + \ell_T = n_i $.
Again, $f^\text{rLN-LN}_{(\ell_1,\ldots,\ell_h,\ldots,\ell_T)}$ is approximated using the method by \citet{fenton_sum_1960}, analogously to the LN-LN model. It is implemented in \code{d.sum.of.mixtures.rLNLN()}.

\item \textbf{EXP-LN:}
 $$f_{(\ell_1,\ell_2,\ldots,\ell_T)}(y_i|\mm{\theta})=f^\text{EXP-LN}_{(\ell_1,\ell_2,\ldots,\ell_T)}(y_i|\lambda,\mu_1,\ldots, \mu_{T-1},\sigma^2) $$
 
is the density of a sum~$Y_i=X_{i1}+\ldots+X_{in_i}$ of~$n_i$ independent random variables with
\[
	X_{ij_i} \sim \begin{cases}
		\gL\gN(\mu_1,\sigma^2) & \text{if }  1\leq j_i\leq J_1\\
		\vdots &\\		 
		 \gL\gN(\mu_h,\sigma^2) & \text{if } J_{h-1}< j_i\leq J_h\\		 
		 \vdots &\\
 		 \gL\gN(\mu_{T-1},\sigma^2) & \text{if } J_{T-2}< j_i\leq J_{T-1}\\
 		 \gE \! \gX \! \gP(\lambda) & \text{if } J_{T-1}< j_i\leq J_T=n_i,
		 \end{cases}
\]
with $J_1 = \ell_1,\ldots, J_h = \ell_1+\ell_2 + \ldots + \ell_h, \ldots, J_T = \ell_1 + \ldots + \ell_T = n_i $.
The sum of independent exponentially distributed random variables with equal intensity parameter follows an Erlang distribution \citep{feldman_applied_2010}, which is a gamma distribution with integer-valued shape parameter that represents the number of exponentially distributed summands. Thus, the PDF for the EXP-LN mixture model is the convolution of one Erlang (or gamma) distribution (namely the sum of all exponentially distributed summands) and one lognormal distribution \citep[namely the sum of all lognormally distributed summands, again using the approximation method by][]{fenton_sum_1960}. The PDF for this convolution is not known in analytically explicit form but expressed in terms of an integral that is solved numerically through the function \code{lognormal.exp.convolution()}. The overall PDF of the EXP-LN model is implemented in \code{d.sum.of.mixtures.EXPLN()}.
\end{enumerate}

\textbf{Example: Mixture of two populations - Part 2.}
In this example of the two-population model, let each observation consist of the same number of~$n=10$ cells. Then~$Y_i$ is a $10$-fold convolution, and the PDF \eqref{multdistr} simplifies to
\begin{equation}
f_{10}\left(y_i|\mm{\theta},\mm{p} \right)=\sum_{\ell=0}^{10} {10 \choose \ell} p^\ell(1-p)^{10-\ell} f_{(\ell,10-\ell)}\left(y_i | \mm{\theta} \right),
\label{convolution_2types}
\end{equation}

where $f_{(\ell,10-\ell)}$ is the PDF of the sum~$Y_i$ of ten independent random variables, that is\linebreak $Y_i=X_{i1}+\ldots+X_{i 10}$. This PDF depends on the particular chosen model:
\begin{enumerate}
\item \textbf{LN-LN} 
 $$f_{(\ell,10-\ell)}(y_i|\mm{\theta})=f^\text{LN-LN}_{(\ell,10-\ell)}(y_i|\mu_1,\mu_2,\sigma^2) $$
 is the PDF of a sum~$Y_i=X_{i1}+\ldots+X_{i 10}$ of ten independent random variables with
 \[
	X_{ij} \sim \begin{cases}
		 \gL\gN(\mu_1,\sigma^2) & \text{if } 1\leq j \leq \ell\\
		 \gL\gN(\mu_2,\sigma^2) & \text{if } \ell< j\leq 10.
		 \end{cases}
\]

\item \textbf{rLN-LN} 
 $$f_{(\ell,10-\ell)}(y_i|\mm{\theta})=f^\text{rLN-LN}_{(\ell,10-\ell)}(y_i|\mu_1,\mu_2,\sigma_1^2, \sigma_2^2) $$
 is the PDF of a sum~$Y_i=X_{i1}+\ldots+X_{i 10}$ of ten independent random variables with
 \[
	X_{ij} \sim \begin{cases}
		 \gL\gN(\mu_1,\sigma_1^2) & \text{if } 1\leq j \leq \ell\\
		 \gL\gN(\mu_2,\sigma_2^2) & \text{if } \ell< j\leq 10.
		 \end{cases}
\]

\item \textbf{EXP-LN} 
 $$f_{(\ell,10-\ell)}(y_i|\mm{\theta})=f^\text{EXP-LN}_{(\ell,10-\ell)}(y_i|\lambda,\mu,\sigma^2) $$
 is the PDF of a sum~$Y_i=X_{i1}+\ldots+X_{i 10}$ of ten independent random variables with
 \[
	X_{ij} \sim \begin{cases}
		 \gL\gN(\mu,\sigma^2) & \text{if } 1\leq j \leq \ell\\
		 	\gE \! \gX \! \gP(\lambda) & \text{if } \ell< j\leq 10.\\
		 \end{cases}
\]   
\end{enumerate}

\subsection{Likelihood function}\label{sec:lf}

Overall, after re-introducing the superscript~$(g)$ for measurements of genes~$g=1,\ldots,m$, we obtain the PDF
\begin{align}
f_{n_i}\left(y_i^{(g)}|  \right. & \left.\mm{\theta}^{(g)}, \mm{p}\right) = \notag \\
 &\,\sum_{\ell_1=0}^{n_i}\sum_{\ell_2=0}^{{n_i}-\ell_1}\cdots\sum_{\ell_{T-1}=0}^{{n_i}-\sum_{h=1}^{T-2} \ell_h} {{n_i} \choose {\ell_1,\ell_2,\ldots,\ell_T}}p_1^{\ell_1}p_2^{\ell_2}\cdots p_T^{\ell_T} f_{ ( \ell_1,\ell_2,\ldots,\ell_T)}\left(y_i^{(g)}|\mm{\theta}^{(g)} \right)
\label{multdistr_with_g}
\end{align}
with model-specific choice of~$f_{ ( \ell_1,\ell_2,\ldots,\ell_T)}$. While~$\bn=(n_1,\ldots,n_k)$ is considered known, we aim to infer the unknown model parameters~$\mm{\theta}=\{\mm{\theta}^{(1)},\ldots,\mm{\theta}^{(m)},\mm{p}\}$ by maximum likelihood estimation. Assuming independent observations~${\mm y}=\{y_i^{(g)}|i=1,\ldots,k;g=1,\ldots,m\}$ of~$Y_i^{(g)}$ for $m$~genes and $k$~tissue samples, where sample~$i$ contains $n_i$ cells, the likelihood function is given by
\[
L(\mm{\theta}|\mm{y}) = \prod_{g=1}^m\prod_{i=1}^k f_{n_i}\left(y_i^{(g)}| \mm{\theta}^{(g)}, \mm{p}\right).
\]
Consequently, the log-likelihood function of the model parameters reads
\begin{equation}
\ell(\mm{\theta}|\mm{y}) = \sum_{g=1}^m \sum_{i=1}^k \log\left[ f_{n_i} \left( y_i^{(g)}|\mm{\theta}^{(g)}, \mm{p}\right) \right].
\label{loglikeli}
\end{equation}

\textbf{Example: Mixture of two populations - Part 3.}
Returning to the two-population example with 10-cell pools, the log-likelihood for $k=100$~tissue samples and $m=5$~genes is given by
\[
\ell(\mm{\theta}|\mm{y}) = \sum_{g=1}^5 \sum_{i=1}^{100} \log \left[ f_{10}\left(y_i^{(g)}|\mm{\theta}^{(g)},\mm{p} \right) \right],
\]
where $f_{10}\left(y_i^{(g)}|\mm{\theta}^{(g)},\mm{p} \right) $ is given by Equation~\eqref{convolution_2types}.

\subsection{Maximum likelihood estimation}\label{subsec:mlpe}

The \pkg{stochprofML} algorithm aims to infer the unknown model parameters using maximum likelihood estimation. As input, we expect an $m\times k$ data matrix of pooled gene expression, known cell numbers~$\vec{n}$, the assumed number of populations $T$ and the choice of single-cell distribution (LN-LN, rLN-LN, EXP-LN). Based on this input, the algorithm aims to find parameter values of~$\mm{\theta}=\{\mm{\theta}^{(1)},\ldots,\mm{\theta}^{(m)},\mm{p}\}$ that maximize~$\ell(\mm{\theta}|\mm{y})$ as given by Equation~\eqref{loglikeli}. This section describes practical aspects of the optimization procedure.

\textbf{Example: Mixture of two populations - Part 4.}
Several challenges occur during parameter estimation. We explain these on the two-population LN-LN example: First, we aim to ensure parameter identifiability. This is achieved for the two-population LN-LN model by constraining the parameters to fulfil either~$p\leq 0.5$ or $\mu_1 > \mu_2$. Otherwise, the two combinations~$(p,\bmu_1,\bmu_2,\sigma)$ and~$(1-p,\bmu_2,\bmu_1,\sigma)$ would yield identical values of the likelihood function and could cause computational problems. For our implementation, we preferred the second possibility, i.\,e.\  $\mu_1 > \mu_2$. The alternative, i.\,e.\  requiring $p\leq 0.5$, led to switchings between~$\mu_1$ and~$\mu_2$ in case of~$p \approx 0.5$. As a second measure, we implement unconstrained rather than constrained optimization: Instead of estimating~$(p,\bmu_1,\bmu_2,\sigma)$ under the constraints~$p\in[0,1]$, $\mu_1 > \mu_2$ and~$\sigma>0$, the parameters are transformed to~$(\text{logit}(p),\bmu_1,\bmu_2,\log(\sigma))$, and an unconstrained optimization method is used. This is substantially faster. 

The aforementioned transformations are likewise employed for all other models (rLN-LN and EXP-LN) and population numbers. In particular, $\sigma$ and~$\lambda$ are log-transformed, and the lognormal populations are ordered according to the log-means~$\mu_h^{(1)}$ of the first gene in the gene list. The population probabilities are transformed to~$\mathbb{R}$ such that they still sum up to one after back-transformation. For details, see Appendix~\ref{app:logit}.

The log-likelihood function is multimodal. Thus, a single application of some gradient-based optimization method does not suffice to find a global maximum. Instead, two approaches are combined which are alternately executed: First, a grid search is performed, where the log-likelihood function is computed at randomly drawn parameter values. This way, high likelihood regions can be identified with low computational cost. In the second step, the Nelder-Mead algorithm \citep{NelderMead65} is repeatedly executed. Its starting values are randomly drawn from the high likelihood regions found before. The following grid search then again explores the regions around the obtained local maxima, and so on.

If a dataset contains gene expressions for~$m$ genes, and if we assume~$T$ populations, there are at minimum~$T(m+1)$ parameters which one seeks to estimate depending on the model framework. This is computationally difficult, because the number of modes of the log-likelihood function increases with the number of parameters. The performance of the numerical optimization crucially depends on the quality of the starting values, and a large number of restarts is required. When analyzing a large gene cluster, it is advantageous to start by considering small clusters and use the derived estimates as initial guesses for larger clusters. 
This is implemented in the function \code{stochprof.loop()} (parameter \code{subgroups}) and demonstrated in \code{analyze.toycluster()}.

Approximate marginal 95\% confidence intervals for the parameter estimates are obtained as follows: We numerically compute the Hessian matrix of the negative log-likelihood function on the unrestricted parameter space and evaluate it at the (transformed) maximum likelihood estimator. Denote by $d_i$ the $i$th diagonal element of the inverse of this matrix. Then the confidence bounds for the $i$th transformed parameter~$\theta_i$ are
\[
\hat{\theta}_i \pm 1.96\sqrt{d_i}.
\]
We obtain respective marginal confidence intervals for the original true parameters by back-transformation of the above bounds. This approximation is especially appropriate in the two-population example for the parameters~$p$ and~$\sigma$ when conditioning on~$\bmu_1$ and~$\bmu_2$. In this case, in practice, the profile likelihood is seemingly unimodal.

\subsection{Model choice} \label{subsec:modelchoice}

By increasing the number~$T$ of populations, we can describe the observed data more precisely, but this comes at the cost of potential overfitting. For example, a three-population LN-LN model may lead to a larger likelihood at the maximum likelihood estimator than a two-population LN-LN model on the same dataset. However, the difference may be small, and the additional third population may not lead to a gain of knowledge. For example, the estimated population probability~$\hat{p}_3$ may be tiny, or the log-means of the second and third population, $\hat\mu_2$ and~$\hat\mu_3$ might hardly be distinguished from each other.

To objectively find a trade-off between necessary complexity and sufficient interpretability, we employ the Bayesian information criterion  \citep[BIC,][]{schwarz_estimating_1978}:
$$
\text{BIC}(\hat{\btheta})=-2\ell(\hat{\btheta})+k\dim(\hat {\btheta}),
$$
where $\hat{\btheta}$ is the maximum likelihood estimate of the respective model, $\dim(\hat{\btheta})$ the number of parameters and $k$~the size of the dataset. From the statistics perspective, the model with smallest BIC is considered most appropriate among all considered models. 

In practice, it is required to estimate all models of interest separately with the  \pkg{stochprofML} algorithm, e.\,g.\ the LN-LN model with one, two and three populations, and/or the respective rLN-LN and EXP-LN models. The BIC values are returned by the function~\code{stochprof.loop()}.

\section{Usage of stochprofML}\label{sec:Il}

This section illustrates the usage of the \pkg{stochprofML} package for simulation and parameter estimation. There are two ways two use the \pkg{stochprofML} package: (i) Two interactive functions\linebreak \code{stochasticProfilingData()} and \code{stochasticProfilingML()} provide low-level access to synthetic data generation and maximum likelihood parameter estimation without requiring advanced programming knowledge. They guide the user through entering the relevant input parameters: Working as question-answer functions, they ask for prompting the data (or file name), the number of cells per sample, the number of genes etc. An example of the use of the interactive functions can be found in Appendix~\ref{App:INteractiveFN}. (ii) The direct usage of the package's \pkg{R} functions allows more flexibility and is illustrated in the following.

\subsection{Synthetic data generation}\label{subsec:sim_data}
We first generate a dataset of~$k=1000$ sample observations, where each sample consists of~$n=10$ cells. We choose a single-cell model with two populations, both of lognormal type, i.e.\ we use the LN-LN model. Let us assume that the overall population of interest is a mixture of $62\%$ of population~$1$ and $38\%$~of population~$2$, i.e.\ $p_1=0.62$. As population parameters we choose $\mu_1 = 0.47$, $\mu_2 = -0.87$ and~$\sigma=0.03$. Synthetic gene expression data for one gene is generated as follows:
%
\begin{Schunk}
\begin{Sinput}
R> library("stochprofML")
R> set.seed(10)
R> k <- 1000
R> n <- 10
R> TY <- 2
R> p <- c(0.62, 0.38)
R> mu <- c(0.47, -0.87)
R> sigma <- 0.03
R> gene_LNLN <- r.sum.of.mixtures.LNLN(k = k, n = n, p.vector = p, 
+    mu.vector = mu, sigma.vector = rep(sigma, TY))
\end{Sinput}
\end{Schunk}
\begin{figure}[!h]
\centering
\includegraphics[width=0.7\textwidth]{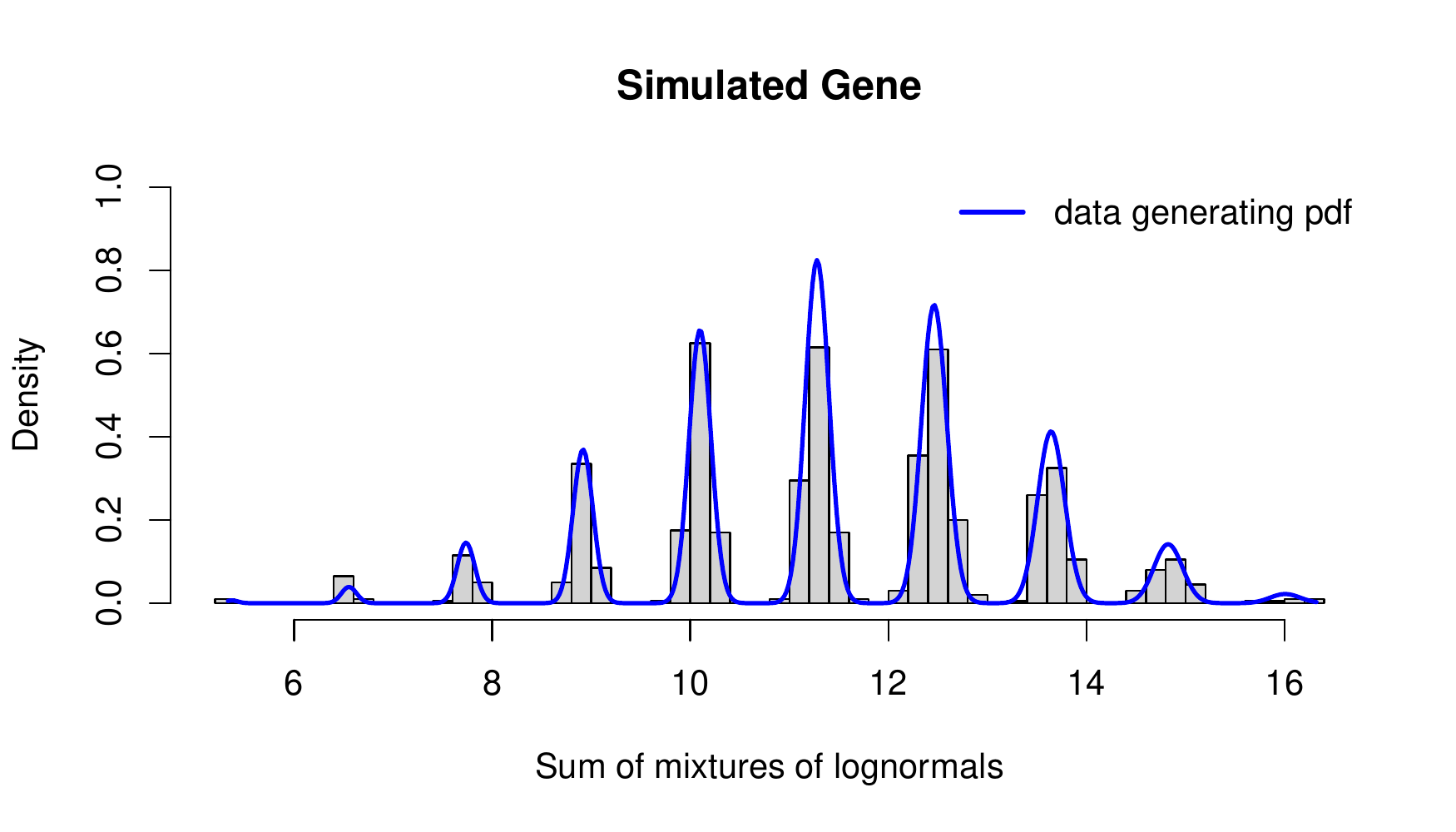}
\caption{Histogram of 1000 synthetic 10-cell observations, together with theoretical PDF. We assumed a two-population LN-LN model with parameters~$p=0.62$, $\mu_1=0.47$, $\mu_2=-0.87$ and~$\sigma=0.03$.}
\label{Fig:Code_Simgene}
\end{figure}

Figure~\ref{Fig:Code_Simgene} shows a histogram of the simulated data as well as the theoretical PDF of the 10-cell mixture. The following code produces this figure:
\begin{Schunk}
\begin{Sinput}
R> x <- seq(from = min(gene_LNLN), to = max(gene_LNLN), length = 500)
R> stochprofML:::set.model.functions("LN-LN")
R> y <- d.sum.of.mixtures(x, n, p, mu,rep(sigma,TY), logdens = FALSE)
R> hist(gene_LNLN, main = paste("Simulated Gene"), breaks = 50,
+    xlab = "Sum of mixtures of lognormals", ylab = "Density",
+    freq = FALSE, col = "lightgrey")
R> lines(x, y, col="blue", lwd = 2)
R> legend("topright", legend = "data generating pdf", col = "blue", 
+    lwd = 2, bty = "n")
\end{Sinput}
\end{Schunk}

\subsection{Parameter estimation}\label{subsec:param_est}
\begin{figure}[!h]
  \centering
 \begin{subfigure}[t]{0.4\textwidth}
 \centering
 \includegraphics[page=1, width=\textwidth]{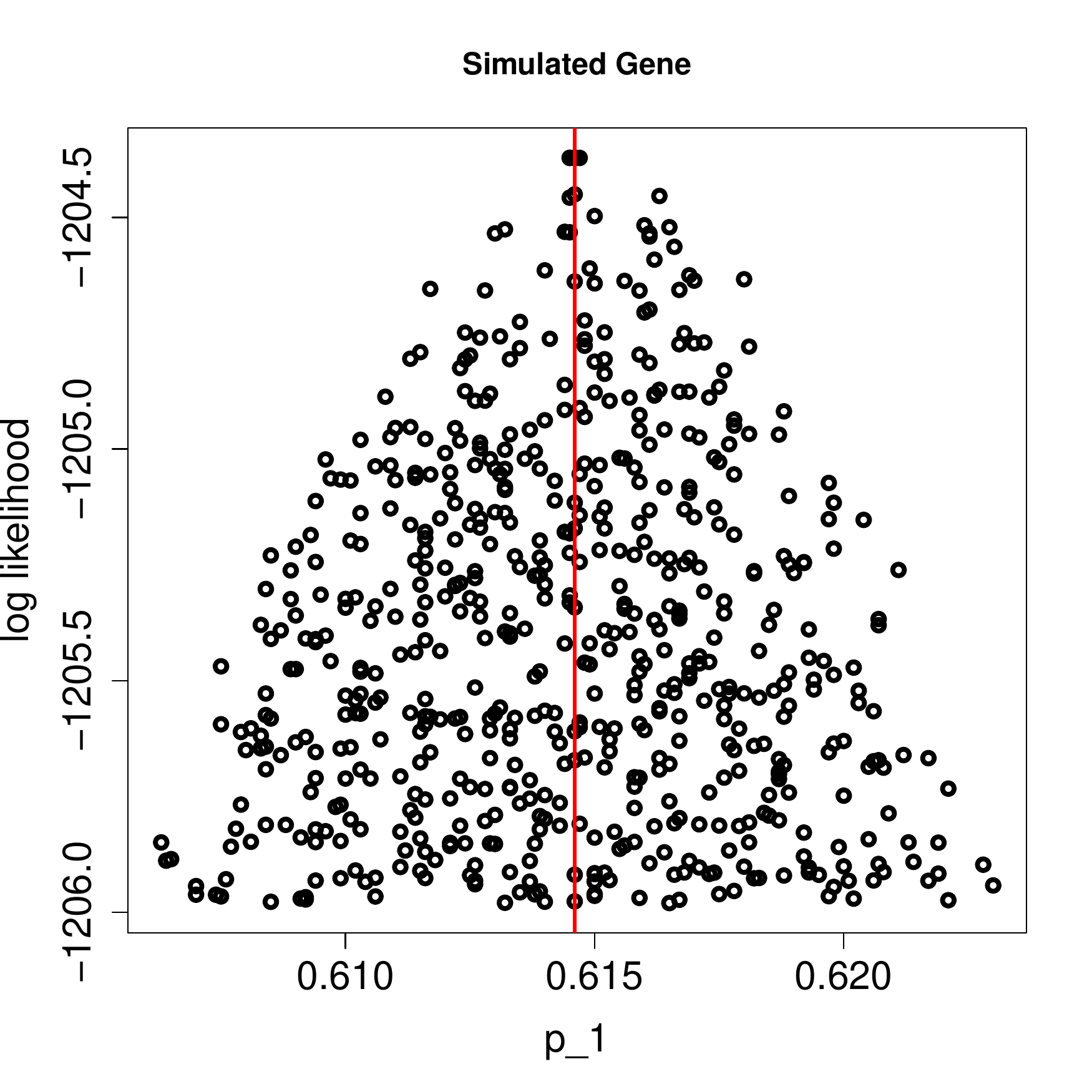}
 \end{subfigure}
 \begin{subfigure}[t]{0.4\textwidth}
 \centering
 \includegraphics[page=2, width=\textwidth]{Simulated_Gene_4.pdf}
 \end{subfigure}%
 ~

 \begin{subfigure}[t]{0.4\textwidth}
 \centering
 \includegraphics[page=3, width=\textwidth]{Simulated_Gene_4.pdf}
 \end{subfigure}
 \begin{subfigure}[t]{0.4\textwidth}
 \centering
 \includegraphics[page=4, width=\textwidth]{Simulated_Gene_4.pdf}
 \end{subfigure}
 \caption{Graphical output of the parameter estimation procedure for $p$, $\mu_1$, $\mu_2$ and $\sigma$ as described in Section~\ref{subsec:param_est}. Each point in the plots corresponds to one combination of values for~$p$, $\mu_1$, $\mu_2$ and~$\sigma$. Each plot depicts the functional relationship between one parameter (e.\,g.\  $p$ in the upper left panel) and the log-likelihood function, whilst the remaining three parameters are integrated out.}
 \label{Fig:stochProfOut_4}
 \end{figure}

Next, we show how the parameters used above can be back-inferred from the generated dataset using maximum likelihood estimation.
%
\begin{Schunk}
\begin{Sinput}
R> set.seed(20)
R> result <- stochprof.loop(model = "LN-LN", 
+    dataset = matrix(gene_LNLN, ncol = 1), n = n, TY = TY, 
+    genenames = "SimGene", fix.mu = FALSE, loops = 10,
+    until.convergence = FALSE, print.output = FALSE, show.plots = TRUE,
+    plot.title = "Simulated Gene", use.constraints = FALSE)
\end{Sinput}
\end{Schunk}
When the fitting is done, pressing <enter> causes \proglang{R} to show plots of the estimation process, see Figure~\ref{Fig:stochProfOut_4}, and displays the results in the following form.
\begin{Schunk}
\begin{Soutput}
Maximum likelihood estimate (MLE):
              p_1 mu_1_gene_SimGene mu_2_gene_SimGene             sigma 
           0.6146            0.4710           -0.8720            0.0310 

Value of negative log-likelihood function at MLE:
1204.371 

Violation of constraints:
none

BIC:
2436.373 

Approx. 95% confidence intervals for MLE:
                        lower      upper
p_1                0.60501813  0.6240938
mu_1_gene_SimGene  0.46972264  0.4722774
mu_2_gene_SimGene -0.87827704 -0.8657230
sigma              0.02967451  0.0323847

Top parameter combinations:
p_1 mu_1_ge_SimGene mu_2_gene_SimGene sigma   target
        p_1 mu_1_gene_SimGene mu_2_gene_SimGene sigma   target
[1,] 0.6146             0.471            -0.872 0.031 1204.371
[2,] 0.6146             0.470            -0.872 0.031 1204.371
[3,] 0.6146             0.471            -0.872 0.031 1204.371
[4,] 0.6146             0.470            -0.872 0.031 1204.371
[5,] 0.6145             0.471            -0.872 0.031 1204.371
[6,] 0.6146             0.471            -0.872 0.031 1204.371

\end{Soutput}
\end{Schunk}
Hence, the marginal confidence intervals cover the true parameter values.

\section{Simulation studies}
\label{sec:simstudies}

This section demonstrates the performance of the estimation depending on pool sizes (Section~\ref{subsec:Est_diff_setting}), true parameter values (Section~\ref{subsec:Est_diff_setting2}) and in case of uncertainty about pool sizes (Section~\ref{subsec:UNc_in_n_nr}). All scripts used in these studies can be found in our open GitHub repository \url{https://github.com/fuchslab/Stochastic_Profiling_in_R}.

\subsection{Simulation study on optimal pool size}\label{subsec:Est_diff_setting}

Stochastic profiling, i.e. the analysis of small-pool gene expression measurements, is a compromise between the analysis of single cells and the consideration of large bulks: Single-cell information is most immediate, but a fixed number~$k$ of samples will only cover~$k$ cells. In pools of cells, on the other hand, information is convoluted, but $k$~pools of size~$n$ cover~$n$ times as much material. An obvious question is the optimal pool size~$n$. The answer is not available in analytically closed form. We hence study this question empirically.

For this simulation study, first, we generate synthetic data for different pool sizes with identical parameter values and settings. Then, we re-infer the model parameters using the \pkg{stochprofML} algorithm. This is repeated 1,000 times for each choice of pool size, enabling us to study the algorithm's performance by simple summary statistics of the replicates.

The fixed settings are as follows: We use the two-population LN-LN model to generate data for one gene with $p_1 = 0.2$, $\mu_1 = 2$, $\mu_2 = 0$ and $\sigma = 0.2$. For each pool size we simulate~$k=50$ observations. The pool sizes are chosen in nine different ways: In seven cases, pool sizes are identical for each sample, namely~$n\in\{1,2,5,10,15,20,50\}$. In two additional cases, pool sizes are mixed, i.e.\  each of the~$k$ samples within one dataset represents a pool of different size~$n_i\in\{1,2,5,10\}$ or~$n_i\in\{10,15,20,50\}$.

\begin{figure}[!h]
\begin{center}
\includegraphics[width = \textwidth]{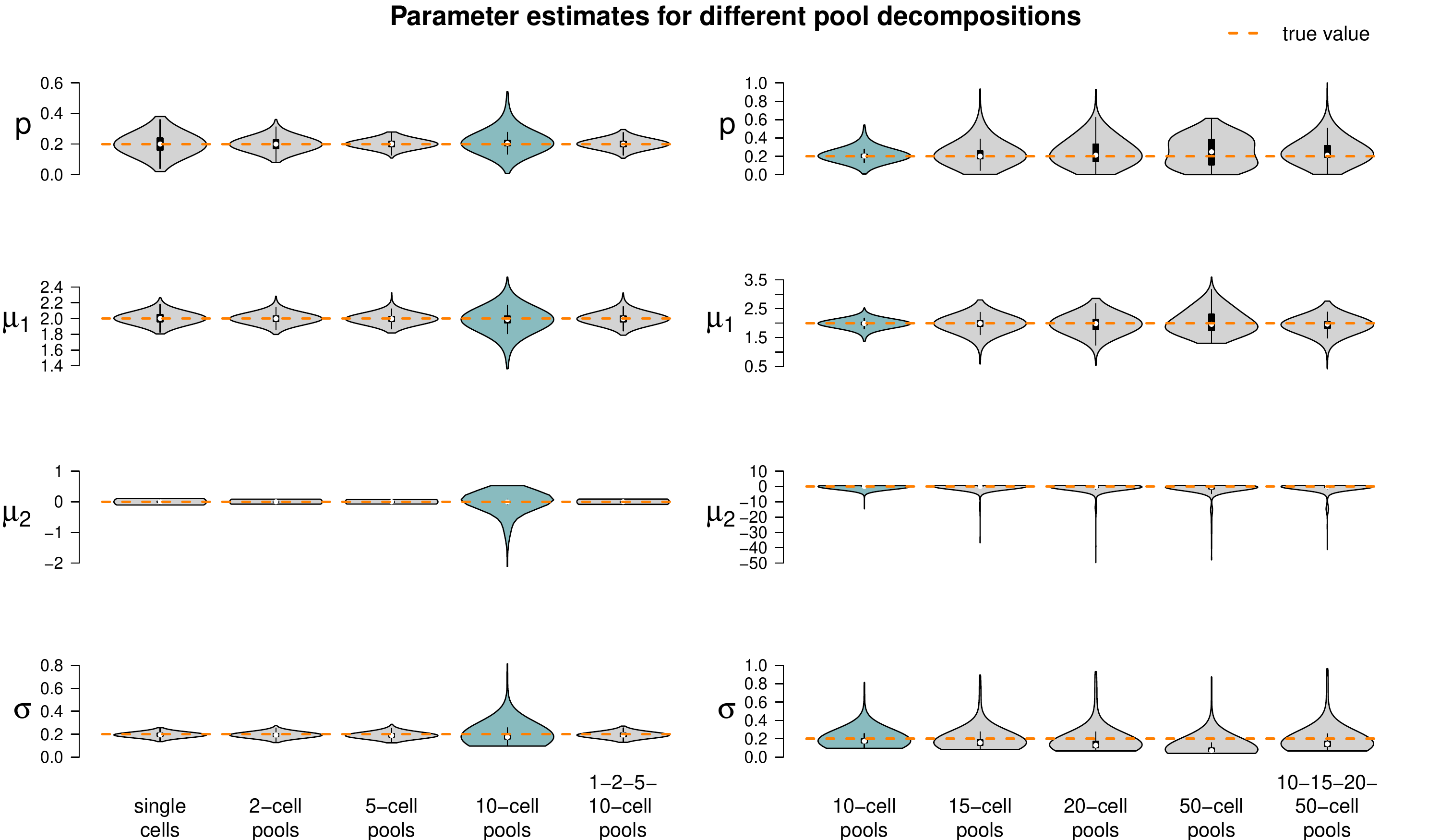} 
\caption{Violin plots of parameter estimates for two-population LN-LN model on 9,000 simulated datasets, i.\,e.\ on 1,000 datasets for each pool size composition. \emph{Left:} Results for single-cell, 2-cell, 5-cell, 10-cell pool and their mixture. \emph{Right:} Results for larger pool sizes, namely 10-, 15-, 20-, 50-cell pools and their mixture. 
\emph{Turquoise:} Results for 10-cell pools; these are repeated across the left and right panels. The true parameters are marked in orange.}
\label{Fig:Simstudy1_Set1_Overview}
\end{center}
\end{figure}

Figure~\ref{Fig:Simstudy1_Set1_Overview} summarizes the point estimates of the 1,000 datasets for each of the nine pool size settings. It seems that (for this particular choice of model parameter values) parameter estimation works reliably for pool sizes up to ten cells, with smaller variance from single-cells to 5-cells. This applies also for the mixture of pool sizes for the small cell numbers. For cell numbers larger than ten, the range of estimated values becomes considerably larger, but without obvious bias, which also applies to the mixture of the larger pool sizes. Appendix~\ref{App:SimStudy1} shows repetitions of this study for different choices of population parameters. The results there confirm the observations made here.

Figure~\ref{Fig:Simstudy1_Set1_Overview} suggests~$n=5$ or varying small pool sizes as ideal choices since its estimates show smaller variance than the other pool sizes. This simulation study, however, has been performed in an idealized \emph{in silico} setting: We did not include any measurement noise. In practice, however, it is well known that single-cells suffer more from such noise than samples with many cells. The ideal choice of pool size may hence be larger in practice.

\subsection{Simulation study on impact of parameter values}\label{subsec:Est_diff_setting2}

The underlying data-generating model obviously influences the ability of the maximum likelihood estimator to re-infer the true parameter values: Values of~$p_1$ close to $0.5$, small differences between~$\mu_1$ and~$\mu_2$ and large~$\sigma$ blur the data and complicate parameter inference in practice. In the simulation study of this section, we investigate the sensitivity of parameter inference and which scenarios could be realistically identified.

We use the same datasets as in the previous simulation study: The parameter choices from Section~\ref{subsec:Est_diff_setting} are considered as the standard and compared to those from Appendix~\ref{App:SimStudy1}. In detail, $p_1$ is reduced from~$0.2$ to~$0.1$ in one setting and increased to~$0.4$ in the next. $\mu_2$ is increased from~$0$ to $1$, and~$\sigma$ increases from~$0.2$ to~$0.5$. $\mu_1$ is kept fixed to~$2$ in all settings. As before, we consider 1,000 data sets for every parameter setting and compare the resulting estimates to the true values. This was done for all pool sizes considered in Section~\ref{subsec:Est_diff_setting}, but here we only comment on the results of the 10-cell pools and refer to Appendix~\ref{App:SimStudy1} for all other pool size settings. 

\begin{figure}[!h]
\begin{center}
\includegraphics[width = \textwidth]{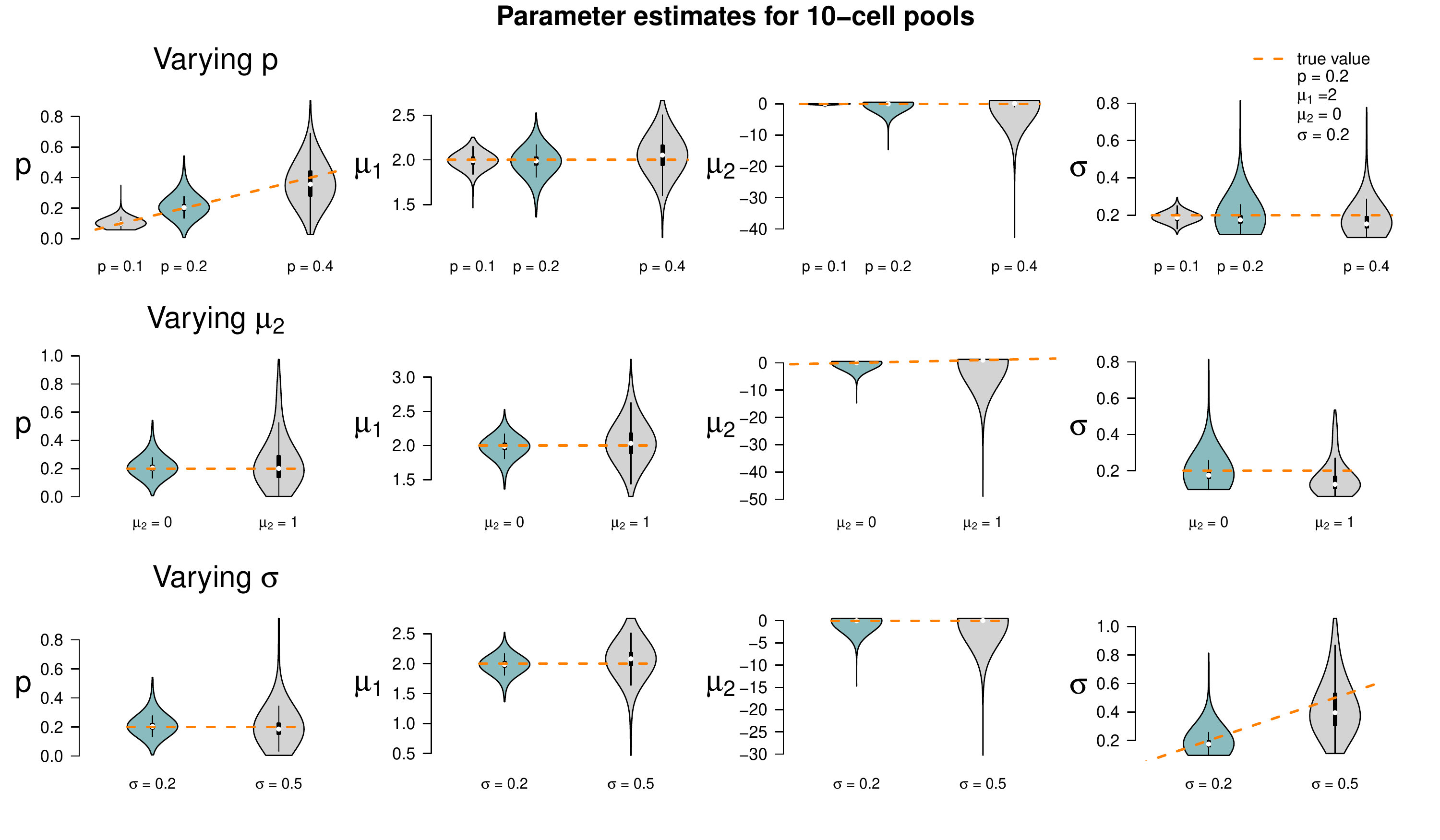} 
\caption{Violin plots of parameter estimates for two-population LN-LN model for varying parameters $p$, $\mu_2$ and~$\sigma$. Five parameter sets (see Appendix~\ref{App:SimStudy1}) were used to simulate 1,000 datasets from each of which they were back-inferred. Violin plots for the standard setting $p=0.2$, $\mu_1 = 2$, $\mu_2 = 0$ and $\sigma = 0.2$ are coloured turquoise. The true parameters used to simulate the data are marked in orange.}
\label{Fig:Simstudy2_10_Overview}
\end{center}
\end{figure}

Figure~\ref{Fig:Simstudy2_10_Overview} shows the results of the study. In each row of the plot, we compare the estimates of the datasets that were simulated with the standard parameters to the estimates of the datasets that were simulated with one of the parameters changed. Even if only one parameter is changed all parameters are estimated. Each violin accumulates the estimates of 1,000 datasets. For easier comparison, each of the twelve tiles shows the standard setting as turquoise violin, which means those are repeated in each row.

When changing the parameter values, they can still be derived without obvious additional bias, but accuracy decreases for increasing~$p$, decreasing~$\mu_2-\mu_1$ and increasing~$\sigma$ (with few exceptions). Result for other pool sizes (see Appendix~\ref{App:SimStudy1}) show that these observations can be transferred to other pool sizes with some additions: Larger pool sizes infer parameters more accurately if~$p$ is smaller. In an increased first population setting ($p = 40\%$), $\mu_1$ can be better inferred if the data set consists of smaller pools. For larger pools, the estimation of~$\mu_1$ and~$\mu_2$ works comparably well after increasing~$\mu_2$. In general, the estimation of $\sigma$ is the most difficult one; already in pools of ten cells with increased~$\mu_2$, $\sigma$ is underestimated. This worsens in larger pools.

\subsection{Simulation study on the uncertainty of pool sizes}\label{subsec:UNc_in_n_nr}

One key assumption of the \pkg{stochprofML} algorithm is that the exact number of cells in each cell pool is known. In \citet{janes_identifying_2010}, accordingly, ten cells were randomly taken from each sample by experimental design. However, different experimental protocols may not reveal the exact cell number: In  \citet{tirier_pheno-seq_2018}, for example, tissue samples were taken as whole cancer spheroids. Here, the cell numbers were experimentally unknown but estimated using light sheet microscopy and 3D image analysis. Since the \pkg{stochprofML} algorithm requires the pool sizes as input parameter, some estimate has to be passed to it. It is intuitively obvious that the better the prior knowledge about the cell pool sizes, the better the final model parameter estimate. In this simulation study, we investigate the consequences of misspecification.

\begin{figure}[!h]
	\begin{center}
		\includegraphics[width = 0.9\textwidth]{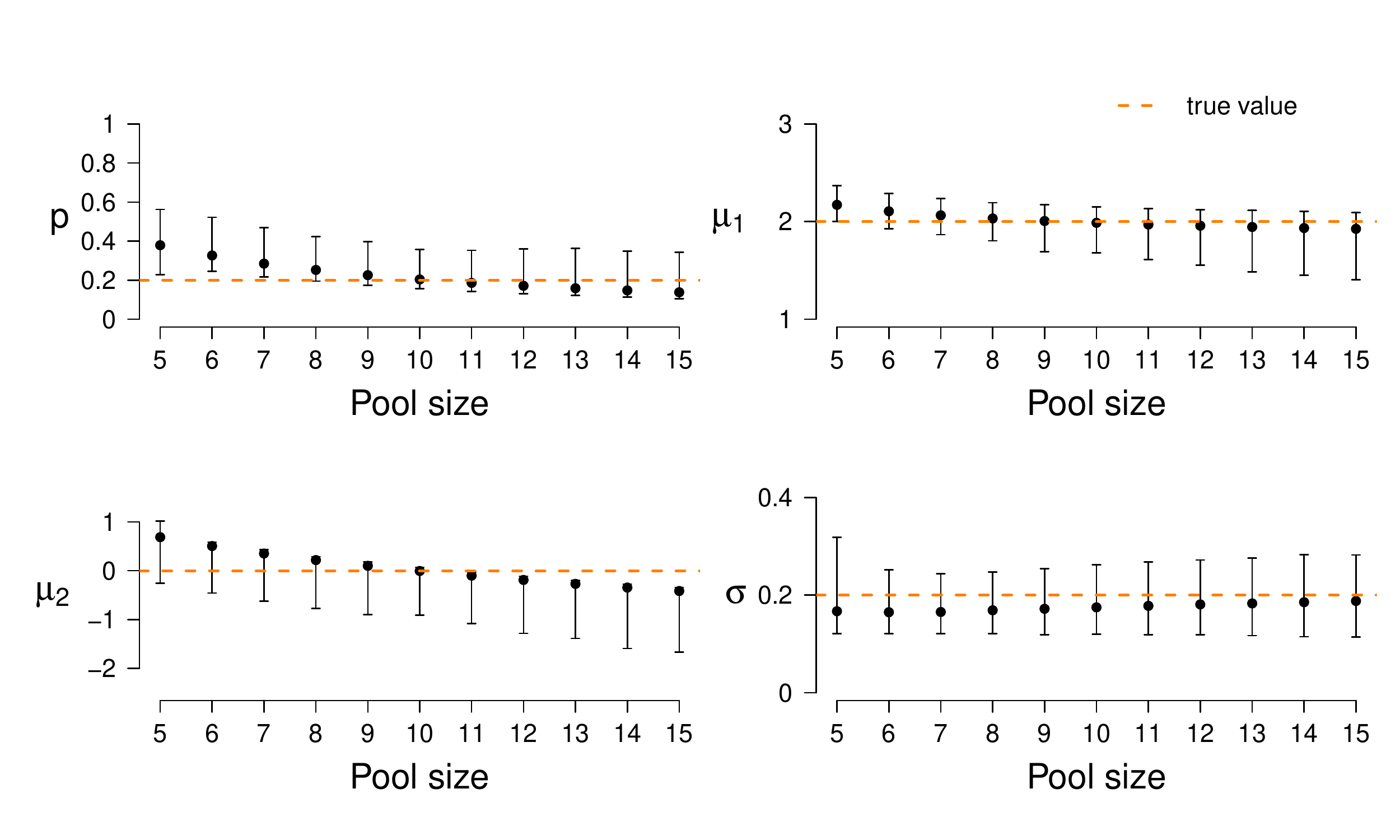} 
		\caption{Parameter estimates for (partly) misspecified pool sizes across 1,000 synthetic datasets: The true pool size is~$10$ in every dataset. The \pkg{stochprofML} algorithm, however, uses values from~$5$ to~$15$ as input parameter. Bars cover the range between the empirical 2.5\%- and 97.5\%-quantiles. The dots mark the empirical median, the orange line the true parameter values used for simulation.}
		\label{Fig:Uncertainty_10}
	\end{center}
\end{figure}

In a first simulation study, we reuse from Section~\ref{subsec:Est_diff_setting} the 1,000 synthetic 10-cell datasets. Each of these contains 50 10-cell samples, simulated with underlying model parameters $p = 0.2$, $\mu_1 = 2$, $\mu_2 = 0$ and $\sigma = 0.2$. As before, we re-infer the population parameters using the   \pkg{stochprofML} algorithm. This time, however, we use varying pool sizes from~$5$ to~$15$ as input parameters of the algorithm. This is a misspecification except for the true value~$10$. The resulting parameter estimates (empirical median and 2.5\%-/97.5\%-quantiles across the 1,000 datasets) are depicted in Figure~\ref{Fig:Uncertainty_10}. Estimates are optimal or at least among the best in terms of empirical bias and variance when using the correct pool size. With increasing assumed cell number, the estimates of~$p$ decrease, i.\,e.\  the fraction of cells from the higher expressed population is assumed to be smaller. This is a reasonable consequence of overestimating~$n$, because in this case the surplus cells are assigned to the second population with lower (or even close-to-zero) expression. Consequently, at the same time the estimates of~$\mu_2$ decrease to be even smaller. 

\begin{figure}[!h]
\begin{center}
\includegraphics[width = \textwidth]{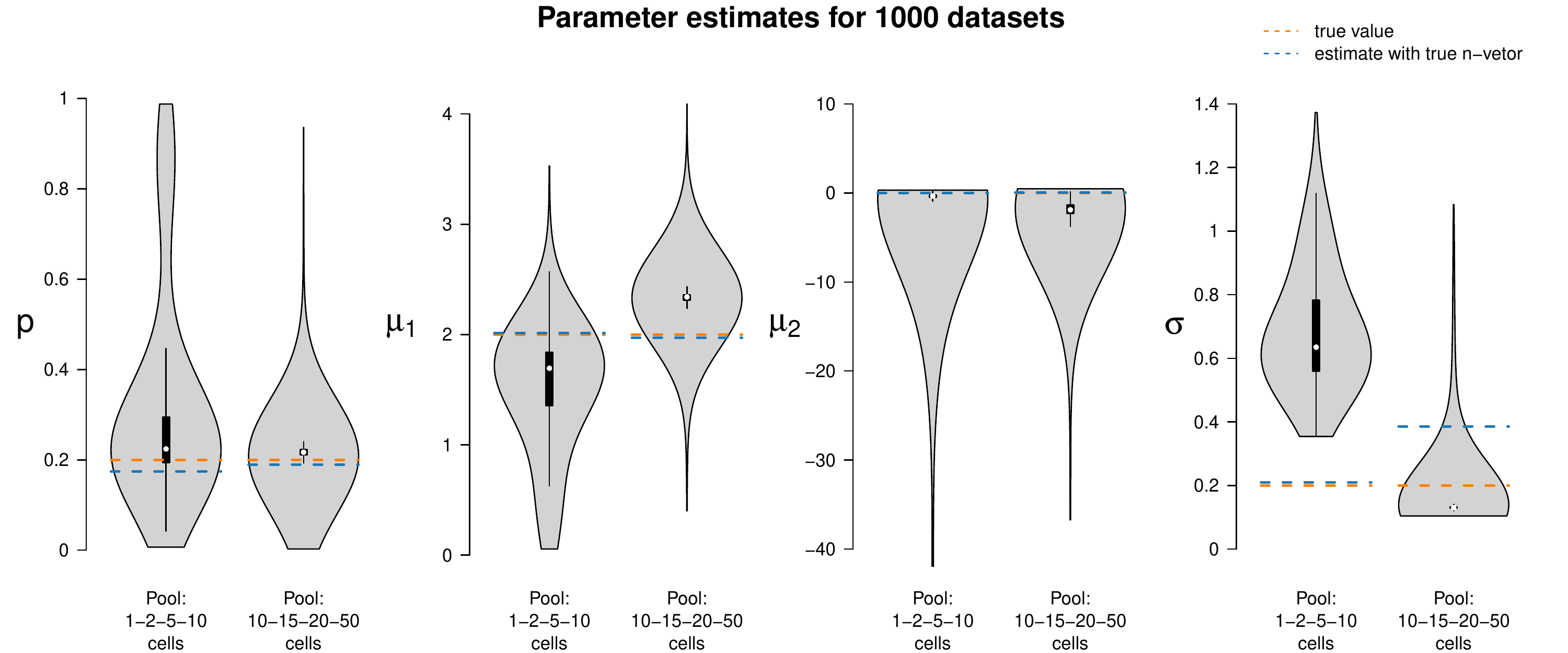} 
\caption{Parameter inference under misspecification of the cell pool size: Parameters are estimated for two datasets, one generated based on a pool size vector~$\vec{n}_\text{small}$ with values between~$1$ and~$10$ (\emph{left violin in each panel}); the other one based on a vector~$\vec{n}_\text{large}$ with values between~$10$ and~$50$ (\emph{right violin in each panel}). \emph{From left to right:} Estimates of~$p$, $\mu_1$, $\mu_2$ and~$\sigma$. The violins depict estimates across 1,000 estimation runs, where each relies on a randomly sampled misspecified pool size vector as described in the main text. \emph{Orange:} True parameters values. \emph{Light blue:} Estimates without misspecification of the pool size vector.}
\label{Fig:Simstudy3_Uncertainty_n}
\end{center}
\end{figure}

In a second simulation study, we use the two settings with mixed cell pool sizes as introduced in Section~\ref{subsec:Est_diff_setting}. One setting embraces cell pools with rather small cell numbers (single-, 2-, 5- and 10-cell samples), the other one pools with larger cell numbers (10-, 15-, 20- and 50-cell samples). For each of the two scenarios, we generate one dataset with 50 samples. We denote the true 50-dimensional pool size vectors by~$\vec{ n}_\text{small}$ and~$\vec{ n}_\text{large}$ and employ these vectors for re-estimating the model parameters~$p$, $\mu_1$, $\mu_2$ and~$\sigma$. Then, we estimate the parameters again for the same two datasets for 1,000 times, but this time using perturbed pool size vectors as input to the algorithm, introducing artificial misspecification. These 50-dimensional pool size vectors are generated as follows: For each component, we draw a Poisson-distributed random variable with intensity parameter equal to the respective component of the true vectors~$\vec{ n}_\text{small}$ or~$\vec{ n}_\text{large}$. Zeros are set to one, the minimum pool size. Figure~\ref{Fig:Simstudy3_Uncertainty_n} shows these $2\times 1,000$ parameter estimates as compared to the true parameter values and those for which the true size vectors~$\vec{ n}_\text{small}$ and~$\vec{ n}_\text{large}$ were used as input.

The violins of the estimates for the smaller cell pools (based on~$\vec{n}_\text{small}$) indicate that the estimates of~$p$ and~$\mu_1$ are fairly accurate, but the estimates of~$\mu_2$ have large variance, and~$\sigma$ is overestimated in all 1,000 runs. This is plausible as population~1 (the one with higher log-mean gene expression) is only present on average in 20\% of the cells; even when misspecifying the pool sizes, the cells of population~1 are still detectable since this is the population responsible for most gene expression. Consequently, all remaining cells are assigned to population~2, which has lower or even almost no expression. If the pool size is assumed too low, this second population will be estimated to have on average a higher expression; if it is assumed too large, the second population will be estimated to have a lower expression. This leads to a broader distribution and thus an overestimation of~$\sigma$.

The results for the larger cell pools (based on~$\vec{n}_\text{large}$) show a similar pattern. In this case, however, the impact of misspecification is less visible, as also confirmed by additional simulations in Appendix~\ref{App:SimStudy1}. For large cell pools, the averaging effect across cells is strong anyway and in that sense more robust. In the study here, the $\sigma$ parameter is often even better estimated when using a misspecified pool size vector than when using the true one.

Taken together, \pkg{stochprofML} can be used even if exact pool sizes are unknown. In that case, the numbers should be approximated as well as possible.

\section{Interpretation of estimated heterogeneity}
\label{sec:interpr}

We investigate what we can learn from the parameter estimates about the heterogeneous populations (Section~\ref{subsec:overlap}) and about sample compositions (Section~\ref{subsec:predictpooldecomposition}).

\subsection{Comparison of inferred populations}
\label{subsec:overlap}

The \pkg{stochprofML} algorithm estimates the assumed parameterized single-cell distributions underlying the samples and; as described in Section \ref{subsec:modelchoice}, we can select the most appropriate number of cell populations using the BIC. Assume we have performed this estimation for samples from two different groups, cases and controls. One may in practice then want to know whether the inferred single-cell populations are substantially different between the two groups, e.g.\  in case the estimated log-means~$\hat\mu_\text{cases}$ and~$\hat\mu_\text{controls}$ are close to each other. A related question is whether the difference is biologically relevant. 

We hence seek a method that can judge statistical significance and potentially reject the null hypothesis that two single-cell populations are the same; and at the same time allow the interpretation of similarity. Direct application of Kolmogorov-Smirnov or likelihood-ratio tests to the observed data is impossible here since the single-cell data is unobserved: We only measure the overall gene expression of pools of cells. Calculation of the Kullback-Leibler divergence of the two distributions would be possible; however, it is not target-oriented for our application where we seek an interpretable measure of similarity rather than a comparison between more than two population densities.

For our purposes, we use a simple intuitive measure of similarity~---~the overlap of two PDFs, that is the intersection of the areas under both PDF curves:
\begin{equation}
\text{OVL}(f,g)=\int_{-\infty}^\infty \min\{f(x),g(x)\}dx
\label{overlap_formula}
\end{equation}
for two continuous one-dimensional PDFs~$f$ and~$g$ \citep[see also][]{pastore_measuring_2019}. The overlap lies between zero and one, with zero indicating maximum dissimilarity and one implying (almost sure) equality. In our case, we are particularly interested in the overlap of two lognormal PDFs:
\begin{Code}
OVL_LN_LN <- function(mu_1, mu_2, sigma_1, sigma_2) {
    f1 <- function(x){dlnorm(x, meanlog = mu_1, sdlog = sigma_1) }
    f2 <- function(x){dlnorm(x, meanlog = mu_2, sdlog = sigma_2) }  
    f3 <- function(x){pmin(f1(x), f2(x))}
    integrate(f3, lower = 0, upper = Inf, abs.tol = 0)$value 
}
\end{Code}
\begin{figure}[!h]
\begin{center}
\includegraphics[width = \textwidth]{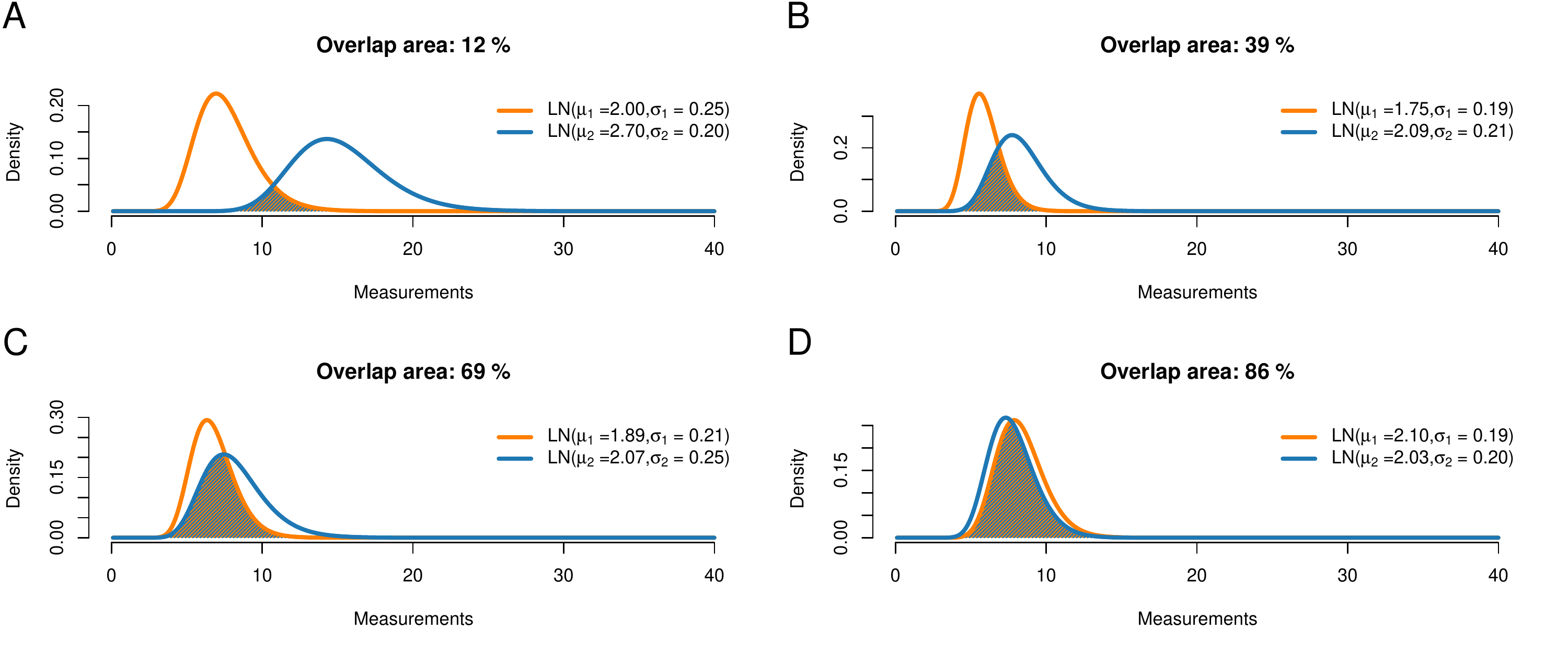} 
\caption{Four examples of overlapping PDFs, together with the overlap area as defined in Equation~\eqref{overlap_formula}.}
\label{Fig:Overlap_Example}
\end{center}
\end{figure}

Figure~\ref{Fig:Overlap_Example} shows examples of such overlaps. Here, the overlap ranges from 12\% for two quite different distributions to 86\% for two seemingly similar distributions. The question is where to draw a cutoff, that is, at what point we decide to label two distributions as different. Current literature considers two cases: Either the parametric case \citep[e.g.][]{inman_overlapping_1989}, where both distributions are given by their distribution families and parameter values; or the nonparametric case \citep[e.g.][]{pastore_measuring_2019}, where observations (but no theoretical distributions) are available for the two populations. Our application builds a third case: On the one hand, we want to compare two parametric distributions, but the model parameters are just given as estimates based on (potentially small) datasets, thus they are uncertain; on the other hand, we do not directly observe the single-cell gene expression but just the pooled one.

\begin{figure}[!h]
\begin{center}
\includegraphics[width = \textwidth]{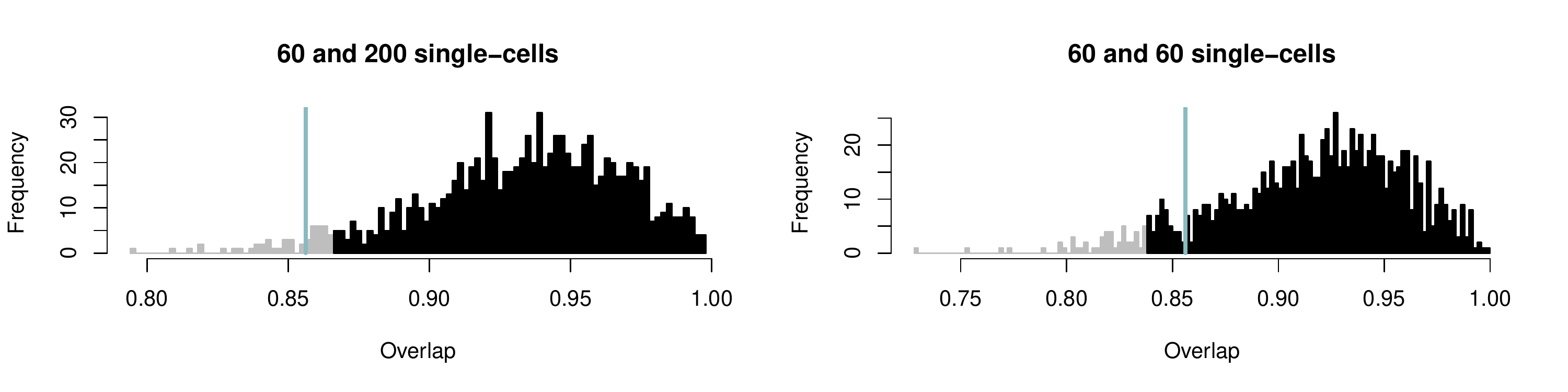} 
\caption{Variability of the overlap between the PDFs of the two distributions described in  Figure~\ref{Fig:Overlap_Example}D. The panels show histograms of~$N=1,000$ simulated overlap values which are simulated as described in the main text. \emph{Left:} We assume that the estimates of the orange distribution relied on 60 single cells and the blue distribution on 200 single cells. \emph{Right:} For both distributions, parameters are assumed to be estimated on 60 single cells. The 86\% overlap of the original PDFs from Figure~\ref{Fig:Overlap_Example}D, i.\,e.\    $\gL\gN(\hat\mu_{1,cases}=2.10,\hat\sigma^2_{cases}=0.19^2)$ and $\gL\gN(\hat\mu_{1,controls}=2.03,\hat\sigma_{controls}^2=0.20^2)$, is marked in turquoise. The light grey bars of the histogram indicate values below the empirical 5\%-quantile. If the original overlap falls into this range, we reject the null hypothesis that both distributions identical.}
\label{Fig:Overlap_Cutoff}
\end{center}
\end{figure}

To address this issue, we suggest to again take into account the original data that led to the estimated parametric PDFs. As an example, assume that we consider two sets of pooled gene expression, one for a group of cases and one for a group of controls. In both groups, pooled gene expression is available as 10-cell measurements, but the two groups differ in sample size. Let's say the cases contain 50 samples and the controls~100. We assume the LN-LN model with two populations and estimate the mixture and population parameters using the \pkg{stochprofML} algorithm separately for each group, leading to estimates $\hat{p}_\text{cases}$, $\hat{\mu}_{1,\text{cases}}$, $\hat{\mu}_{2,\text{cases}}$, $\hat{\sigma}_\text{cases}$ and $\hat{p}_\text{controls}$, $\hat{\mu}_{1,\text{controls}}$, $\hat{\mu}_{2,\text{controls}}$, $\hat{\sigma}_\text{controls}$. We now aim to assess whether the first populations in both groups have identical characteristics, i.\,e.\  whether $\gL\gN(\hat{\mu}_{1,\text{cases}},\hat{\sigma}^2_\text{cases})$ and~$\gL\gN(\hat{\mu}_{1,\text{controls}},\hat{\sigma}^2_\text{controls})$ are the same. 

Figure~\ref{Fig:Overlap_Example} displays the single-cell PDFs of the first population and their overlaps for various values of the estimates. For example, in Figure~\ref{Fig:Overlap_Example}D, the orange curve shows the single-cell PDF of population~1 inferred from the cases, yielding $\gL\gN(\hat{\mu}_{1,\text{cases}}=2.10,\hat{\sigma}^2_\text{cases}=0.19^2)$, and the blue one shows the inferred single-cell PDF of population~1 from the controls, $\gL\gN(\hat{\mu}_{1,\text{controls}}=2.03,\hat{\sigma}^2_\text{controls}=0.20^2)$. The overlap of these two inferred PDFs equals 86\%.

We now aim to test the null hypothesis that these populations are the same versus the experimental hypothesis that they are different. We perform a sampling-based test: Taking into account the inferred population probabilities $\hat{p}_\text{cases}$ and~$\hat{p}_\text{controls}$ and the number of samples and cells in the data, we can estimate the number of cells which the estimates~$\hat{\mm{\theta}}_\text{cases}$ and~$\hat{\mm{\theta}}_\text{controls}$ relied on. The larger this cell number, the less expected uncertainty about the estimated population distributions~$\gL\gN(\hat{\mu}_{1,\text{cases}},\hat{\sigma}^2_\text{cases})$ and~$\gL\gN(\hat{\mu}_{1,\text{controls}},\hat{\sigma}^2_\text{controls})$ (neglecting the impact of pool sizes).

In our example, let $\hat{p}_\text{cases}=12\%$. Then, approximately 12\% of the 500 cells from the cases group ($50$ $\times$ 10-cell samples) belonged to population~1, that is 60 cells. For $\hat{p}_\text{controls}=20\%$, 200 cells were expected to be from the first population (that is 20\% of 1,000 cells, coming from the 100 $\times$ 10-cell measurements for the controls). In our procedure, we compare parameter estimates that are based on the respective numbers of single cells, i.\,e.\  60 cells for cases and 200 cells for controls. We perform the following steps:
\begin{itemize}
	\item Calculate $\text{OVL}_\text{original}$, the overlap of the PDFs of $\gL\gN(\hat{\mu}_{1,\text{cases}}=2.10,\hat{\sigma}^2_\text{cases}=0.19^2)$ and $\gL\gN(\hat{\mu}_{1,\text{controls}}=2.03,\hat{\sigma}_\text{controls}^2=0.20^2)$.
	\item Under the null hypothesis, the two distributions are identical. We approximate the parameters of this identical distribution as $\tilde{\mu}_{1,\text{mean}}= (\hat{\mu}_{1,\text{cases}}+\hat{\mu}_{1,\text{controls}})/2$ and \linebreak $\tilde{\sigma}_\text{mean}= (\hat{\sigma}_\text{cases}+\hat{\sigma}_\text{controls})/2$. 
	\item Repeat $N = 1,000$ times:
	\begin{itemize}
		\item Draw dataset~$A$ of size~$60$ from $\gL\gN(\tilde{\mu}_{1,\text{mean}},\tilde{\sigma}_\text{mean}^2)$.
		\item Draw dataset~$B$ of size~$200$ from $\gL\gN(\tilde{\mu}_{1,\text{mean}},\tilde{\sigma}_\text{mean}^2)$.
		\item Estimate the log-mean and log-sd for these two datasets using the method of maximum likelihood, yielding~$\hat{\mu}_A$, $\hat{\sigma}_A$, $\hat{\mu}_B$ and~$\hat{\sigma}_B$.
		\item Calculate $\text{OVL}\left(f_{\gL\gN(\hat{\mu}_A,\hat{\sigma}_A^2)},f_{\gL\gN(\hat{\mu}_B,\hat{\sigma}_B^2)}\right)$.
	\end{itemize}
	\item Sort the $N$ overlap values and select the empirical 5\% quantile~$\text{OVL}_{0.05}$.
	\item Compare the overlap from the original data to this quantile:
	\begin{itemize}
		\item If $\text{OVL}_{original}\leq\text{OVL}_{0.05}$, the null hypothesis that both populations are the same can be rejected.
		\item If $\text{OVL}_{original}>\text{OVL}_{0.05}$, the null hypothesis cannot be rejected.
	\end{itemize}
\end{itemize}
This proceeding is related to the idea of parametric bootstrap with the difference that our original data is on the $n$-cell level and the parametrically simulated data is on the single-cell level.

The left panel of Figure~\ref{Fig:Overlap_Cutoff} shows one outcome of the above-described procedure (i.\,e.\  the stochastic, sampling-based algorithm was run once) with the above-specified values of the parameter estimates. Here, $\text{OVL}_{original}$ lies in the critical range such that we reject the null hypothesis that the gene expression of the populations in question stem from the same lognormal distribution. We thus assume a difference here. The right panel of Figure~\ref{Fig:Overlap_Cutoff} demonstrates the importance of taking into account the number of cells which the original estimates were based on: Here, we show one outcome of the above described steps, but this time we assume that for the control group there were only 30 10-cell samples (i.e. 300 cells in total). With the same population fraction as before ($\hat{p}_\text{controls}=20\%$), the datasets~$B$ now contain only 60 cells. Here, the value $\text{OVL}_{original}$ does not fall into the critical range, and therefore we would not reject the null hypothesis that the two populations of interest are the same.

When testing for heterogeneity for several genes simultaneously, multiple testing issues should be taken into account. However, genes will in general not be independent from each other.

\subsection{Prediction of sample compositions}\label{subsec:predictpooldecomposition}

The \pkg{stochprofML} algorithm estimates the parameters of the mixture model, i.\,e.~---~in case of at least two populations~---~the probability for each cell within a pool to fall into the specific populations. It does \emph{not} reveal the individual pool compositions. In some applications, however, exactly this information is of particular interest. Here, we present how one can infer likely population compositions of a particular cell pool.

\begin{figure}[!b]
\begin{center}
\includegraphics[width=\textwidth]{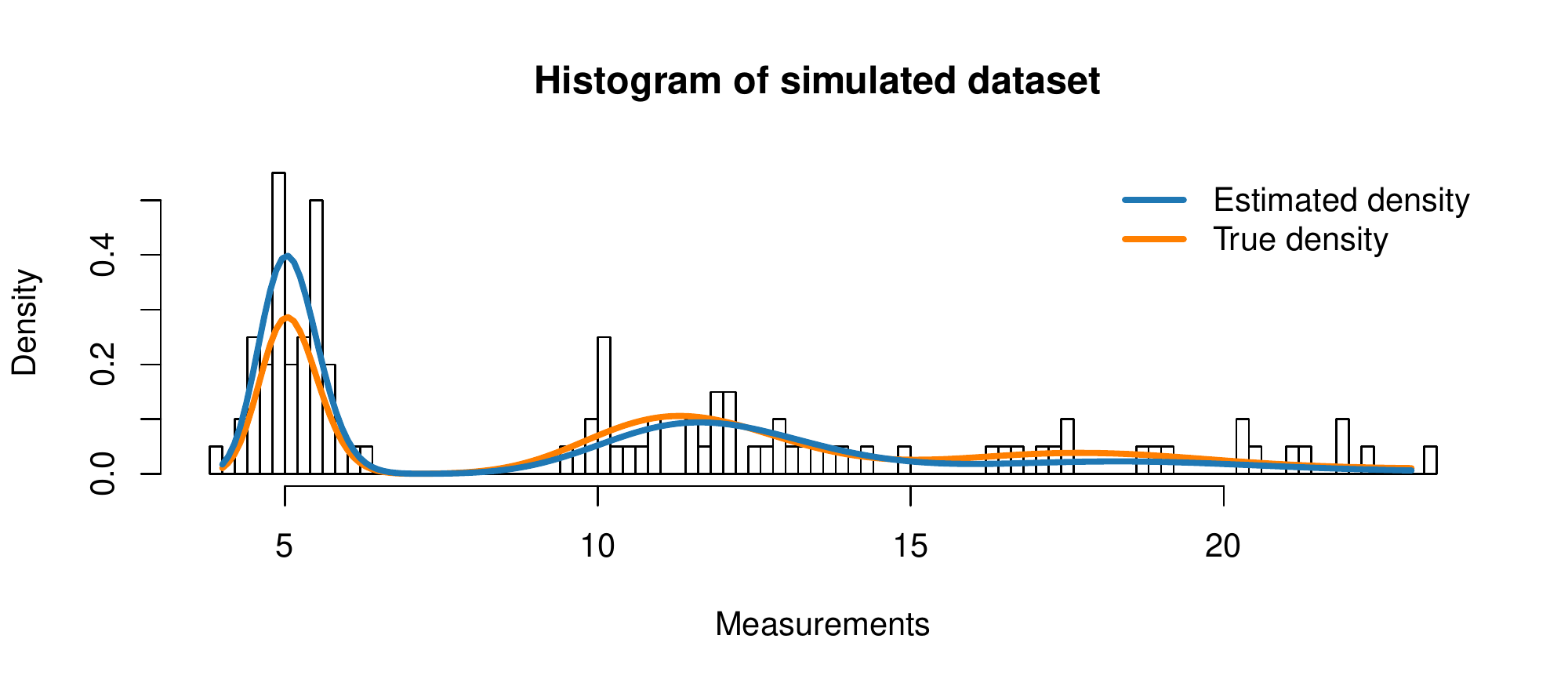} 
\caption{Histogram of simulated data underlying the prediction of cell pool compositions in Figure~\ref{Fig:WellPred_ProbGene2}A: 100~synthetic 5-cell measurements arising from the LN-LN model with two populations with parameters $\mm{p}=(0.2,0.8)$, $\mm{\mu}=(2,0)$ and~$\sigma=0.2$. The PDF with true model parameters is shown in orange, the PDF with estimated parameters $\mm{\hat{p}}=(0.14, 0.86)$, $\mm{\hat{\mu}}=( 2.04, 0)$ and $\hat{\sigma}= 0.20$ in blue.}
\label{Fig:WellPred_HistGene2}
\end{center}
\end{figure}

This is done in a two-step approach via conditional prediction: First, one estimates the model parameters from the observed pooled gene expression, i.\,e.\  one obtains an estimate~$\hat{\mm{\theta}}$ of~${\mm{\theta}}$. Then, one assumes that~$\mm{\theta}$ equals~$\hat{\mm{\theta}}$ and derives the most probable population composition via maximizing the conditional probability of a specific composition given the pooled gene expression (for calculations, see Appendix~\ref{App:WellPred}).

We evaluate this procedure via a simulation study. As before, we simulate data using the \pkg{stochprofML} package. In particular, we use the LN-LN model with two populations with parameters $\mm{p}=(0.2,0.8)$, $\mm{\mu}=(2,0)$ and~$\sigma=0.2$. Each simulated measurement shall contain the pooled expression of~$n=5$ cells, and we sample~$k=100$ such measurements. We store the original true cell pool compositions from the data simulation step in order to later compare the composition predictions to the ground truth.

\begin{figure}[!t]
\begin{center}
\includegraphics[width=0.95\textwidth]{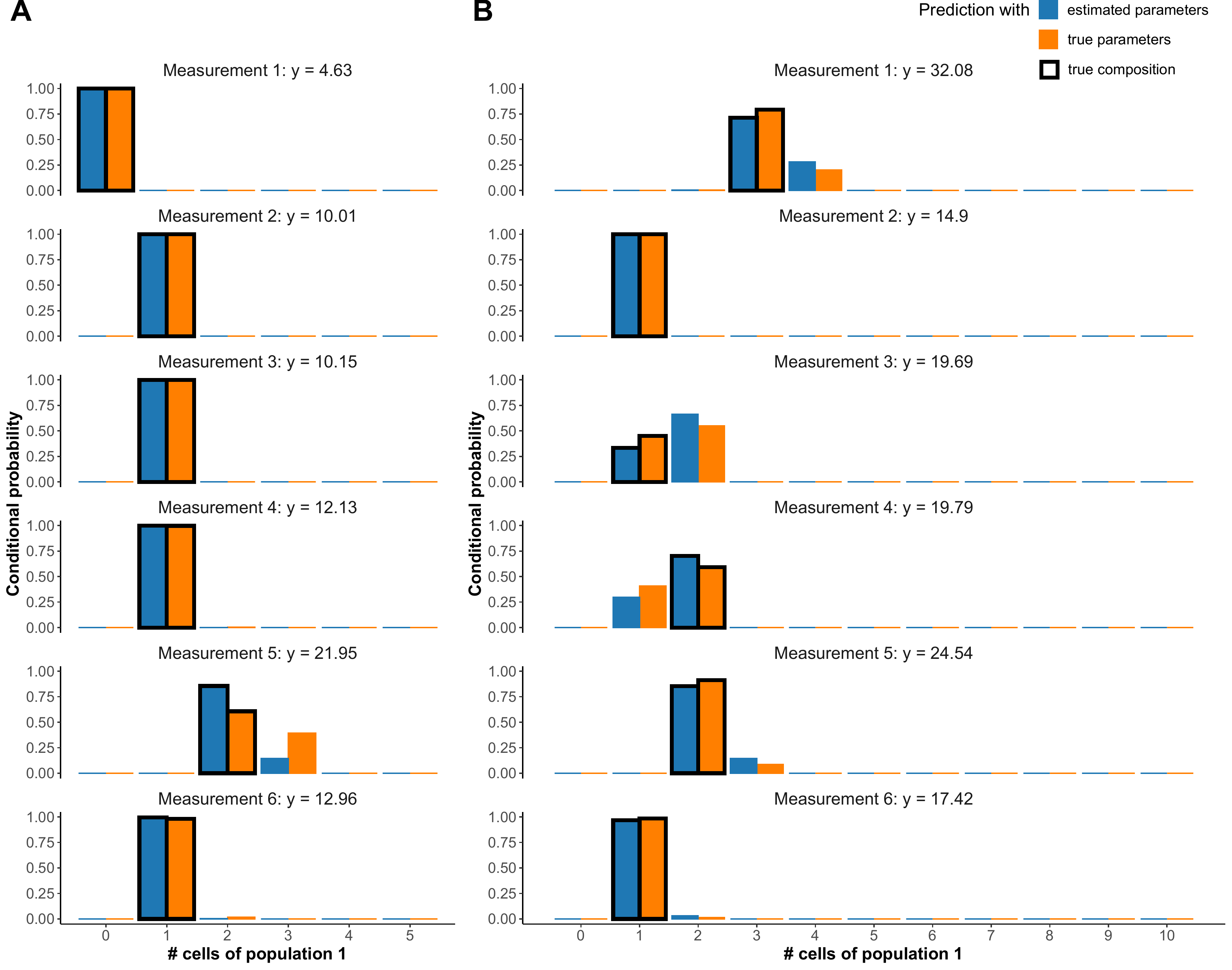}
\caption{Estimation of cell pool compositions in the two-population LN-LN model: Conditional probabilities of numbers of cells from the first population in the first six measurements of the synthetic datasets described in the main text and in Figure~\ref{Fig:WellPred_HistGene2}, given the respective pooled gene expression measurement. Blue bars show the conditional probabilities using estimated model parameters, and orange bars show those when using the true parameters. True cell numbers from the first population are marked with a black box around the bars. Results for (A) simulated 5-cell data and, (B) 10-cell data.}
\label{Fig:WellPred_ProbGene2}
 \end{center}
\end{figure} 
Having generated the synthetic data, we apply \pkg{stochprofML} to estimate the model parameters~$\mm{p}$, $\mm{\mu}$ and~$\sigma$. Figure~\ref{Fig:WellPred_HistGene2} shows a histogram of one simulated data set along with the PDF of the true population mixture and the PDF of the estimated population mixture (that is the LN-LN model with parameters $\mm{\hat{p}}=(0.14, 0.86)$, $\mm{\hat{\mu}}=( 2.04, 0)$ and $\hat{\sigma}= 0.20$).

Next, we calculate the conditional probability mass function (PMF; see Appendix \ref{App:WellPred} for details) for each possible population composition conditioned on the particular pooled gene expression measurement. Figure~\ref{Fig:WellPred_ProbGene2}A and Table~\ref{Tab:WellPred} show results for the first six (out of 100) pooled measurements.

In particular, Figure~\ref{Fig:WellPred_ProbGene2}A displays the conditional PMF of all possible  compositions (i.\,e.\  $k$ times population~1 and~$5-k$ times population 2 for $k\in\{0,1,\ldots,5\}$). Blue bars stand for these probabilities when $\hat{\mm{\theta}}$ is used as model parameter value. Orange stands for the hypothetical case  where the true value~$\mm{\theta}$ is known and used. These two scenarios are in good agreement with each other.


\begin{table}[t!]
	\centering
	\begin{tabular}{c | l | c c c c c c |c}
		
		\multicolumn{2}{c|}{\multirow{2}{*}{ \makecell{\bfseries Estimator for \# \\ \bfseries of cells in pop.\  1 }}} & 
		\multicolumn{6}{c|}{\bfseries Measurement index} &
		\multirow{2}{*}{ \makecell{\bfseries\# of hits}}\\ 
		\cline{3-8}
		\multicolumn{2}{c|}{}&\bf 1& \bf 2& \bf 3& \bf 4&\bf 5&\bf 6&\\ 
		\hline
		\multirow{2}{*}{{\makecell{Estimated \\ parameters}}}& Mean & 0.00&1.00&1.00&1.00&2.14&1.01& 98\\
		&  MLE (CI)& 0 (0,0)&1 (1,1)& 1 (1,1) &1 (1,1)& 2 (2,3)&1 (1,1)& 98 (100)\\   
		\hline
		\multirow{2}{*}{{\makecell{True \\ parameters}}}& Mean & 0.00&1.00&1.00&1.00&2.39&1.02&97\\
		&  MLE (CI)& 0 (0,0)&1 (1,1)& 1 (1,1) &1 (1,1)& 2 (2,3)&1 (1,1)& 97 (100)\\
		\hline
		\multicolumn{2}{c|}{ \multirow{2}{*}{ \makecell{True \# of cells  \\  from population 1}}} & \multirow{2}{*}{0} & \multirow{2}{*}{1} & \multirow{2}{*}{1} & \multirow{2}{*}{1}& \multirow{2}{*}{2}& \multirow{2}{*}{1} & \\
		\multicolumn{2}{c|}{}& &  & & & & &
	\end{tabular}
\caption{Estimates of numbers of cells from the first population in the simulated 5-cell data described in Figures~\ref{Fig:WellPred_HistGene2} and~\ref{Fig:WellPred_ProbGene2}A and in the main text. \emph{Columns:} Estimation results for the first six measurements from the datasets and (last column) summary across all 100~samples. \emph{Rows:} Estimation of cell numbers are based on conditional probabilities that use either the estimated model parameters (rows~1 and~2, corresponding to blue bars in Figure~\ref{Fig:WellPred_ProbGene2}A) or the true values (rows~3 and~4, orange bars). Within each of these two choices one can consider the mean number of cells from population~1 as determined by the conditional probabilities (rows~1 and~3) or the MLE that maximizes the conditional probabilities (rows~2 and~4, first value) including a 95\% confidence interval that covers at least 95\% of the conditional probability mass (rows~2 and~4, in parentheses). The last row shows the true pool composition. The last column shows for each estimator how many of the 100 cell numbers were inferred correctly (defined as follows: rounded mean is exact match; MLE is exact match; CI includes correct number).}
\label{Tab:WellPred}
\end{table}

We regard the most likely sample composition to be the one that maximizes the conditional PMF (maximum likelihood principle). The true composition (ground truth) is marked with a black box around the blue and orange bars. We observe in Figure~\ref{Fig:WellPred_ProbGene2}A that the composition is in all six cases inferred correctly and mostly unambiguously. Only for the fifth measurement, there is visible probability mass on a composition other than the true one. In fact, it is the only pool (out of the six considered ones) with two cells from the first population. Alternatively to the maximum likelihood estimator, one can also regard the expected composition~---~the empirical weighted mean of numbers of cells in the first population~---~or confidence intervals for this number. The respective estimates for the first six measurements of the dataset are shown in Table~\ref{Tab:WellPred}. The results are consistent with the interpretation of Figure~\ref{Fig:WellPred_ProbGene2}A.

Certainly, the precision of the prediction depends on the employed pool sizes, the underlying true model parameters and how reliably these were inferred during the first step. We showed in Section~\ref{sec:simstudies} that larger cell pools lead to less precise parameter inference. Hence, we repeat the prediction of sample compositions on another dataset, this time based on $10$-cell pools. All other parameters remain unchanged. The resulting conditional probabilities are depicted in Figure~\ref{Fig:WellPred_ProbGene2}B. Since $p=0.2$, one expects on average two cells to be from the first population in each 10-cell pool. As in the previous $5$-cell case, most predictions show a clear pattern. However, probability masses are spread more widely. Measurements~3 and~4 exemplify that almost identical gene expression measurements ($y=19.69$ and $y=19.79$) can arise from different underlying pool compositions (two times population~1 in measurement~3 vs.\  three times population~1 in measurement 4). For more similar population parameters, the estimation will get worse, which will then propagate to the well composition prediction. In such cases, to predict the pool compositions, one may use additional parallel measurements of other genes that might separate the population better by their different expression profiles while the pool composition stays the same across genes.


\section{Summary and Discussion}\label{sec:sum_dis}

With the \pkg{stochprofML} package, we provide an environment to profile gene expression measurements obtained from small pools of cells. Experimentalists may choose this approach if single-cell measurements are impossible in their lab (e.\,g.\ for bacteria), if the drop-out rate is high in single-cell libraries, if budget or time are limited, or if one prefers to avoid the stress which is put on the cells during separation. The latest implementation even allows to combine information from different pool sizes, in particular, to simultaneously analyze single-cell and and n-cell data. 

We demonstrated the usage and performance of the \pkg{stochprofML} algorithm in various examples and simulation studies. These have been performed in an idealized \emph{in silico} environment. This should be kept in mind when incorporating the results into experimental planning and analysis. The interpretation of heterogeneity will be informative if based on a good model estimate. The assumption of independent expression across genes within the same tissue sample is a simplification of nature that leads to less complex parameter estimation. The optimal pool size with respect to bias and variance of the corresponding parameter estimators will depend on unknown properties such as numbers of populations and their characteristics, and also on the relationship between the pool size and the amount of technical measurement noise. The latter aspect has been excluded from the studies here but further supports the application of stochastic profiling.


\section*{Computational details}
We used \proglang{R} version 3.5.3 \citep{Rversion}. In addition to our \pkg{stochprofML} version 2.0.1 \citep{stochProfPkg}, we attached the following \proglang{R} packages \pkg{MASS} version 7.3-51.1 \citep{Masspkg}, \pkg{numDeriv} version 2016.8-1.1 \citep{numDerivPkg}, \pkg{EnvStats} version 2.3.1 \citep{EnvStats-book}, 
\pkg{vioplot} version 0.3.4. \citep{vioplotpkg}, 
\pkg{zoo} version 1.8-7 \citep{zoo_pkg},
\pkg{sm} version 2.2-5.6 \citep{sm_pkg},
\pkg{cowplot} version 1.0.0 \citep{cowplot_pkg}, 
\pkg{ggplot2} version 3.2.1 \citep{ggplot2Package}, \pkg{knitr} version 1.27 \citep{knitrPackage}, and \pkg{RcolorBrewer} version 1.1-2 \citep{RColorBrewerPackage}. All calculations were performed on a 64-bit x86\_64-redhat-linux-gnu platform running under Fedora 28. 



\section*{Acknowledgments}

\noindent Our research was supported by the German Research Foundation within the SFB 1243, Subproject A17, by the German Federal Ministry of Education and Research under grant number 01DH17024, by the Helmholtz Initiating and Networking Funds, Pilot Project Uncertainty Quantification, and by the National Institutes of Health under grant number U01-CA215794. We thank Susanne Amrhein and Xiaoling Zhang for code contributions to the simulation studies and Merc\`e Gar\'i for feedback. The paper was designed by LA and CF. CF developed the first version of the software, LA developed the second version and performed the simulation studies. LA and CF wrote the paper.


\bibliography{stochprofMLBib}

\newpage
\begin{appendix}
\section{PDF of $n$-cell measurements of $T$ cell populations}\label{App:Trinom}
Equation~\eqref{multdistr} in Section~\ref{subsubsec:spmodel} displays the PDF of the overall gene expression of a cell pool of size~$n_i$, where each single cell from the pool can originate from any of~$T$ cell populations. We derive this PDF here. To make it easier to follow the lengthy calculations, we build up the formulas in four steps: We start with the simplest case of 2-cell measurements in the presence of two populations. Then, we continue with 2-cell samples and three populations. Next, we increase the cell number to~$n$ and finally raise the population number to~$T$.

\subsection*{PDF of 2-cell measurements of two populations ($n=2$, $T=2$)}
First, we derive the PDF of a measurement~$y$ of a 2-cell pool, i.e. of~$Y=X_1+X_2$. Assume we know that two cell populations are present in the tissue, and each of them is described by an individual distribution. In this section, we denote the univariate population distributions by $\gD_h$, $h=1,\ldots,T=2$. In Section~\ref{sec:models}, they are replaced by the distributions that were presented in Section~\ref{sec:statdistr}: $\gL\gN(\mu,\sigma^2)$ or $\gE \! \gX \! \gP(\lambda)$ with population-specific parameter values. For now, we consider for~$j=1,2$
\begin{align*}
X_j \overset{iid}{ \sim}  \begin{cases}
    \gD_1 & \text{with probability } p_1\\
    \gD_2 & \text{with probability } 1-p_1,
    \end{cases}
\end{align*}
where~$p_1 \in [0,1]$. Hence, the PDF of each $X_j$ is
\begin{align*}    
f_X(x) = p_1 f_{\gD_1}(x) + p_2 f_{\gD_2}(x)
\end{align*}
with $p_2 = 1-p_1$. To determine the distribution of $Y$, we use the convolution of the single-cell PDFs, which are the same functions~$f_X$ for both~$X_1$ and~$X_2$:
\begin{align*}
f_Y(y) = \int_0^y &f_{X}(x_1) f_{X}(y-x_1) dx_1 \\
= \int_0^y &\Biggl( \Bigl[ p_1 f_{\gD_1}(x_1) + p_2 f_{\gD_2}(x_1)\Bigr] \Bigl[ p_1 f_{\gD_1}(y-x_1) + p_2 f_{\gD_2}(y-x_1)\Bigr] \Biggr) dx_1 \\
= \int _0^y & \Biggl(  p_1^2 f_{\gD_1}(x_1) f_{\gD_1}(y-x_1)+ p_2^2 f_{\gD_2}(x_1) f_{\gD_2}(y-x_1)\\
&+ p_1 p_2 f_{\gD_1}(x_1) f_{\gD_2}(y-x_1) + p_2 p_1 f_{\gD_2}(x_1) f_{\gD_1}(y-x_1)\Biggr)dx_1\\
= p_1^2 & \int_0^y f_{\gD_1}(x_1) f_{\gD_1}(y-x_1) dx_1 + p_2^2 \int_0^y f_{\gD_2}(x_1)f_{\gD_2}(y-x_1) dx_1 \\
&+ p_1 p_2 \int_0^y f_{\gD_1}(x_1) f_{\gD_2}(y-x_1) dx_1 + p_2 p_1 \int_0^y f_{\gD_2}(x_1) f_{\gD_1}(y-x_1)dx_1.
\end{align*}
Each of these integrals $\int_0^y f_{\gD_i}(x_1) f_{\gD_j}(y-x_1) dx_1$ is the PDF of a random variable~$Z_1+Z_2$ evaluated at~$y$, where $Z_1 \sim \gD_i$ and $Z_2 \sim \gD_j$ are independent. This holds for both~$i\neq j$ and~$i=j$. We denote this density by~$f_{i,j}$. All together, we get
\begin{align*}
f_Y(y) = \sum_{i=1}^2 \sum_{j=1}^2 p_i p_j f_{i,j}(y).
\end{align*}
An alternative formulation is
\begin{align}
f_Y(y) = \sum_{\ell_1=0}^2 \binom{2}{\ell_1} p_1^{\ell_1} p_2^{\ell_2} f_{(\ell_1,\ell_2)}(y),
\label{2cell2poppdf}
\end{align}
where $\ell_1$ and~$\ell_2= 2-\ell_1$ show how often a cell of population~1 and~2 is present in the pool. 
The two PDFs $f_{(\ell_1,\ell_2)}$ and~$f_{i,j}$ are directly connected: $f_{(\ell_1,\ell_2)}$ considers \emph{how often} populations~1 and~2 are represented, and~$f_{i,j}$ denotes \emph{which} populations are present. For example, $f_{(1,1)}(y)=f_{1,2}(y)$ and $f_{(0,2)}(y)=f_{2,2}(y)$.

\subsection*{PDF of 2-cell measurements of three populations ($n=2$, $T=3$)}
Next, we derive the PDF of a measurement~$y$ of a 2-cell pool, i.e. of~$Y=X_1+X_2$. Now, we assume three cell populations to be present in the tissue. Again, each of them is described by an individual distribution~$\gD_h$ for $h = 1,\ldots,T=3$:
\begin{align*}
X_j \overset{iid}{ \sim}  \begin{cases}
    \gD_1 & \text{w.p. } p_1\\
    \gD_2 & \text{w.p. } p_2 \\
	\gD_3 & \text{w.p. } 1-p_1-p_2,
    \end{cases}
\end{align*}
for~$j=1,2$ where $p_1,p_2 \in [0,1]$ and~$p_1+p_2 \leq 1$. Hence, the PDF of each $X_j$ is
\begin{align*}    
f_X(x) = p_1 f_{\gD_1}(x) + p_2 f_{\gD_2}(x) + p_3 f_{\gD_3} (x)
\end{align*}
with $p_3 = 1-p_1-p_2$. To determine the distribution of~$Y=X_1+X_2$, we again use the convolution of the single-cell PDFs:
\color{black}
\begin{align*}
f_Y(y) = \int_0^y &f_{X}(x_1) f_{X}(y-x_1) dx_1 \\
= \int_0^y &\Biggl( \Bigl[ p_1 f_{\gD_1}(x_1) + p_2 f_{\gD_2}(x_1) + p_3 f_{\gD_3}(x_1)\Bigr] \\
& \times\Bigl[ p_1 f_{\gD_1}(y-x_1) + p_2 f_{\gD_2}(y-x_1) + p_3 f_{\gD_3}(y-x_1) \Bigr] \Biggr) dx_1 \\
= \int _0^y & \Biggl(  p_1^2 f_{\gD_1}(x_1) f_{\gD_1}(y-x_1) \\
&+ p_2^2 f_{\gD_2}(x_1) f_{\gD_2}(y-x_1)+ p_3^2 f_{\gD_3}(x_1) f_{\gD_3}(y-x_1) \\
&+ p_1 p_2 f_{\gD_1}(x_1) f_{\gD_2}(y-x_1) + p_2 p_1 f_{\gD_2}(x_1) f_{\gD_1}(y-x_1)\\
&+ p_1 p_3 f_{\gD_1}(x_1) f_{\gD_3}(y-x_1) + p_3 p_1 f_{\gD_3}(x_1) f_{\gD_1}(y-x_1) \\
&+ p_2 p_3 f_{\gD_2}(x_1) f_{\gD_3}(y-x_1) + p_3 p_2 f_{\gD_3}(x_1) f_{\gD_2}(y-x_1) \Biggr)  dx_1,
\end{align*}
leading to
\begin{align*}
f_Y(y)
= p_1^2 & \int_0^y f_{\gD_1}(x_1) f_{\gD_1}(y-x_1) dx_1 \\
&+ p_2^2 \int_0^y f_{\gD_2}(x_1)f_{\gD_2}(y-x_1) dx_1 + p_3^2 \int_0^y f_{\gD_3}(x_1) f_{\gD_3}(y-x_1) dx_1\\
&+ p_1 p_2 \int_0^y f_{\gD_1}(x_1) f_{\gD_2}(y-x_1) dx_1 + p_2 p_1 \int_0^y f_{\gD_2}(x_1) f_{\gD_1}(y-x_1)dx_1\\
 &+ p_1 p_3 \int_0^y f_{\gD_1}(x_1) f_{\gD_3}(y-x_1)dx_1+ p_3 p_1 \int_0^y f_{\gD_3}(x_1) f_{\gD_1}(y-x_1) dx_1\\
&+ p_2 p_3 \int_0^y f_{\gD_2}(x_1) f_{\gD_3}(y-x_1)dx_1 + p_3 p_2 \int_0^y f_{\gD_3}(x_1) f_{\gD_2}(y-x_1) ) dx_1.
\end{align*}
Once more, we make use of the fact that $\int_0^y f_{\gD_i}(x_1) f_{\gD_j}(y-x_1) dx_1$ is the PDF of the sum~$Z_1+Z_2$ of two independent random variables, where $Z_1 \sim \gD_i$ and $Z_2 \sim \gD_j$ (now with~$i,j\in\{1,2,3\}$). As before, we denote this density by~$f_{i,j}$. Overall, we obtain
\begin{align*}
f_Y(y) = \sum_{i=1}^3 \sum_{j=1}^3 p_i p_j f_{i,j}(y), 
\end{align*}
or alternatively
\begin{align}
f_Y(y) = \sum_{\ell_1=0}^2 \sum_{\ell_2=0}^{2-\ell_1} \binom{2}{\ell_1} \binom{2-\ell_1}{\ell_2} p_1^{\ell_1} p_2^{\ell_2} p_3^{\ell_3} f_{(\ell_1,\ell_2,\ell_3)}(y),
\label{2cell3poppdf}
\end{align}
where $\ell_1, \ell_2, \ell_3= 2-\ell_1-\ell_2$ show how often cells of population 1, 2 and 3 are present in the pool. Again, $f_{(\ell_1,\ell_2,2-\ell_1-\ell_2)}(y)$ is connected to $f_{i,j}$.
For example, $f_{(0,1,1)}(y)=f_{2,3}(y)$ and~$f_{(2,0,0)}(y)=f_{1,1}(y)$.
\\

\subsection*{PDF of $n$-cell measurements of three populations ($n$ arbitrary, $T=3$)}
Next, we suppose that we measure pools of $n$~cells originating from three cell populations. Let~$Y=X_1+\ldots+X_n$. Then Equation~\eqref{2cell3poppdf} turns into
\begin{align}
f_Y(y) = \sum_{\ell_1=0}^n \sum_{\ell_2=0}^{n-\ell_1} \binom{n}{\ell_1} \binom{n-\ell_1}{\ell_2} p_1^{\ell_1} p_2^{\ell_2} p_3^{\ell_3} f_{(\ell_1,\ell_2,\ell_3)}(y), 
\label{ncell3poppdf}
\end{align}
where $p_3=1-p_1-p_2$ and $\ell_3=n-\ell_1-\ell_2$.

\subsection*{PDF of $n$-cell measurements of $T$ populations ($n$ and $T$ arbitrary)}
Finally, we extend Equation~\eqref{ncell3poppdf} to the most general case, where $n$-cell pools are measured from a tissue that consists of $T$~cell populations. Here, we obtain
\begin{align*}
f_Y(y)= \sum_{\ell_1=0}^n \sum_{\ell_2=0}^{n-\ell_1} \ldots \sum_{\ell_{T-1}=0}^{n-\ell_1-\ldots-\ell_{T-1}} \binom{n}{\ell_1} \binom{n-\ell_1}{\ell_2} \ldots \binom{n-\ell_1-\ldots-\ell_{T-2}}{\ell_{T-1}} p_1^{\ell_1}\ldots p_T^{\ell_T} f_{(\ell_1,\ldots,\ell_T)}(y), 
\end{align*}
where $p_T=1-p_1-\ldots-p_{T-1}$ and~$\ell_T= n-\ell_1-\ldots-\ell_{T-1}$. 
The binomial coefficients form together the multinomial coefficient
\begin{align*}
\binom{n}{\ell_1} \binom{n-\ell_1}{\ell_2}& \ldots \binom{n-\ell_1-\ldots-\ell_{T-2}}{\ell_{T-1}} \\
&= \frac{n! (n-\ell_1)! \cdots (n-\ell_1-\ldots-\ell_{T-2})!}{\ell_1! \ell_2! ... \ell_{T-1}!(n-\ell_1)!(n-\ell_1-\ell_2)!\cdots (n-\ell_1-\ldots-\ell_{T-1})!} \\
&= \frac{n!}{\ell_1! \ell_2! \cdots \ell_T!} = \binom{n}{\ell_1,\ldots,\ell_T}.
\end{align*}
Taken together, this leads to the final PDF~\eqref{multdistr} from Section~\ref{subsubsec:spmodel}:
\begin{align*}
f_Y(y)= \sum_{\ell_1=0}^n \sum_{\ell_2=0}^{n-\ell_1} \cdots \sum_{\ell_{T-1}=0}^{n-\ell_1-\ldots-\ell_{T-1}} \binom{n}{\ell_1,\ldots,\ell_T} p_1^{\ell_1}\cdots p_T^{\ell_T} f_{(\ell_1,\ldots,\ell_T)}(y). 
\end{align*}
The terms $\binom{n}{\ell_1,\ldots,\ell_T} p_1^{\ell_1}\cdots p_T^{\ell_T}$ are probabilities arising from the multinomial distribution and can be seen as multinomial weights of the densities~$f_{(\ell_1,\ldots,\ell_T)}(y)$.

\section{PDF of pooled gene expression for mixed pool size vectors} \label{mixed n_densities_dens}

When estimating a gene expression model from data, one may want to verify whether the estimated model adequately describes the data. In Figure~\ref{Fig:WellPred_HistGene2}, we did this by comparing the estimated PDF to the histogram of the data and to the true PDF: The orange curve was known since we treated the case of synthetic data. For the blue curve, we first estimated the model parameters and then plugged these in into the general model PDF. In case of a uniform pool size across all measurements, this procedure is straightforward. For a vector of pool sizes, i.\,e.\ a mix of e.\,g.\ 1-cell, 2-cell and 10-cell data, the PDF (see e.\,g.\  Figure~\ref{Fig:OverviewSPML}B) is less obvious. We calculate this function as follows:

\begin{itemize}
\item For each cell number contained in the $n$-vector, calculate the PDF of the respective pool size and plug in the parameter estimates.
\item Calculate the weighted sum of these PDFs~---~weighted according to the times the respective pool size occurs in the $n$-vector.
\end{itemize}
The resulting PDF approximates the PDF of a sample where the observations are based on the pool sizes of the considered $n$-vector. While this PDF describes a mixture distribution with randomly drawn pool sizes (according to the weights used), we in our applications assume the pool sizes to be known for each measurement.
\begin{Code}
mix.d.sum.of.mixtures.LNLN <- function(y, n.vector, p.vector, mu.vector, 
+    sigma.vector){
+        densmix <- matrix(0, ncol = length(y), nrow = length(n.vector))
+        for(i in 1:length(n.vector)){
+            densmix[i, ] <- d.sum.of.mixtures.LNLN(y = y, n = n.vector[i], 
+                p.vector = p.vector, mu.vector = mu.vector,
+                sigma.vector = sigma.vector, logdens = FALSE)
+        }
+        Dens<-colSums(densmix)/length(n.vector)
+    }
\end{Code}

\section{Transformation of population probabilities} \label{app:logit}

As described in Section~\ref{subsec:mlpe}, we transform the model parameters before optimization of the likelihood function such that no constraints of the parameter space have to be accounted for. Here, we provide details about the transformation of the population probabilities.

In case of two populations, there is only one parameter~$p\in[0,1]$ that determines the probabilities~$p$ and~$1-p$ of populations~1 and~2.
We transform~$p$ to
$$
w = \text{logit}(p)= \log \left( \frac{p}{1-p} \right) 
\in\mathbb{R}
$$
and later back-transform this via
$$
p = \text{logit}^{-1}(w)=\text{expit}(w)=\frac{\exp(w)}{1+\exp(w)} \in[0,1]\ .
$$
The advantage of $w$ as compared to $p$ is the unrestricted range $\mathbb{R}$ instead of~$[0,1]$.

In case of $T>2$ populations, the probabilities~$p_1,\ldots,p_T$ have to fulfill~$p_h\in[0,1]$ for \linebreak all~$h=1,\ldots,T$ and~$\sum_{h=1}^T p_h=1$.
We set $\tilde{p}_h= p_1 + \cdots + p_h$ and use the following transformations 
$$
w_h = \text{logit} \left( \frac{p_1 +\cdots + p_h}{p_1 +\cdots + p_{h+1}} \right) =  \text{logit} \left( \frac{\tilde{p}_h}{\tilde{p}_{h+1}} \right) \in\mathbb{R}  \qquad \text{for all } h \in 1,\ldots, {T-1}.
$$

For the back-transformations, we start at $h={T-1}$ and calculate 
$$
\tilde{p}_h = \text{expit}(w_h) \, \tilde{p}_{h+1}  \in[0,1]\,  \qquad \text{for all } h \in {T-1},\ldots, 1
$$
in reverse order. We set $\tilde{p}_T=1$ to ensure that the probabilities sum up to one. Additionally, one has~$\tilde{p}_h \leq \tilde{p}_{h+1}$ as $\text{expit}(w_h) \in [0,1]$ for all $h \in 1,\ldots, {T-1}$. Obviously, $p_1 = \tilde{p}_1$, and the remaining population probabilities are given by
$$
p_h = \tilde{p}_h -\tilde{p}_{h-1}\in[0,1]\,  \qquad \text{for all } h \in 2,\ldots, {T}.
$$

The (back-)transformations are implemented in \code{transform.par()} and~\code{backtransform.par()}.

\section[Derivation of sample composition probabilities]{Derivation of sample composition probabilities}\label{App:WellPred}

We describe how to predict the population composition of a cell pool, as applied in Section~\ref{subsec:predictpooldecomposition}. A key formula here is the conditional probability of a cell composition given the measured gene expression, which we derive here. We use the following notations and assumptions:
\begin{itemize}
\item The overall gene expression of a cell pool is denoted by~$Y$ and assumed a continuous random variable with PDF~$f_Y (y)$ .
\item $L=(L_1,\ldots,L_T)$ denotes the specific cell population combinations, i.\,e.\ $L_i$ is the number of cells of population~$i$ for all $i=1,\ldots,T$, within a pool of~$L_1+\ldots+L_T$ cells. $L$~is a discrete random vector with PMF $P(L=\ell)$.
\item $f_{Y|L=\ell}(y)$ is the conditional PDF of the overall gene expression in a cell pool whose composition is known to equal~$\ell$. For shorter notation, this was referred to as $f_{ ( \ell_1,\ell_2,\ldots,\ell_T)}\left(y_i|\mm{\theta} \right)$ in Section \ref{subsubsec:spmodel}.
\item In turn, $P(L=\ell|Y=y)$ is the conditional PMF of the cell pool composition given the pool gene expression measurement~$Y=y$. 
\end{itemize}
We use Bayes' theorem to derive the latter PMF:
\begin{equation}
P(L=\ell|Y=y) = \frac{f_{Y|L=\ell}(y) P(L=\ell)}{f_Y (y)}=\frac{f_{Y|L=\ell} (y)P(L=\ell)}{\sum_{j \in J} f_{Y|L=j}(y) P(L=j)},
\label{eq:condPMF}
\end{equation}
where $J$ is the set of all possible compositions of the cell pool, i.\,e.\  the set of all vectors~$(j_1,\ldots,j_T)$ with~$j_i\in\mathbb{N}_0$ and~$j_1+\ldots+j_T=\ell_1+\ldots+\ell_T$. 
\newline The terms in Equation~\eqref{eq:condPMF} depend on the population probabilities~$\mm{p}=(p_1,\ldots,p_T)$ and the gene expression model (in this work: LN-LN, rLN-LN, or EXP-LN), characterized by its respective parameters. We assume the expression model to be fixed and denote all model parameters (including~$\mm{p}$) by~$\mm{\theta}$. In practice, $\mm{\theta}$ is unknown, and hence we use its maximum likelihood estimates here.
\newline

Given the estimate $\hat{\mm{p}}$ of $\mm{p}$, $L=\ell=(\ell_1,\ldots,\ell_T)$ approximately follows a multinomial distribution with parameters $n=\ell_{1}+\ldots+\ell_T$ and~$\hat{\mm{p}}$. The PMF of the cell pool composition~$(\ell_1,\ldots,\ell_T)$ hence reads
$$P(L=(\ell_1,\ldots,\ell_T))=  { n  \choose  \ell_{1},\ell_{2},\ldots, \ell_{T} } \hat{p}_1^{\ell_1}\hat{p}_2^{\ell_2} \cdots \hat{p}_T^{\ell_T},$$
where ${ n  \choose  \ell_{1},\ell_{2},\ldots, \ell_T }  = {\frac {n!}{\ell_1!\,\ell_2!\cdots \ell_T!}}$ is the multinomial coefficient. With this, the conditional PMF of the cell pool composition given the pooled gene expression measurement~$Y$ reads:
\begin{align}
P(L=\ell|Y=y) &= \frac{f_{Y|L=\ell}(y;\hat{\mm{\theta}})  
 { n  \choose  \ell_{1},\ell_{2},\ldots, \ell_{T} } \hat{p}_1^{\ell_1}\hat{p}_2^{\ell_2} \cdots \hat{p}_T^{\ell_T}}{f_Y (y;\hat{\mm{\theta}})}\\ \nonumber
 &=\frac{f_{Y|L=\ell} (y;\hat{\mm{\theta}}) { n  \choose  \ell_{1},\ell_{2},\ldots, \ell_{T} } \hat{p}_1^{\ell_1}\hat{p}_2^{\ell_2} \cdots \hat{p}_T^{\ell_T}}{\sum_{j \in J} f_{Y|L=j}(y;\hat{\mm{\theta}})
 { n  \choose  j_{1},j_{2},\ldots, j_{T} } \hat{p}_1^{j_1}\hat{p}_2^{j_2} \cdots \hat{p}_T^{j_T}}.
\end{align}

\color{black}

\section[Interactive Functions]{Interactive Functions}\label{App:INteractiveFN}
As indicated in Section \ref{sec:Il}, we show an example of the interactive functions for data generation, \code{stochasticProfilingData()}, and parameter estimation, \code{stochasticProfilingML()}.

\subsection*{Synthetic data generation: \code{stochasticProfilingData()}}
\begin{Schunk}
\begin{Sinput}
R> library("stochprofML")
R> set.seed(10)
R> stochprofML::stochasticProfilingData()
\end{Sinput}
\begin{Soutput}
This function generates synthetic data from the stochastic profiling model.
In the following, you are asked to specify all settings. By pressing 'enter',
you choose the default option.

---------
Please choose the model you would like to generate data from:
 1: LN-LN
 2: rLN-LN
 3: EXP-LN
(default: 1)
\end{Soutput}
\begin{Sinput}
R> 1
\end{Sinput}
\begin{Soutput}
---------
Please enter the number of different populations you would like to consider:
(default: 2)
\end{Soutput}
\begin{Sinput}
R> 2
\end{Sinput}
\begin{Soutput}
---------
Please enter the number of stochastic profiling observations you wish to generate:
(default: 100)
\end{Soutput}
\begin{Sinput}
R> 1000
\end{Sinput}
\begin{Soutput}
---------
Next we enter the number of cells that should enter each observation, 
which case do you want:
 1: all observations should contain the same number of cells, or 
 2: each observation contains a different number of cells 
(default: 1).
\end{Soutput}
\begin{Sinput}
R> 1
\end{Sinput}
\begin{Soutput}
---------
Please enter the number of cells that should enter each sample:
(default: 10)
\end{Soutput}
\begin{Sinput}
R> 10
\end{Sinput}
\begin{Soutput}
---------
Please enter the number of co-expressed genes you would like to collect
in one cluster
(default: 1)
\end{Soutput}
\begin{Sinput}
R> 1
\end{Sinput}
\begin{Soutput}
---------
Please enter the probabilities for each of the 2 populations, e.g. type
0.62, 0.38
or
0.62 0.38.
It is recommended to choose the order of the populations such that
(for the first gene, if there is more than one)
log-mean for population 1 >= log-mean for population 2 >= ...
\end{Soutput}
\begin{Sinput}
R> 0.62, 0.38
\end{Sinput}
\begin{Soutput}
---------
Please enter the log-means for each of the 2 populations, e.g. type
0.47, -0.87.
\end{Soutput}
\begin{Sinput}
R> 0.47, -0.87
\end{Sinput}
\begin{Soutput}
---------
Please enter the log-standard deviation, which is the same for all
populations, i.e. type e.g.
0.03
irrespectively of the number of populations.
\end{Soutput}
\begin{Sinput}
R> 0.03
\end{Sinput}
\begin{Soutput}
---------
Would you like to write the generated dataset to a file? (Be careful not to
overwrite any existing file!) Please type 'yes' or 'no'.
\end{Soutput}
\begin{Sinput}
R> yes
\end{Sinput}
\begin{Soutput}
Please enter a valid path and filename, either a full path, e.g.
D:/Users/lisa.amrhein/Desktop/mydata.txt
or just a file name, e.g.
mydata.txt.
The current directory is
D:/Users/lisa.amrhein/Desktop.
\end{Soutput}
\begin{Sinput}
test.txt
\end{Sinput}
\begin{Soutput}
Hit <Return> to see next plot:
\end{Soutput}
\begin{Sinput}
R> <Return>
\end{Sinput}
\includegraphics[width=0.8\textwidth]{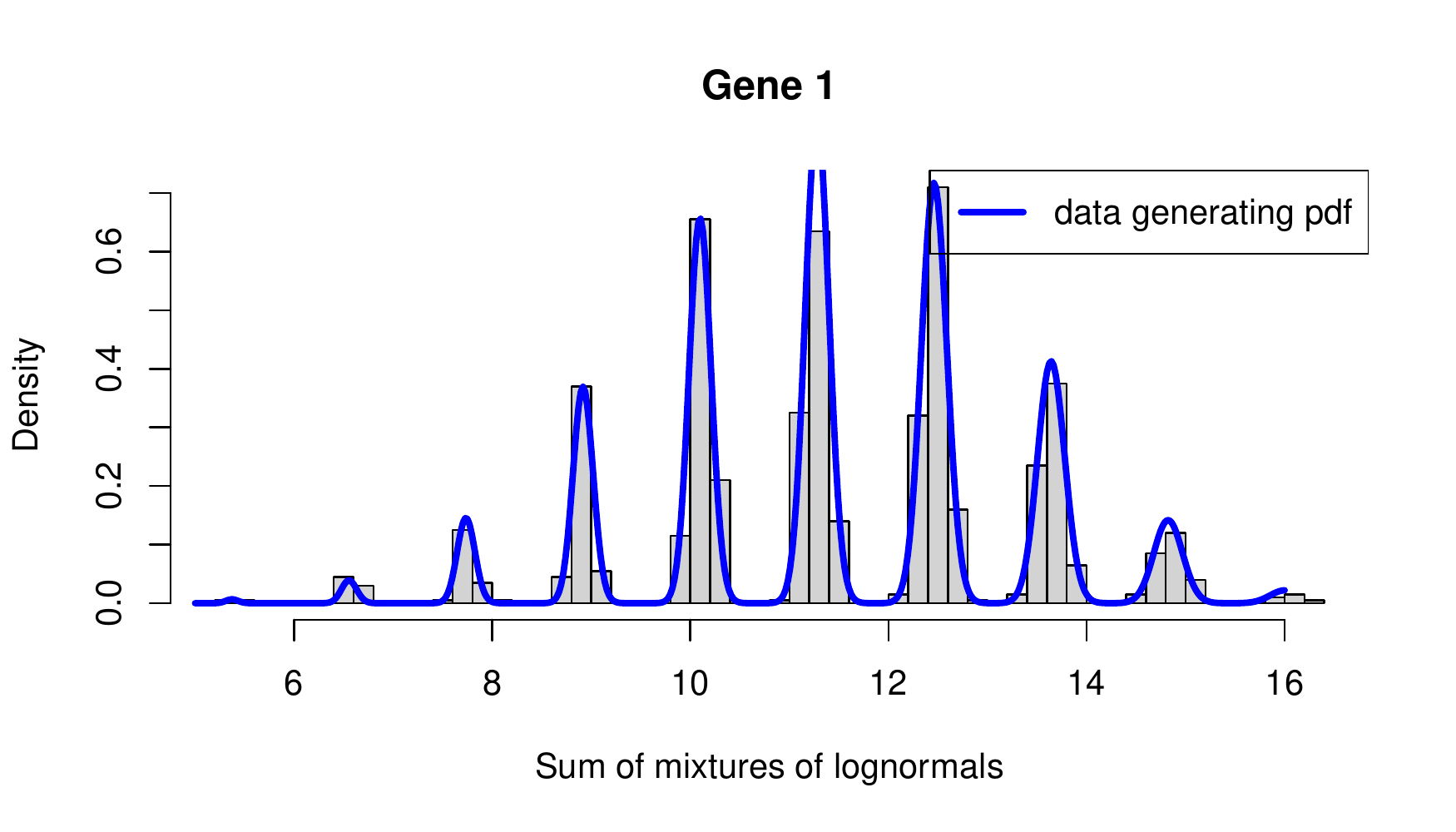}
\begin{Soutput}
The dataset has been generated. The first 50 observations are:

gene 1
observation 1  12.444051
observation 2  12.406390
observation 3  12.412508
observation 4  11.274005
observation 5  12.455432
observation 6  11.255553
observation 7  11.281351
observation 8  13.587420
observation 9  11.222902
observation 10 12.586923
observation 11 12.278757
observation 12 14.746502
observation 13 12.313032
observation 14 12.387004
observation 15 10.007876
observation 16  9.978140
observation 17 11.461335
observation 18 10.015642
observation 19 12.575700
observation 20 12.411795
observation 21  9.965103
observation 22 10.139115
observation 23 12.437228
observation 24 10.039271
observation 25 12.489048
observation 26 11.218177
observation 27 13.217671
observation 28 13.683550
observation 29  8.877630
observation 30 12.550239
observation 31  9.922716
observation 32 10.110123
observation 33  8.720011
observation 34 10.237112
observation 35 11.223407
observation 36 12.468208
observation 37 12.561782
observation 38 15.013314
observation 39 10.151065
observation 40 12.405660
observation 41 13.490263
observation 42 14.675084
observation 43 11.432289
observation 44 13.565312
observation 45 10.076739
observation 46 12.414651
observation 47  8.861460
observation 48 12.401696
observation 49 11.271324
observation 50 13.750237


The full dataset has been written to
test.txt.
It is also stored in the .Last.value variable.
\end{Soutput}
\end{Schunk}

\subsection*{Parameter estimation: \code{stochasticProfilingML()}}

\begin{Schunk}
\begin{Sinput}
R> library("stochprofML")
R> set.seed(20)
R> stochprofML::stochasticProfilingML()
\end{Sinput}
\begin{Soutput}
This function performs maximum likelihood estimation for the stochastic
profiling model. In the following, you are asked to enter your data and
specify some settings. By pressing 'enter', you choose the default option.

---------
How would you like to input your data?
 1: enter manually
 2: read from file
 3: enter the name of a variable
(default: 1).
\end{Soutput}
\begin{Sinput}
R> 2
\end{Sinput}
\begin{Soutput}
---------
The file should contain a data matrix with one dimension standing for genes
and the other one for observations. Fields have to be separated by tabs or white
spaces, but not by commas. If necessary, please delete the commas in the
text file using the 'replace all' function of your text editor.

Please enter a valid path and filename, either a full path, e.g.
D:/Users/lisa.amrhein/Desktop/mydata.txt
or just a file name, e.g.
mydata.txt.
The current directory is
D:/Users/lisa.amrhein/Desktop.
\end{Soutput}
\begin{Sinput}
R> test.txt
\end{Sinput}
\begin{Soutput} 
Does the file contain column names? Please enter 'yes' or 'no'.
\end{Soutput}
\begin{Sinput}
R> yes
\end{Sinput}
\begin{Soutput}
Does the file contain row names? Please enter 'yes' or 'no'.
\end{Soutput}
\begin{Sinput}
R> no
\end{Sinput}
\begin{Soutput}
Do the columns stand for different genes or different observations?
 1: genes
 2: observations.
\end{Soutput}
\begin{Sinput}
R> 1
\end{Sinput}
\begin{Soutput}
This is the head of the dataset (columns contain different genes):
       gene.1
[1,] 12.44405
[2,] 12.40639
[3,] 12.41251
[4,] 11.27400
[5,] 12.45543
[6,] 11.25555


If the matrix does not look correct to you, there must have been an error
in the answers above. In this case, please quit by pressing 'escape' and call
stochasticProfilingML() again.

The file contained the following gene names:
gene.1
\end{Soutput}
\begin{Sinput}
R> no
\end{Sinput}
\begin{Soutput}
---------
Please choose the model you would like to estimate:
 1: LN-LN
 2: rLN-LN
 3: EXP-LN
(default: 1)
\end{Soutput}
\begin{Sinput}
R> 1
\end{Sinput}
\begin{Soutput}
---------
Please enter the number of different populations you would like to estimate:
(default: 2)
\end{Soutput}
\begin{Sinput}
R> 2
\end{Sinput}
\begin{Soutput}
---------
Please enter the number of cells that entered each observation, either
 1: all observations contain the same number of cells, or 
 2: each observation contains a different number of cells 
(default: 1).
\end{Soutput}
\begin{Sinput}
R> 1
\end{Sinput}
\begin{Soutput}
---------
Please enter the number of cells that should enter each observation:
(default: 10)
\end{Soutput}
\begin{Sinput}
R> 10
\end{Sinput}
\begin{Soutput}

***** Estimation started! *****


Maximum likelihood estimate (MLE):
             p_1 mu_1_gene_gene.1 mu_2_gene_gene.1            sigma 
          0.6142           0.4690          -0.8650           0.0290 

Value of negative log-likelihood function at MLE:
1124.932

Violation of constraints:
none

BIC:
2277.496

Approx. 95% confidence intervals for MLE:
                       lower       upper
p_1               0.60461644  0.62369584
mu_1_gene_gene.1  0.46779634  0.47020366
mu_2_gene_gene.1 -0.87085629 -0.85914371
sigma             0.02774407  0.03031279

Top parameter combinations:
        p_1 mu_1_gene_gene.1 mu_2_gene_gene.1 sigma   target
[1,] 0.6142            0.469           -0.865 0.029 1124.932
[2,] 0.6142            0.469           -0.865 0.029 1124.932
[3,] 0.6141            0.469           -0.865 0.029 1124.932
[4,] 0.6142            0.469           -0.865 0.029 1124.932
[5,] 0.6142            0.469           -0.865 0.029 1124.933
[6,] 0.6142            0.469           -0.865 0.029 1124.933

\end{Soutput}
\end{Schunk}

\section[Details on Simulation Studies]{Details on Simulation Studies}\label{App:SimStudy1}
The general procedure of the simulation studies shown in Section~\ref{sec:simstudies} is to first generate synthetic datasets with some predefined population parameters and frequencies using\linebreak \code{r.sum.of.mixtures()}. Thereby datasets with either fixed or varying pool sizes are generated, i.\,e.\ the numbers of cells contained in one pool are fixed or vary from cell pool to cell pool within a dataset. 
Next, we assume that we do not know the predefined model parameters and estimate them using \code{stochprof.loop()}. Then we compare the estimates of the parameters in different ways, e.\,g.\ how they are influenced by increasing cell numbers or how their variance differs when the dataset was generated with differing population parameters. 

Here, we give an overview about the different model parameter settings and pool sizes used in data generation: We use datasets with fixed pool sizes that contain single-cells, 2 cells,
5 cells, 10 cells, 15 cells, 20 cells or 50 cells. Additionally, we chose two types of datasets with varying pool sizes. The first contains small cell pools with 1, 2, 5 and 10 cells, the second contains larger cell pools with 10, 15, 20 and 50 cells. Thus, in total we have nine different cell pool settings that we use for data generation. 

In all simulation studies, we use the LN-LN model with five different parameter settings, given in Table \ref{tab:OverviewParameter}. While the first set is considered to be the default, each of the other parameter sets differs from it in one of the population parameters.

\begin{table}[h!]
\centering
\begin{tabular}{lrrrr}
\toprule
  & $p$ & $\mu_1$ & $\mu_2$ & $\sigma$\\
\midrule
 Set 1 & 0.2 & 2 & 0 & 0.2\\
Set 2 & 0.1 & 2 & 0 & 0.2\\
  Set 3 & 0.4 & 2 & 0 & 0.2\\
Set 4 & 0.2 & 2 & 1 & 0.2\\
 Set 5 & 0.2 & 2 & 0 & 0.5\\
\bottomrule
\end{tabular}
\caption{Overview of the five model parameter settings used in the simulation studies in Section~\ref{sec:simstudies}.}
\label{tab:OverviewParameter}
\end{table}

Taken together, for each of the nine cell pool settings and each of the five parameter settings 1,000 datasets are generated using \code{r.sum.of.mixtures.LNLN()}, so that in total we
have \texttt{5*9*1000\ =} \ensuremath{4.5\times 10^{4}} simulated
datasets.

\subsection*{Impact of pool sizes}\label{impact-of-pool-sizes}

In the first simulation study (Section~\ref{subsec:Est_diff_setting}), we investigate how parameter estimation is influenced by increasing cell numbers within the cell pools. The results for parameter set~1 are displayed in the main part of the manuscript. Here, we show the corresponding results for the remaining four parameter settings. 

\begin{figure}[!h]
\begin{center}
\includegraphics[width=0.85\textwidth]{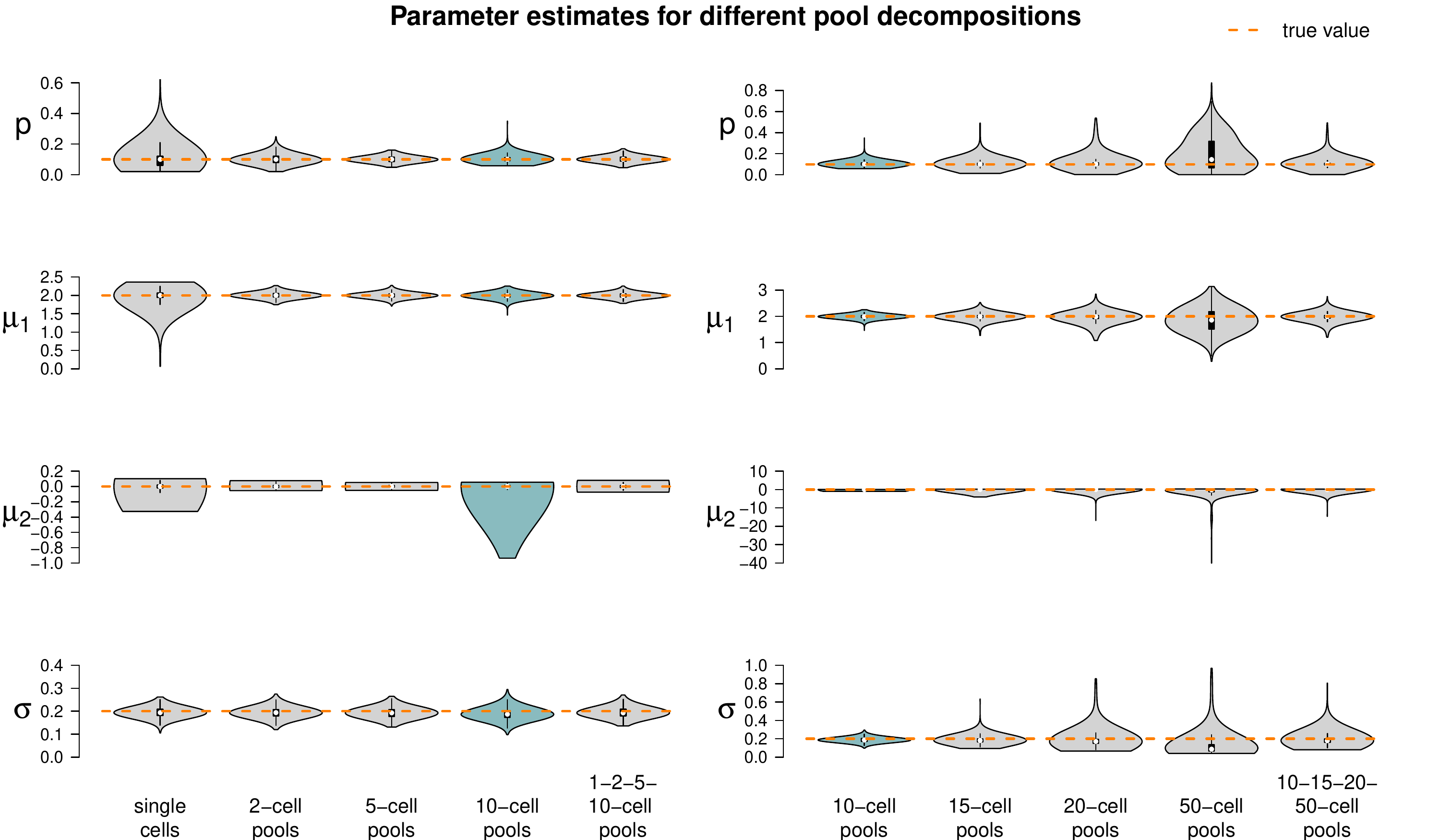} 
\caption{Estimated parameters of LN-LN-model on 9,000 simulated datasets, i.\,e.\ 1,000 datasets of each pool composition generated with parameter set~2 (see Table~\ref{tab:OverviewParameter}). \emph{Left:} Accumulated parameter fits of the single-cell, 2-cell, 5-cell, 10-cell and mixture of single-, 2-, 5- and 10-cell pools. \emph{Right:} Results of the 10-cell pools are repeated (turquoise violins), next to those of the larger pool sizes, namely 15-, 20-, 50-cells and their mixture. Each violin is based on 1,000 parameter estimates. The true parameter values are marked in orange.}
\label{Fig:Simstudy1_Set2_Overview}
\end{center}
\end{figure}

In the second parameter setting, the fraction of the first population was reduced to \(10\%\) as compared to the first parameter setting. The results are shown in Figure~\ref{Fig:Simstudy1_Set2_Overview}. They are similar to the results of the first parameter set in Figure~\ref{Fig:Simstudy1_Set1_Overview}. For set~2, however, single cells lead to large variance of estimates, supposedly due to the small sample size of 50 in combination with the small probability (10\%) of the first population: We can only expect five single cells of the first population to be measured on average. In some datasets, this will be too low to estimate the parameters of the first population and/or their proportion satisfactorily. Consequently, the violins of the single-cell estimates show a higher variance, especially for the estimates of the parameters of the first population.

\begin{figure}[!h]
\begin{center}
\includegraphics[width=0.85\textwidth]{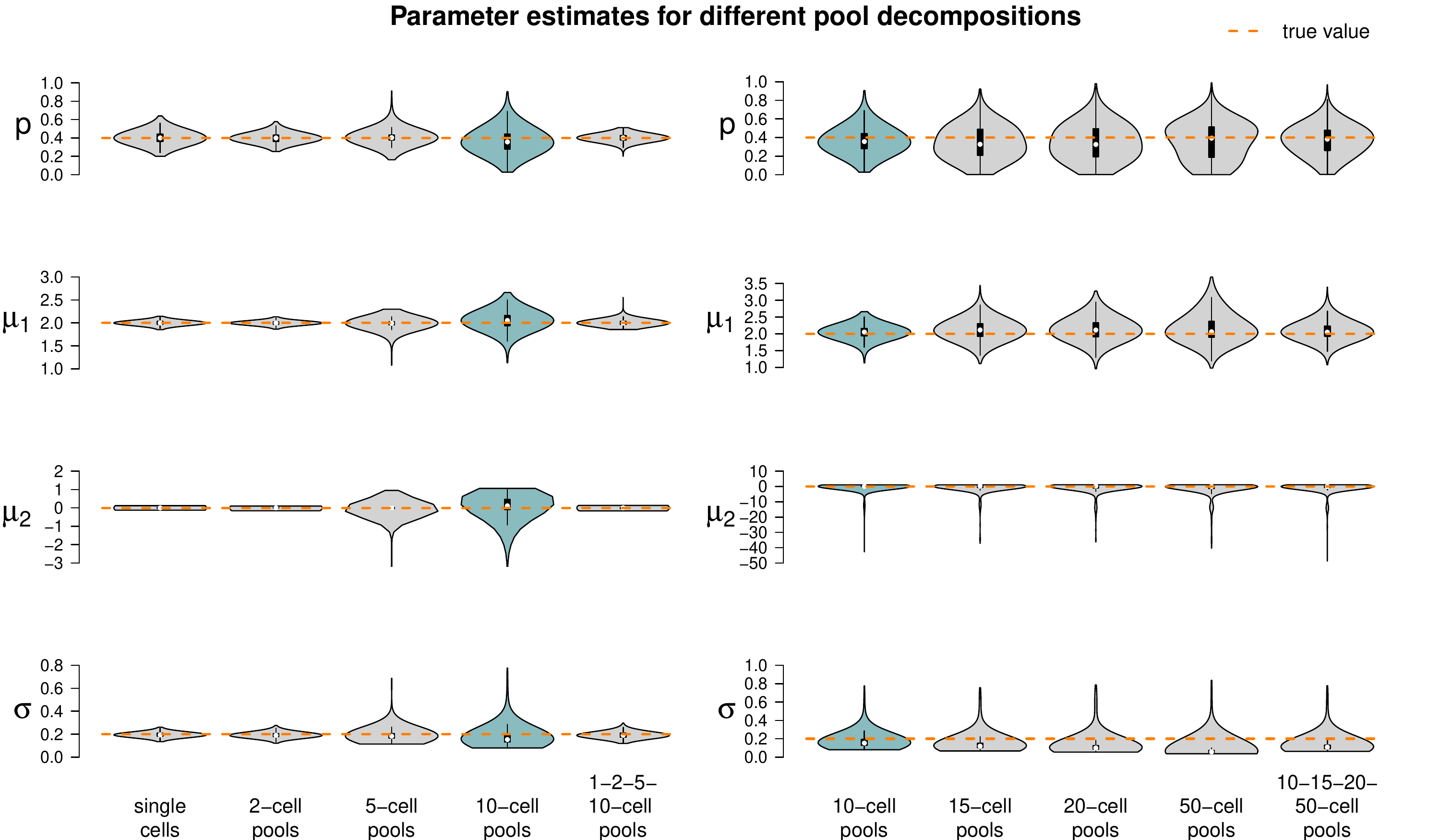} 
\caption{Parameter estimates as in Figure~\ref{Fig:Simstudy1_Set3_Overview} but for parameter set~3 (see Table~\ref{tab:OverviewParameter}).}
\label{Fig:Simstudy1_Set3_Overview}
\end{center}
\end{figure}

In the third parameter setting, the fraction of the first population was increased to \(40\%\). The resulting estimates are shown in Figure~\ref{Fig:Simstudy1_Set3_Overview}. In this setting, both populations are similarly frequent; hence, it seems plausible that the single-cell estimates show similar variability as for example the 2-cell estimates. The estimates of the mixed pools of the lower cell numbers provide estimates that are as accurate as the ones for single-cell and 2-cell data. From a pool size of five cells on, the estimates vary strongly. Apparently, low cell numbers are advisable if a tissue is not dominated by one cell population.

\begin{figure}[!h]
\begin{center}
\includegraphics[width=0.85\textwidth]{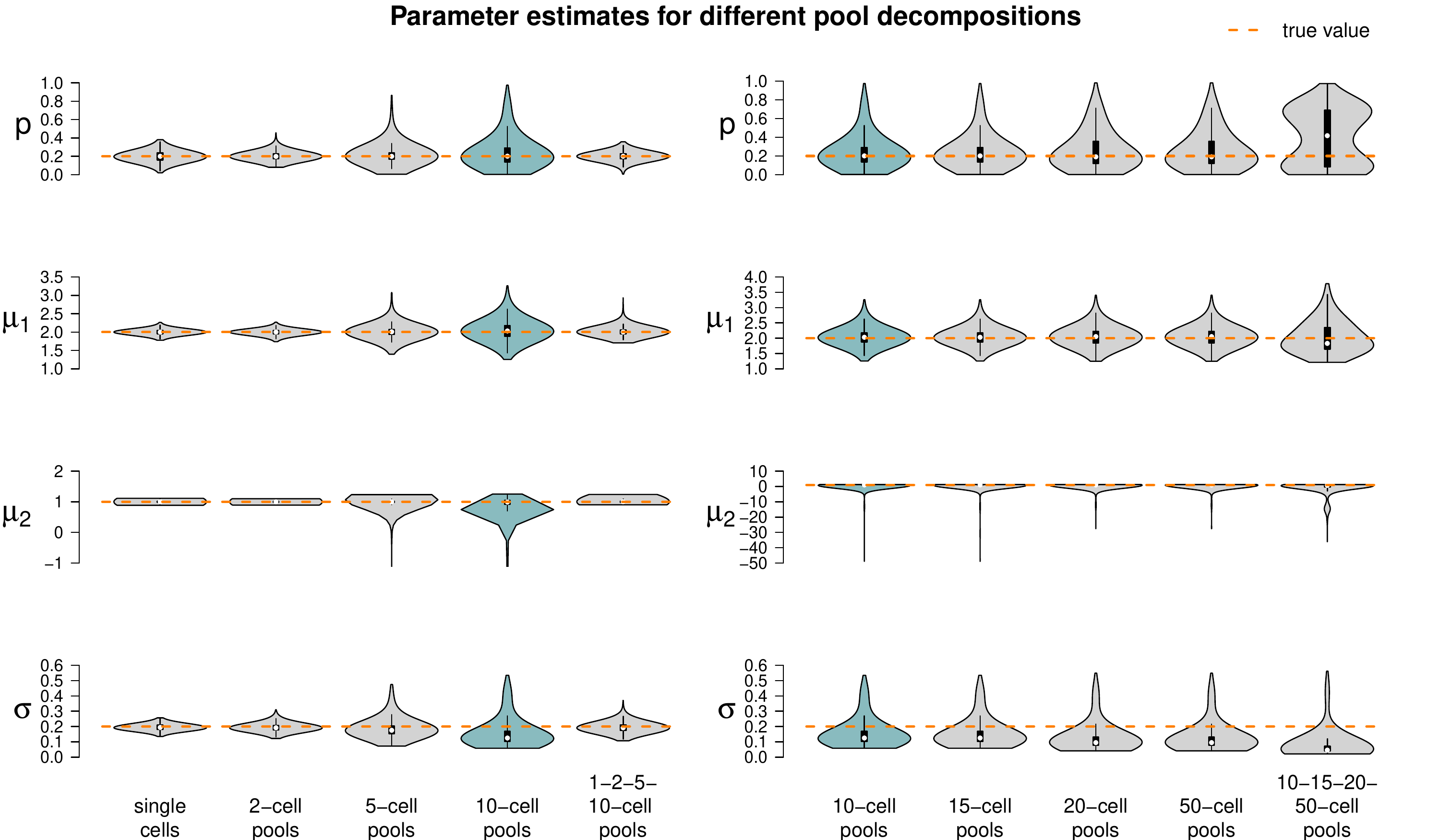} 
\caption{Parameter estimates as in Figure~\ref{Fig:Simstudy1_Set3_Overview} but for parameter set~4 (see Table~\ref{tab:OverviewParameter}).}
\label{Fig:Simstudy1_Set4_Overview}
\end{center}
\end{figure}

In the fourth parameter setting, \(\mu_2\) is increased to \(1\) and thus larger than in the first parameter setting. The two populations are more similar. The resulting estimates are shown in Figure~\ref{Fig:Simstudy1_Set4_Overview}. Starting from a pool size of 10 cells, it seems as if the variance of the estimates did not increase any more. The estimates for the mixed pools with larger cell numbers can sometimes not distinguish the populations, therefore the violin of~$p$ is bi-modal. We draw the same conclusion as for two populations with similar frequencies that more similar populations should be investigated in pools with lower cell numbers because their individual expression profile is blurred for small pool sizes already.

\begin{figure}[!h]
\begin{center}
\includegraphics[width=0.85\textwidth]{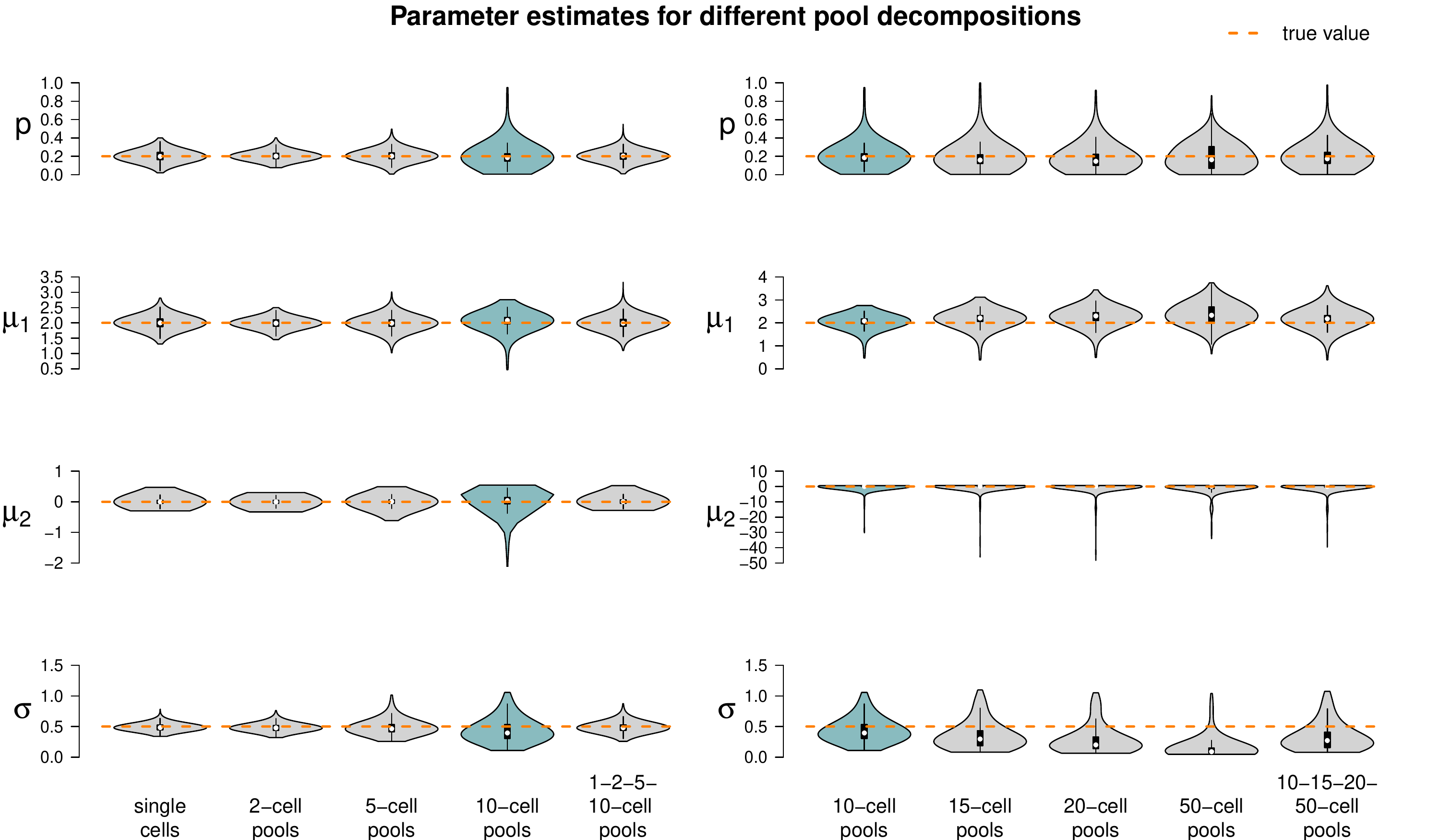} 
\caption{Parameter estimates as in Figure~\ref{Fig:Simstudy1_Set3_Overview} but for parameter set~5 (see Table~\ref{tab:OverviewParameter}).}
\label{Fig:Simstudy1_Set5_Overview}
\end{center}
\end{figure}

Finally, we investigate the effect of different pool sizes in the fifth parameter set, where the log-sd \(\sigma\) of both populations is increased to \(0.5\). The resulting estimates of the model parameters are shown in Figure~\ref{Fig:Simstudy1_Set5_Overview}. With an increase of~$\sigma$, both populations have broader distributions. It appears that there is an increase in variance in the estimates between the 5-cell and the 10-cell measurements. Increasing cell numbers in the pools mainly influences the estimate of~$\sigma$, which is increasingly underestimated. 

\clearpage
\subsection*{Impact of parameter values}\label{impact-of-parameter changes}

In Section~\ref{subsec:Est_diff_setting2}, we investigate the influence of the model parameter values on the estimation performance while fixing the pool size. In the main part of the manuscript, we presented results for 10-cell pools (see Figure~\ref{Fig:Simstudy2_10_Overview}). Here, corresponding analyses for the remaining eight cell pool sizes ($n\in\{1,2,5,15,20,50\}$ and two mixtures) are shown. 

\begin{figure}
\begin{center}
\includegraphics[width=0.85\textwidth]{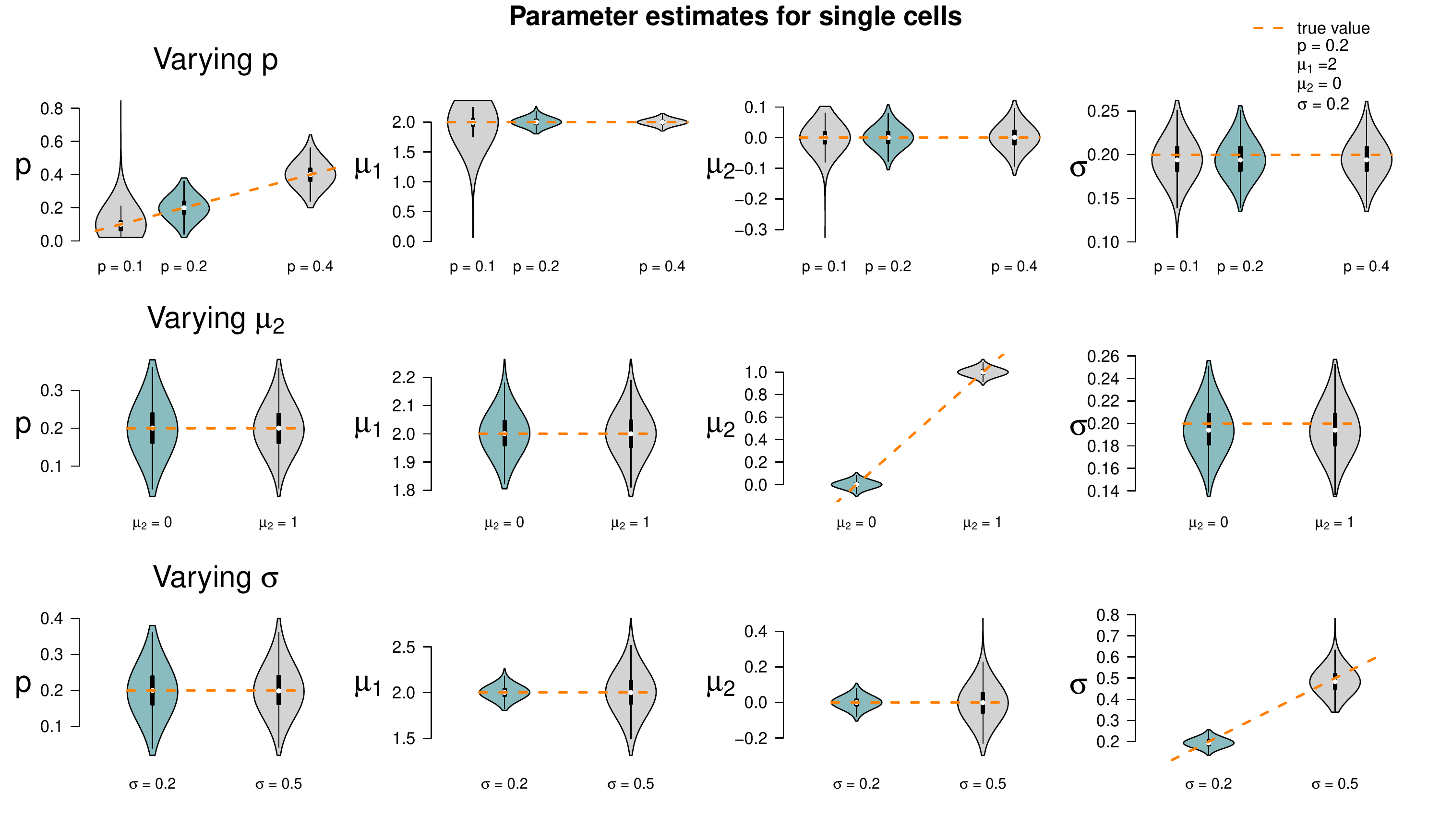} 
\caption{Parameter estimates for single-cell data and varying parameter values: Synthetic data is generated using the LN-LN model for varying values of~$p$, $\mu_2$ and~$\sigma$. Results for the standard setting $p=0.2$, $\mu_1 = 2$, $\mu_2 = 0$ and~$\sigma = 0.2$ are shown in turquoise, results for four more settings in grey. For each setting, we generate 1,000 synthetic datasets and back-infer the model parameters. Violin plots summarize the 1,000 estimates. The underlying true parameter values are marked in orange.}
\label{Fig:Simstudy2_1_Overview}
\end{center}
\end{figure}

\begin{figure}
\begin{center}
\includegraphics[width=0.85\textwidth]{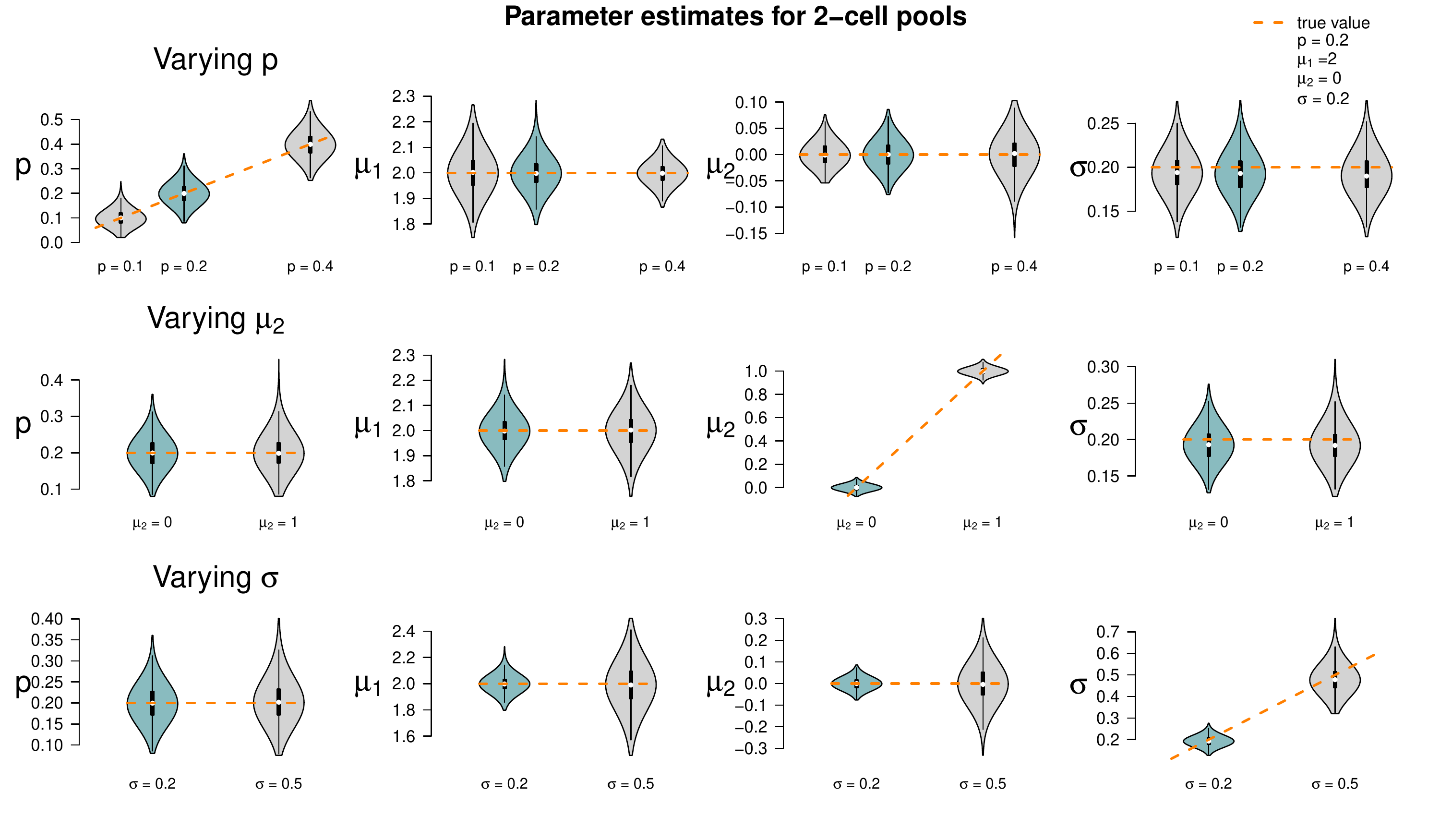} 
\caption{As Figure~\ref{Fig:Simstudy2_1_Overview}, but for 2-cell data.}
\label{Fig:Simstudy2_2_Overview}
\end{center}
\end{figure}

Results for single-cell and 2-cell pools look alike (Figures~\ref{Fig:Simstudy2_1_Overview} and~\ref{Fig:Simstudy2_2_Overview}). As discussed before, the variance of the estimates become large for a small value of~$p$ in combination with the small pool sizes. For both single-cell and 2-cell data, varying~$\mu_2$ does not affect the estimation accuracy of the estimation, whereas a larger value of~$\sigma$ leads to higher variance of all parameter estimates but for~$p$.

\begin{figure}
\begin{center}
\includegraphics[width=0.85\textwidth]{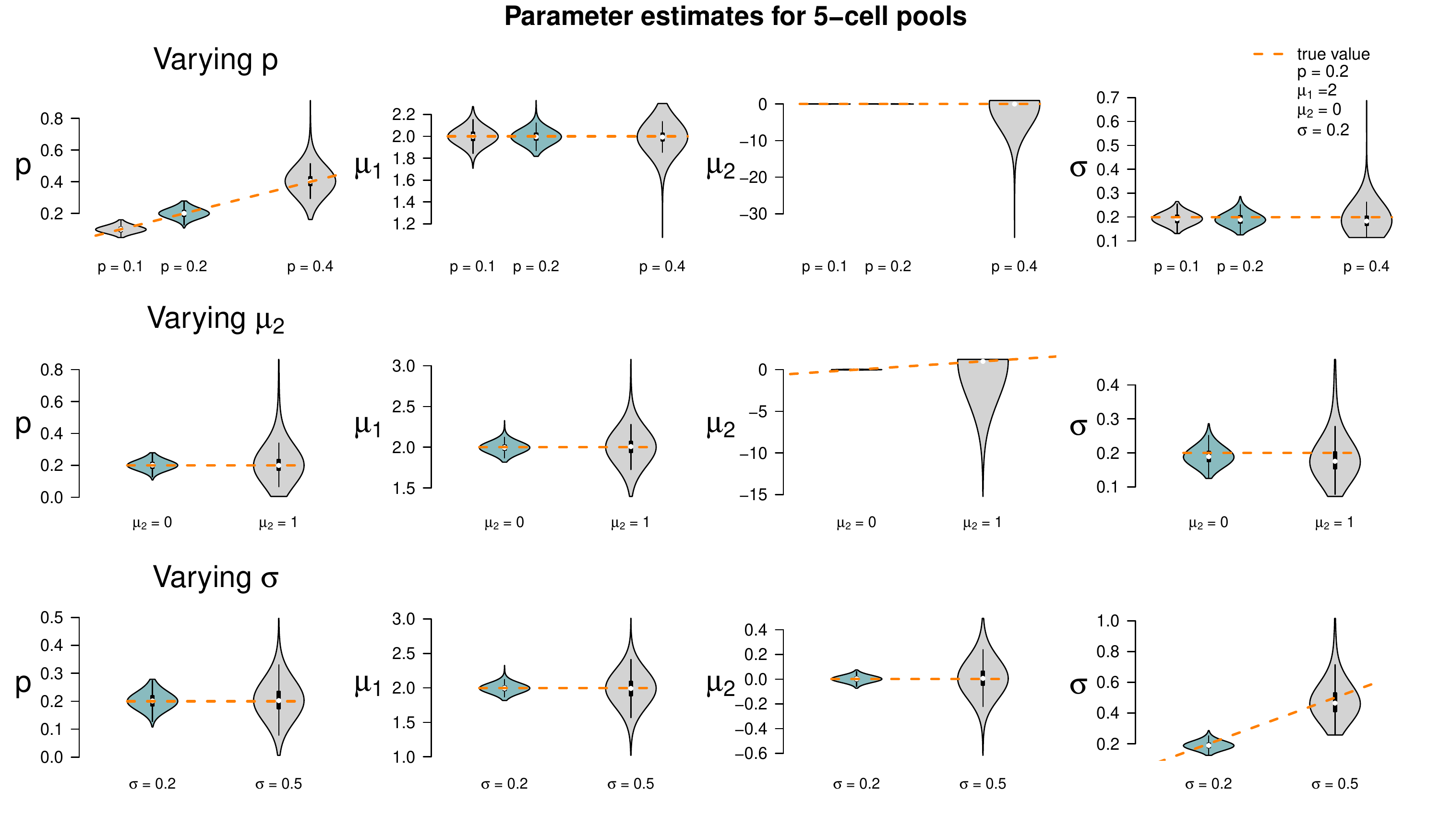} 
\caption{As Figure~\ref{Fig:Simstudy2_1_Overview}, but for 5-cell data.}
\label{Fig:Simstudy2_5_Overview}
\end{center}
\end{figure}

\begin{figure}
\begin{center}
\includegraphics[width=0.85\textwidth]{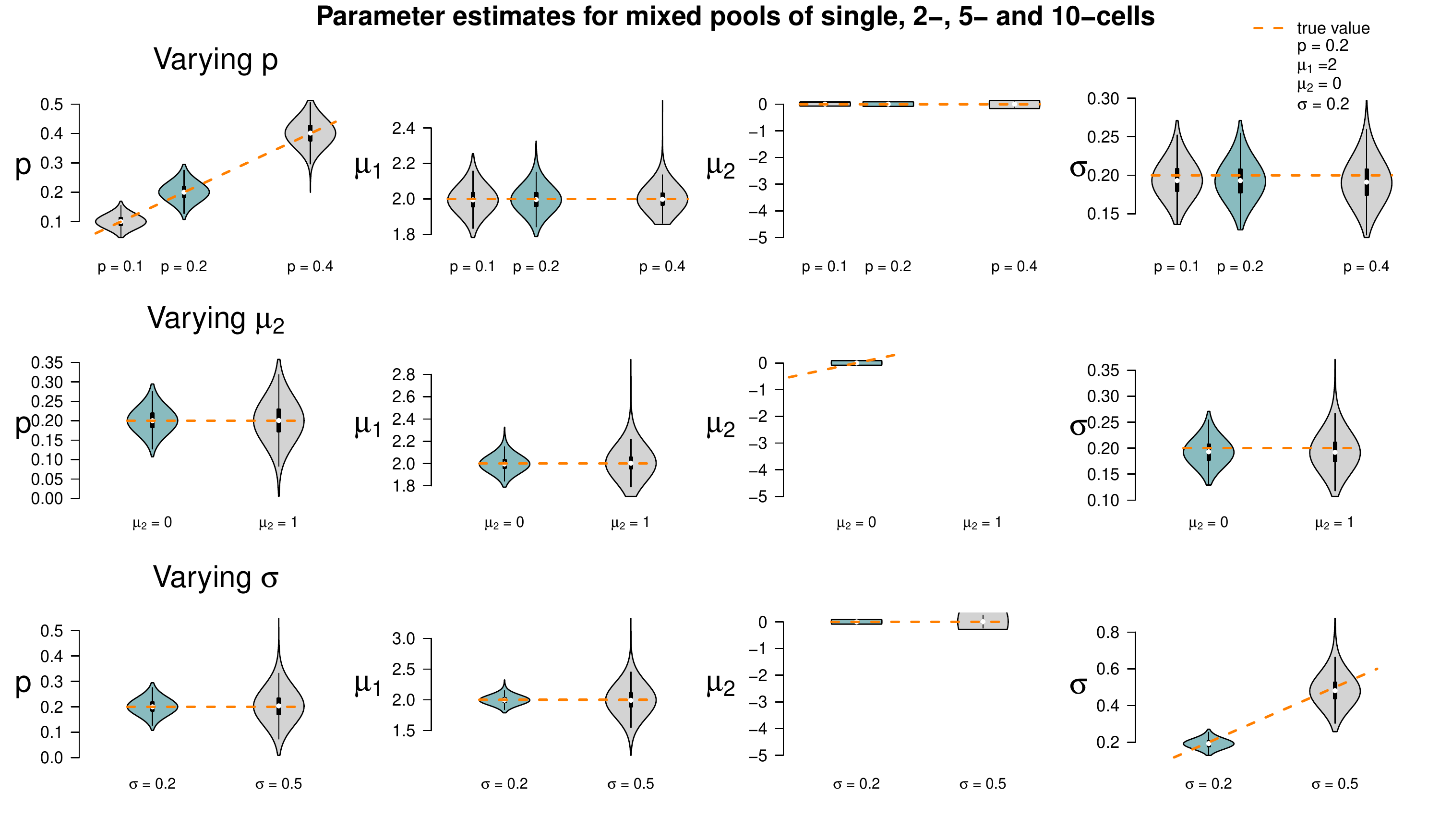} 
\caption{As Figure~\ref{Fig:Simstudy2_1_Overview}, but for a mixture of single-, 2-, 5- and 10-cell data.}
\label{Fig:Simstudy2_mix1_Overview}
\end{center}
\end{figure}

In contrast to this, the 5-cell data results in a different pattern (Figure~\ref{Fig:Simstudy2_5_Overview}): As compared to the estimates from the standard setting, the estimates show a larger variance.
The mixture of small cell pool numbers (Figure~\ref{Fig:Simstudy2_mix1_Overview}), however, lead to similar results as the pure 2-cell datasets. 

\begin{figure}
\begin{center}
\includegraphics[width=0.85\textwidth]{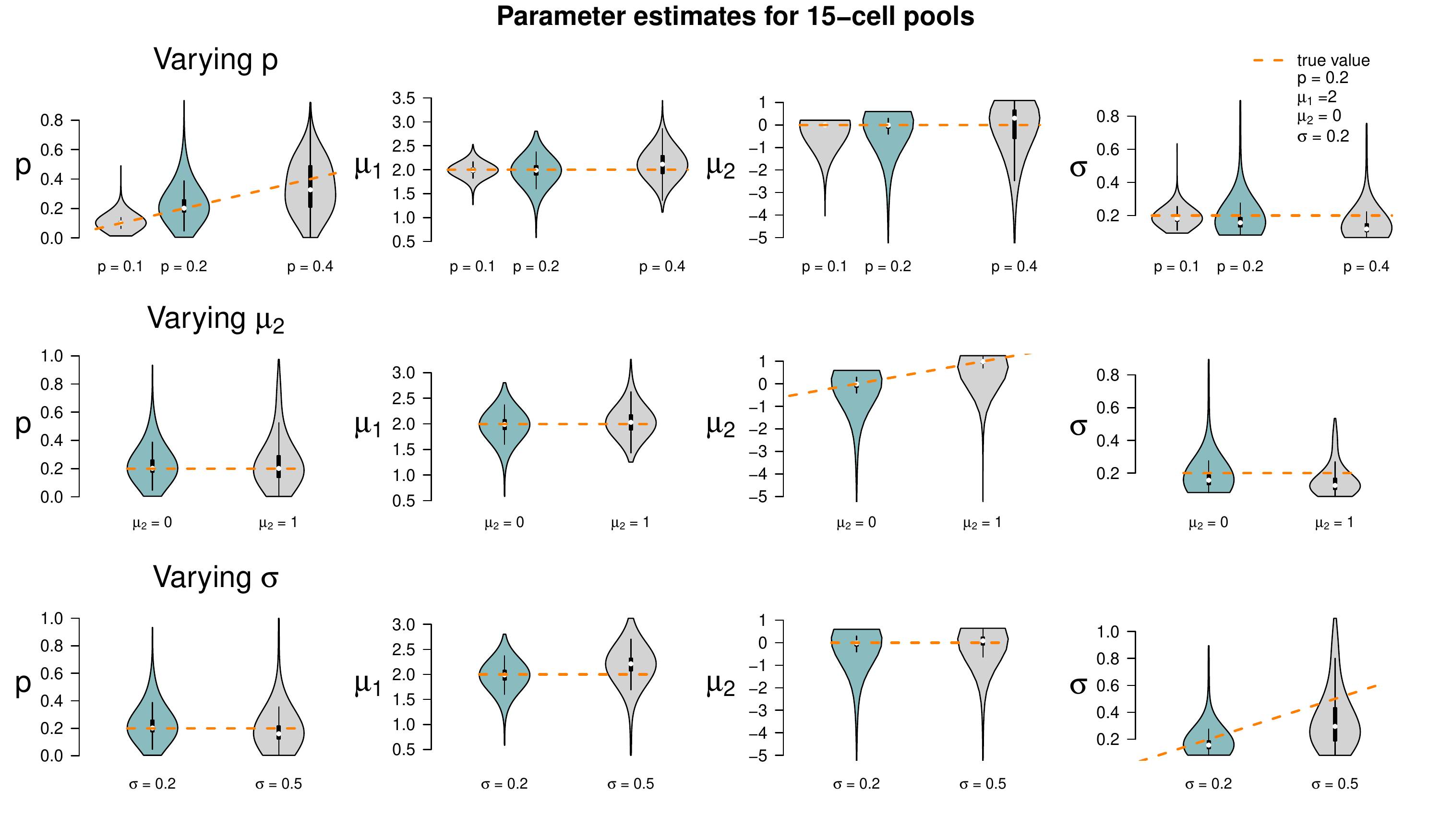} 
\caption{As Figure~\ref{Fig:Simstudy2_1_Overview}, but for 15-cell data.}
\label{Fig:Simstudy2_15_Overview}
\end{center}
\end{figure}

\begin{figure}
\begin{center}
\includegraphics[width=0.85\textwidth]{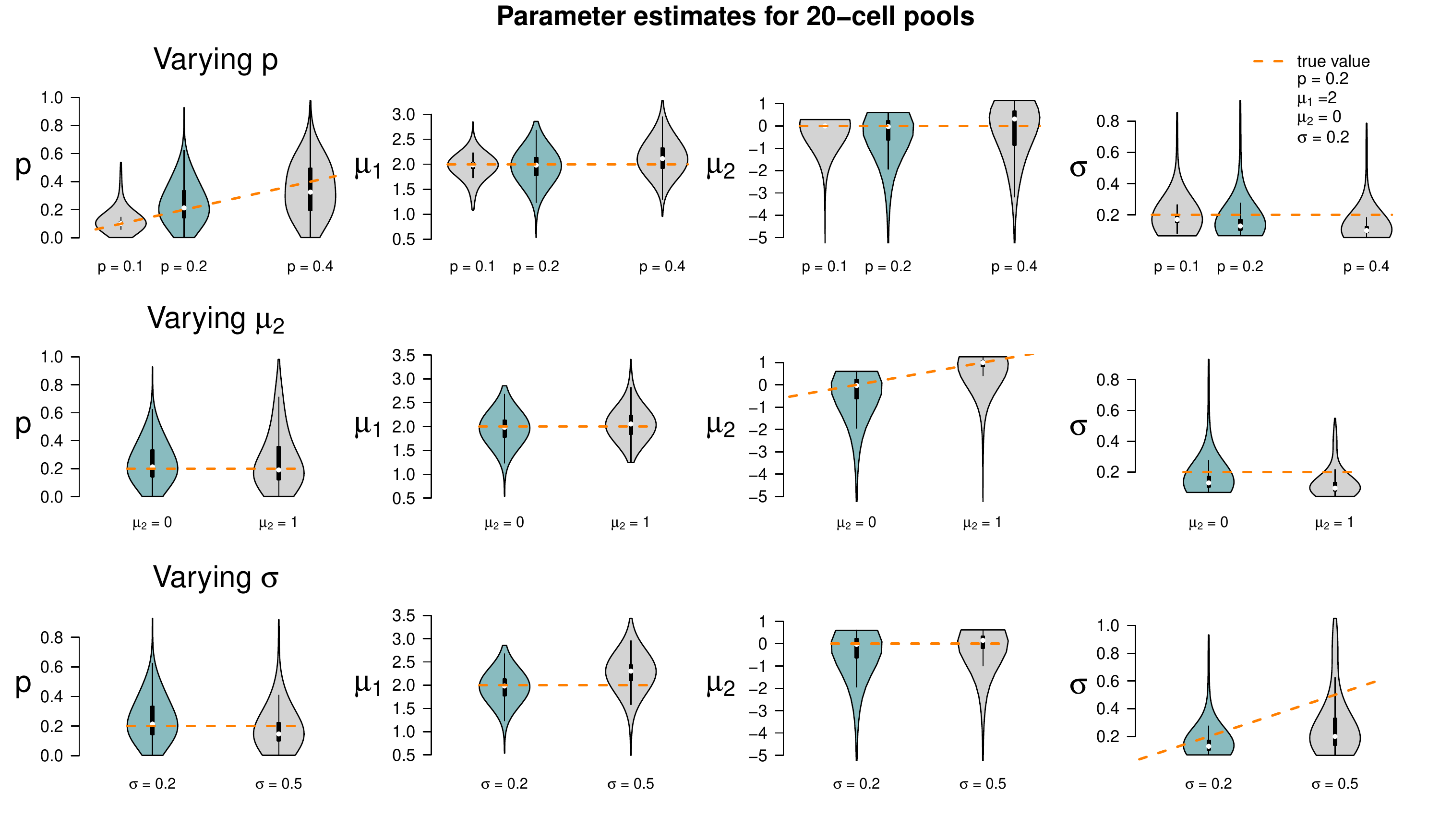} 
\caption{As Figure~\ref{Fig:Simstudy2_1_Overview}, but for 20-cell data.}
\label{Fig:Simstudy2_20_Overview}
\end{center}
\end{figure}

\begin{figure}
\begin{center}
\includegraphics[width=0.85\textwidth]{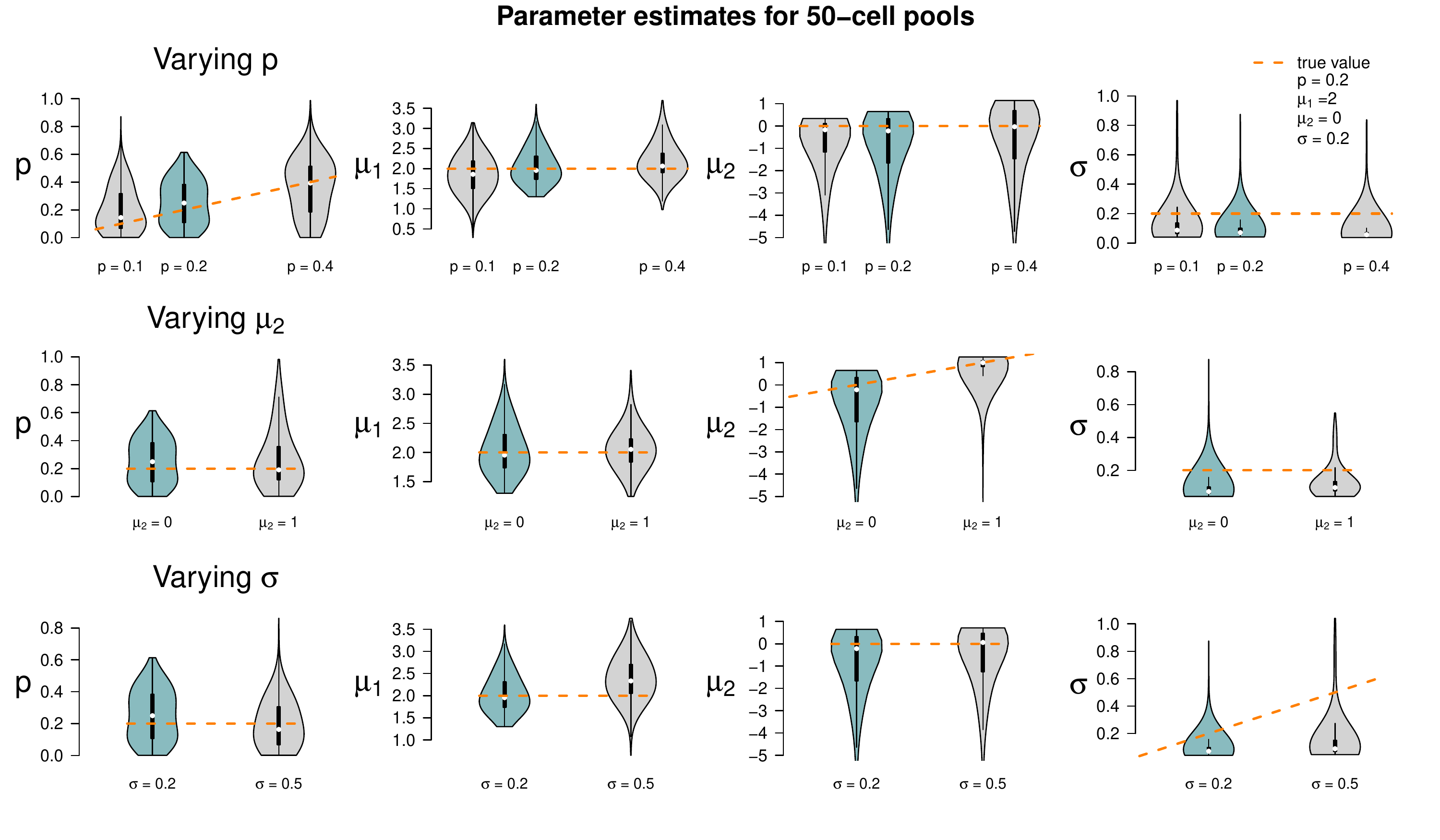} 
\caption{As Figure~\ref{Fig:Simstudy2_1_Overview}, but for 50-cell data.}
\label{Fig:Simstudy2_50_Overview}
\end{center}
\end{figure}

\begin{figure}
\begin{center}
\includegraphics[width=0.85\textwidth]{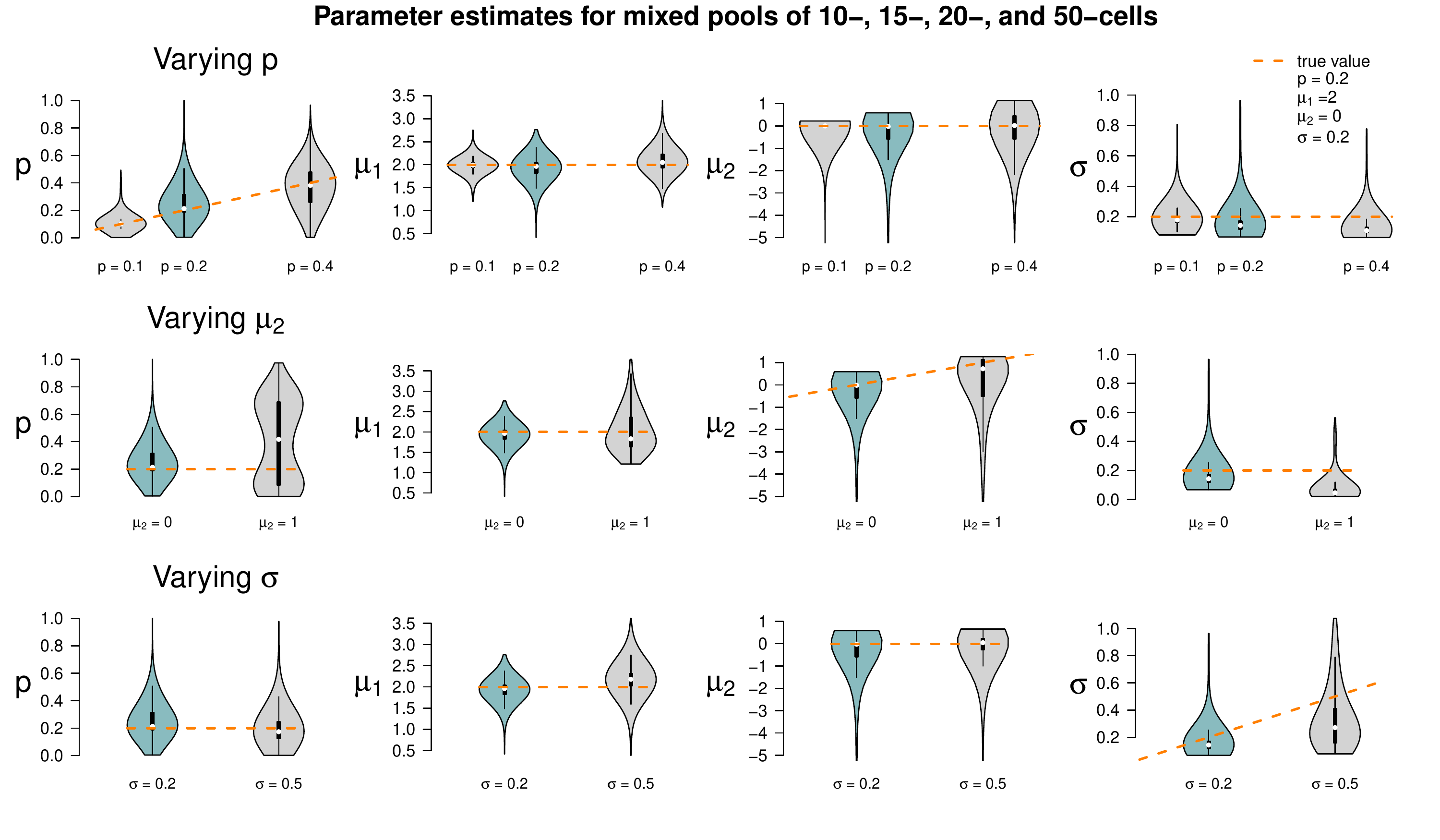} 
\caption{As Figure~\ref{Fig:Simstudy2_1_Overview}, but for a mixture of 10-, 15-, 20- and 50-cell data.}
\label{Fig:Simstudy2_mix2_Overview}
\end{center}
\end{figure}

Figure~\ref{Fig:Simstudy2_15_Overview} displays the results for the 15-cell data. For most parameter combinations, the variance of the estimates does not change dramatically. The most accurate estimates are achieved for small~$p$, the least accurate ones for large~$\sigma$, in which case~$\sigma$ gets underestimated. The same holds true for the 20- and 50-cell datasets (Figures~\ref{Fig:Simstudy2_20_Overview} and~\ref{Fig:Simstudy2_50_Overview}), with even larger variance.
For the mixture of large cell pools (Figure~\ref{Fig:Simstudy2_mix2_Overview}), estimation performance is comparable to the one for the pure 50-cell measurements.

\end{appendix}

\clearpage
\end{document}